\RequirePackage{lineno}
\documentclass[twocolumn,nofootinbib,floatfix]{revtex4-1}

\usepackage{amsmath,amssymb}
\usepackage{cancel}
\usepackage{graphicx}
\usepackage{epsfig}
\usepackage{amsmath,amsfonts,amssymb}
\usepackage{colortbl}
\usepackage{hhline}
\usepackage{pifont}
\usepackage{multirow}

\newcommand{\ptj}{p_{TJ}}

\begin{document}

\newcommand{\Gent}{{\tt GenT}$^3$}
\newcommand{\gentx}{{\tt GenT}$^x$}

\title{Anomaly detection from mass unspecific jet tagging}

\author{J.~A.~Aguilar-Saavedra}
\affiliation{Departamento de F\'{\i}sica Te\'{o}rica y del Cosmos, Universidad de Granada, E-18071 Granada, Spain}
\affiliation{Instituto de F\'{\i}sica Te\'{o}rica UAM-CSIC, Campus de Cantoblanco, E-28049 Madrid, Spain}

\begin{abstract}
We introduce a novel anomaly search method based on (i) jet tagging to select interesting events, which are less likely to be produced by background processes; (ii) comparison of the untagged and tagged samples to single out features (such as bumps produced by the decay of new particles) in the latter. We demonstrate the usefulness of this method by applying it to a final state with two massive boosted jets: for the new physics benchmarks considered, the signal significance increases an order of magnitude, up to a factor of 40. We compare to other anomaly detection methods in the literature and discuss possible generalisations. 
\end{abstract}

\maketitle

\tableofcontents

\section{Introduction}

Naturalness arguments suggest that models beyond the Standard Model (SM) of particle interactions involve mass scales not much larger than the electroweak scale. If this is indeed the case, they can be tested at the Large Hadron Collider (LHC) and its future upgrades. However, a complication results from the fact that not all particles in the models must necessarily lie at about the same mass scale, and even if they do, mass hierarchies are expected. Consequently, the production and decay of new particles may take place in multiple processes and cascade decay chains, leading to a plethora of experimental signatures, for which dedicated searches are simply not possible.

Anomaly detection methods may become very useful to fill this gap. With this approach, one considers a specific final state characterised by a number of detector `objects' (electrons, muons, photons, jets) and some tool --- often involving multivariate analysis --- that tries to find whether, for example, in that final state there is a specific phase space region unusually populated. Or if there is overabundance of objects whose aspect is not as typically expected from SM production processes. Several anomaly detection methods have been proposed in the literature, either multivariate generic taggers for the detection of `exotic' jets~\cite{Aguilar-Saavedra:2017rzt,Heimel:2018mkt,Farina:2018fyg,Dillon:2019cqt,Aguilar-Saavedra:2020sxp,Amram:2020ykb,Cheng:2020dal,Aguilar-Saavedra:2020uhm,Dillon:2021nxw,Atkinson:2021nlt} or anomaly-detection tools that consider the kinematics of the event, in some cases including also some jet substructure observables~\cite{Collins:2018epr,DAgnolo:2018cun,DeSimone:2018efk,Hajer:2018kqm,Cerri:2018anq,Blance:2019ibf,DAgnolo:2019vbw,Nachman:2020lpy,Andreassen:2020nkr,Knapp:2020dde,Dillon:2020quc,Khosa:2020qrz,Bortolato:2021zic,Hallin:2021wme,Ostdiek:2021bem}.
An anomaly detection method selects some specific final state and compares data with predictions (either from theory or even from data itself, extrapolating from other phase space region) to test the likelihood for the presence of some unknown signal.

We introduce in this paper a new concept for anomaly detection, which follows two steps:
\begin{enumerate}
\item[(i)] Select a sample of {\em interesting events}, based on the presence of detector objects that are less likely to be produced by background processes. In other words, a tagging is performed to spot events with (one or more) less-likely objects, and that tagged sample is retained for further analysis.
\item[(ii)] Single out features such as mass bumps in that sample, by comparing with the untagged one. 
\end{enumerate}
We dub this method as Single Out Features in Interesting Events (SOFIE). The concept is quite general and relies on some magnitude to quantify the `SM-likeliness' of an object, so that `interesting' and `reference' samples can be built based on the value of that magnitude, for one or more objects. Although not all final states of interest at the LHC may benefit from this approach, there is a wide variety of signals for which it can be useful. In this paper we will develop and investigate the SOFIE anomaly detection method specifically for final states with hadronic boosted jets. These jets are tagged to select `interesting' events containing jets that are less likely to be produced by QCD interactions. Subsequently, the sample of events with tagged jets is compared to the sample of untagged jets to detect the presence of mass bumps. The latter sample is expected to have a negligible amount of signal, and thus serves as reference.

The method presented here is complementary to the ones that use kinematical sidebands~\cite{Collins:2018epr,Nachman:2020lpy,Hallin:2021wme} and has different uncertainties, in particular the calibration of the tagger. The high sensitivity in situations where the signal to background ratio is small, of the order of $10^{-3}$ and below, make the SOFIE method especially useful. Besides, we note that an anomaly search by looking for different kinematical features in data and Monte Carlo evaluated with statistical estimators has been proposed in Ref.~\cite{DeSimone:2018efk}.

The first step in the SOFIE anomaly detection method can be performed with either a supervised~\cite{Aguilar-Saavedra:2017rzt,Aguilar-Saavedra:2020sxp,Aguilar-Saavedra:2020uhm} or non-supervised~\cite{Heimel:2018mkt,Farina:2018fyg,Dillon:2019cqt, Amram:2020ykb,Cheng:2020dal,Dillon:2021nxw} jet tagger. We choose the first option, and use the Mass Unspecific Supervised Tagging (MUST) method developed in Ref.~\cite{Aguilar-Saavedra:2020uhm}. The tagger built using this approach, which is presented in section~\ref{sec:2}, is able to identify multi-pronged jets across a wide range of jet mass and transverse momentum. Compared to previous work~\cite{Aguilar-Saavedra:2020uhm}, the tagger is generalised to discriminate between quarks and gluons, and its kinematical range of application extended. We also note that in Ref.~\cite{Aguilar-Saavedra:2020uhm} taggers for jets with a specific prongness were built. They have a better performance than fully-generic taggers and might also be used for less general anomaly searches.

The second step involves the comparison of tagged and untagged samples. We use a mixed-sample estimator, which we present and discuss in detail in section~\ref{sec:3}. Despite being quite a simple statistical estimator, it is quite powerful to detect bumps in density distributions. We note that multivariate methods could also be used in this step. They might offer some advantage at high dimensionalities of the feature space. Further explorations in this direction are out of the scope of this work.

The jet mass is an excellent discriminant between jets originated by (background) quarks or gluons, and by (signal) massive particles. In section~\ref{sec:4} we discuss in detail two methods for mass decorrelation of the tagger. We note that the need for mass decorrelation is specific for this development of the SOFIE method for final states with boosted jets. In applications of the SOFIE method to objects other than boosted jets, an analogous decorrelation might be necessary as well.

 An application is presented in section~\ref{sec:5} for final states with two boosted jets, and a comparison with other methods in section~\ref{sec:6}. Given the generality of the concept, straightforward generalisations to multi-jet final states are possible. A complication may arise when there are non-neglibible SM backgrounds involving boosted weak bosons $W,Z$. We discuss in section~\ref{sec:7} how to proceed in this case, and show that the method is robust against systematic uncertainties in the sample composition. Our results are discussed in section~\ref{sec:8}. Additional details of the analysis are given in three apendices.

\section{Mass unspecific supervised jet tagging}
\label{sec:2}

In Ref.~\cite{Aguilar-Saavedra:2020uhm} we developed a supervised generic jet tagger, dubbed as {\tt GenT}, to discriminate between quark/gluon one-pronged jets\footnote{For brevity, we often refer to jets initiated by quarks and gluons simply by quark and gluon jets, though at higher orders additional hard coloured partons are present.} and multi-pronged jets from boosted `signal' massive particles, using a neural network (NN). The jet mass range in the training was limited to $m_J \in [50,250]$ GeV, and the transverse momentum to $\ptj \in [200,2200]$ GeV. Here we train two different taggers:
\begin{itemize}
\item A three-class tagger \Gent\ that discriminates among gluon jets, quark jets, and multi-pronged jets, trained in the range  $m_J \in[10,500]$ GeV, $\ptj \in [200,2200]$ GeV.
\item A two-class tagger \gentx\ with the same design and architecture as that in Ref.~\cite{Aguilar-Saavedra:2020uhm} but with the jet mass range extended to $m_J \in[10,500]$ GeV.
\end{itemize}

QCD jets, which constitute the background for the tagger, are generated with {\scshape MadGraph}~\cite{Alwall:2014hca}. For \Gent\ the gluon and quark jets are separately generated in the processes $pp \to gg$, $pp \to qq$, and for \gentx\ in the inclusive process $pp \to jj$. Event samples are generated in 100 GeV bins of $p_T$, starting at $[200,300]$ GeV and up to $[2.2,2.3]$ TeV or $p_T \geq 2.2$ TeV, depending on the sample. For \Gent\ the training and validation of the NN is done with events from sets GS1 and QS1, and for \gentx\ the events are taken from set JS1 (see appendix~\ref{sec:a} for notation and details on the different background sets). We note that large event samples are required in order to have sufficient events at high $m_J$. 

The model independent (MI) data considered as signal are generated with {\scshape Protos}~\cite{protos} in the process $pp \to ZS$, with $Z \to \nu \nu$ and $S$ a scalar. We consider the six decay modes
\begin{align}
&  \text{4-pronged (4P):} && S \to u \bar u u \bar u \,,~ S \to b \bar{b} b \bar{b} \, \notag \\
& \text{3-pronged (3P):} && S \to F \,\nu \,; \quad F \to u d d \,,~ F \to u d b \, \notag \\
& \text{2-pronged (2P):} && S \to u \bar u \,,~ S \to b \bar b \,,
\label{ec:MIdata}
\end{align}
to generate multi-pronged jets ($F$ is a colour-singlet fermion). To remain as model-agnostic as possible, the $S$ and $F$ decays are implemented with a flat matrix element, so that the decay weight of the different kinematical configurations only corresponds to the four-, three- or two-body phase space. Signal jet samples are also generated in 100 GeV bins of $p_T$. To cover different jet masses, the mass of $S$ (and of $F$ for 3-pronged decays) is randomly chosen event by event within the interval $[10,800]$~GeV, and setting an upper limit $M_S \leq p_T R/2$ to ensure that all decay products are contained in a jet of radius $R=0.8$.

The parton-level event samples are hadronised with {\scshape Pythia}~\cite{Sjostrand:2007gs} and a fast detector simulation is performed with {\scshape Delphes}~\cite{deFavereau:2013fsa}, using the CMS card. Jets are reconstructed with {\scshape FastJet}~\cite{Cacciari:2011ma} applying the anti-$k_T$ algorithm~\cite{Cacciari:2008gp} with $R=0.8$, and groomed with Recursive Soft Drop~\cite{Dreyer:2018tjj}. 
Jet substructure is characterised by a set of subjettiness variables proposed in~\cite{Thaler:2010tr,Datta:2017rhs},
 \begin{equation}
 \left\{ \tau_1^{(1/2)}, \tau_1^{(1)}, \tau_1^{(2)}, \dots , \tau_{5}^{(1/2)}, \tau_{5}^{(1)}, \tau_{5}^{(2)}, \tau_{6}^{(1)}, \tau_{6}^{(2)} \right\} \,,
 \label{ec:taulist}
 \end{equation}
computed for ungroomed jets. Together with the groomed jet mass and $p_T$, they constitute the input to the NN.

The training set is obtained by dividing the $m_J$ range in ten bins, all of 50 GeV except the first one $[10,50]$ GeV, and the $p_T$ range in 100 GeV bins, starting at $[200,300]$ GeV and up to $[2100,2200]$ GeV. In the lower $p_T$ samples we drop the higher mass bins, considering the full $m_J$ range only for the $p_T$ bins above 1200 GeV. In each (two-dimensional) bin we select for the training set 3000 events from each of the six types of signal jets in (\ref{ec:MIdata}), and 18000 background events, in order to have a balanced sample. For \Gent\ these 18000 background events are 9000 quark and 9000 gluon jets, whereas for \gentx\ the set is inclusive, with a proportion of quark and gluon jets that is $p_T$-dependent. The validation sets used to monitor the NN performance are similar to the training ones.

The NNs are implemented using {\scshape Keras}~\cite{keras} with a {\scshape TensorFlow} backend~\cite{tensorflow}. For the training, a standardisation of the 19 inputs, based on the SM background distributions, is performed. The NNs contain two hidden layers of 2048 and 128 nodes, with Rectified Linear Unit (ReLU) activation for the hidden layers. For the output layer:
\begin{itemize}
\item For \Gent\ a softmax function is used, which yields the probabilities $P_g$, $P_q$, $P_s$ that the jet is initiated by a gluon, a quark, or a massive signal particle. In order to discriminate signal from QCD background (quarks or gluons) one can simply use the signal probability $P_s$.
\item For \gentx\ a sigmoid function is used, yielding the NN score, i.e. the signal probability, that can be used to discriminate signal jets from QCD jets.
\end{itemize}
The NNs are optimised by minimising the categorical cross-entropy (\Gent) and binary cross-entropy (\gentx) loss functions, using the Adam~\cite{Adam} algorithm.

For the evaluation of the tagger performance and comparison with previous work we use five selected signals from the cascade decay of a heavy $Z'$ boson into neutral scalars $S$ and $A$ (we generically use $S$ and $A$ for scalars yielding four-pronged and two-pronged jets, respectively), or top quarks:
\begin{itemize}
\item four-pronged: $Z' \to SS$, with (i) $S \to AA \to 4b$, taking $M_{Z'} = 2.2$ TeV, $M_S = 80$ GeV, $M_A = 30$ GeV; (ii) $S \to WW \to 4q$, taking $M_{Z'} = 3.3$ TeV, $M_S = 200$ GeV;
\item three-pronged: $Z' \to t \bar t$, with $t \to W b \to q \bar q b$, with $M_{Z'} = 3.3$ TeV;
\item two-pronged: (iii) $Z' \to AA$, $A \to b \bar b$, with $M_{Z'} = 2.2$ TeV, $M_A = 80$ GeV; (iv) $Z' \to WW$, $W \to q \bar q$, with $M_{Z'} = 3.3$ TeV.
\end{itemize}
The background sets are composed of quark and gluon jets with nearly the same $p_T$ as the signals. For $M_{Z'} = 2.2$ TeV we use the sets G1.0\_80, Q1.0\_80, and for $M_{Z'} = 3.3$ TeV the sets G1.5\_175, Q1.5\_175, G1.5\_200, Q1.5\_200  (see appendix~\ref{sec:a}).

\begin{figure}[t]
\begin{center}
\begin{tabular}{c}
\includegraphics[width=8.2cm,clip=]{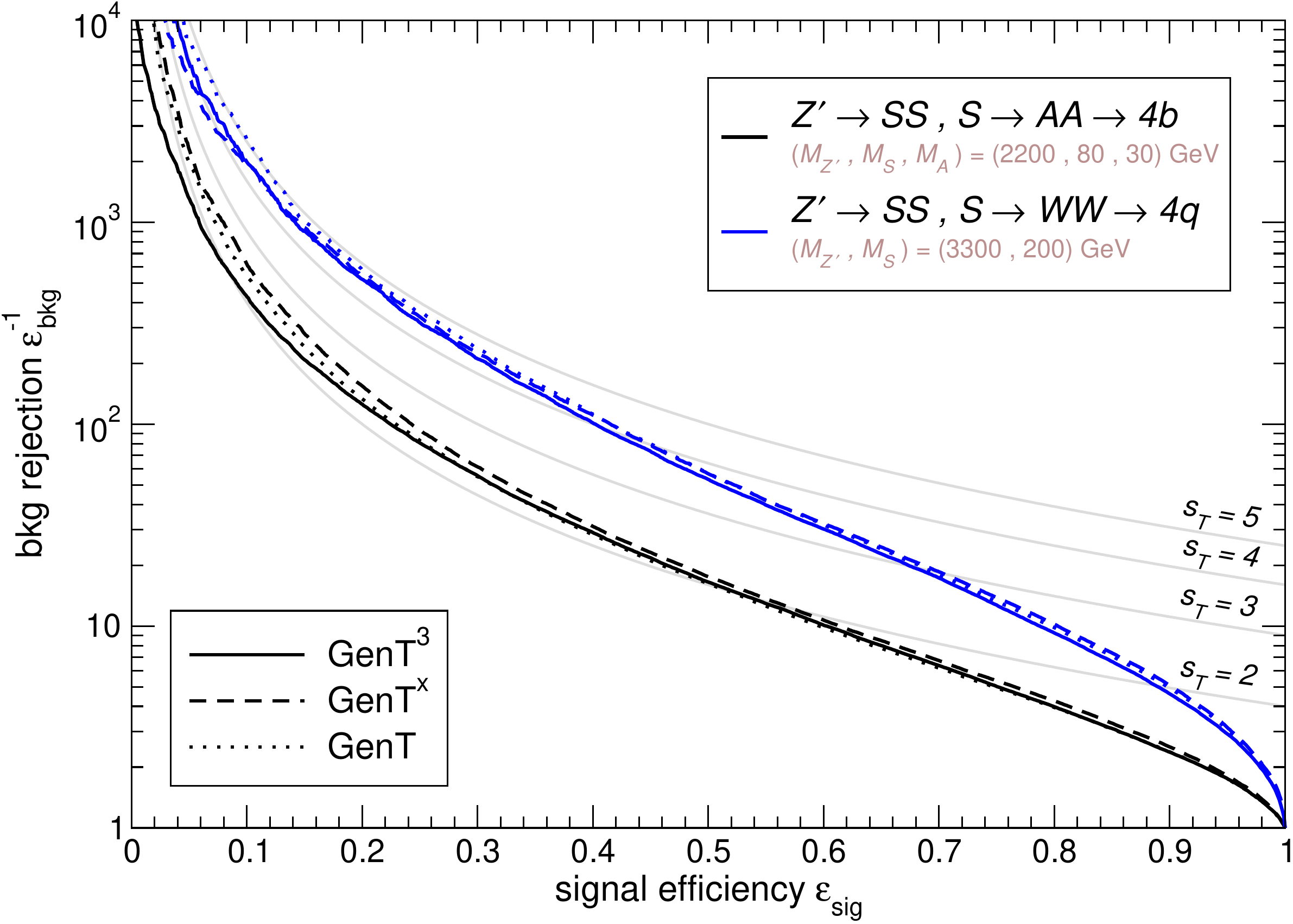} \\
\includegraphics[width=8.2cm,clip=]{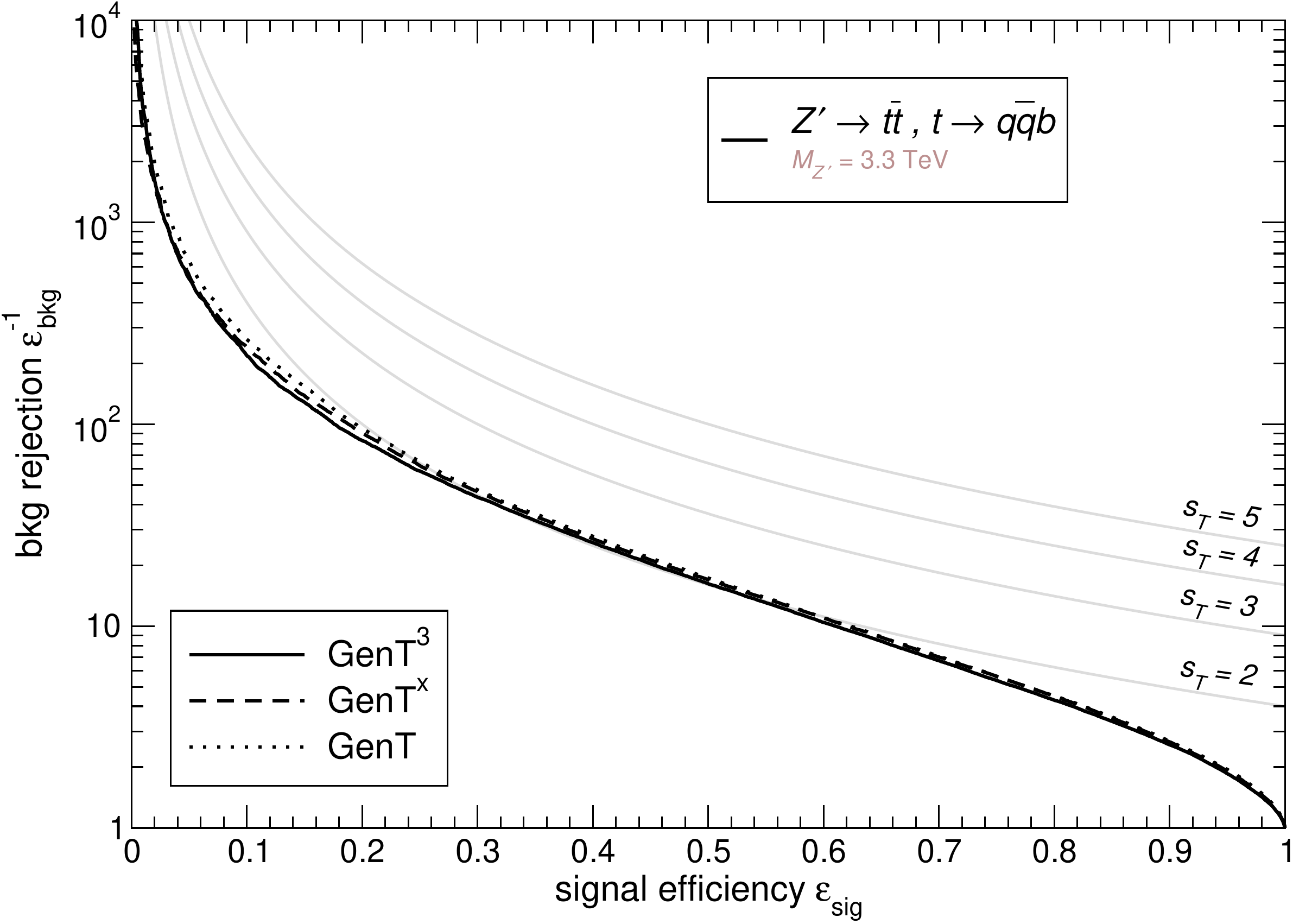} \\
\includegraphics[width=8.2cm,clip=]{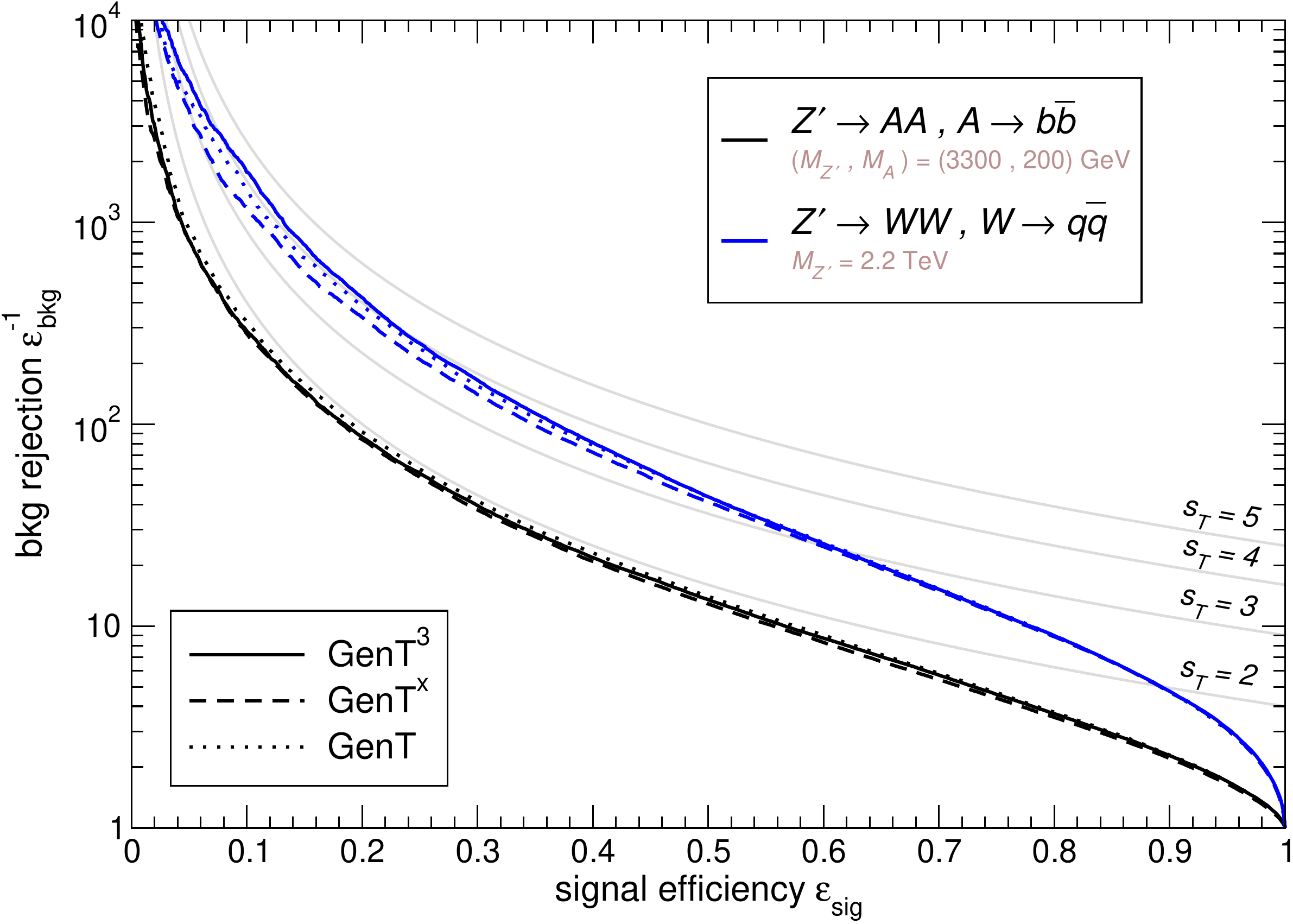}
\end{tabular}
\caption{ROC curves for selected four-pronged (top), three-pronged (middle) and two-pronged (bottom) jet signals.}
\label{fig:ROC}
\end{center}
\end{figure}

We show in Fig.~\ref{fig:ROC} the receiver operating characteristic (ROC) curves for the aforementioned signals, for the two taggers developed here as well as for {\tt GenT} in Ref.~\cite{Aguilar-Saavedra:2020uhm}. As it can be seen, the differences are minimal and do not favour one tagger over another. The performance for other mass values and for the rest of signals studied in Ref.~\cite{Aguilar-Saavedra:2020uhm} is quite similar as well, and detailed results on the tagger performance can be found in that reference. Together with the tagger results, we include lines corresponding to several values of the tagger significance improvement, defined as
\begin{equation}
s_T = \frac{\varepsilon_\text{sig}}{\sqrt{\varepsilon_\text{bkg}}} \,.
\end{equation}
This quantity measures the improvement in the signal significance $n_\text{sig}/\sqrt{n_\text{bkg}}$ (with $n_\text{sig}$ the expected number of signal events and $n_\text{bkg}$ the expected background) achieved with the tagging of one jet. Overall, supervised taggers offer a better performance than unsupervised methods. We can for example compare with the graph neural network (GNN) autoencoder of Ref.~\cite{Atkinson:2021nlt}.
\begin{itemize}
\item For four-pronged jets, a direct comparison is not possible because of the different masses.  Ref.~\cite{Atkinson:2021nlt} takes a benchmark with a 600 GeV scalar decaying into two $W$ bosons, while we only consider benchmarks up to $M_S = 200$ GeV and the tagger is trained with jet masses up to 500 GeV.
\item For top quarks (three-pronged), the autoencoder yields an AUC of 0.75, with a background rejection $\varepsilon_\text{bkg}^{-1} = 5.4$ for $\varepsilon_\text{sig} = 0.5$. \gentx\ has an AUC of 0.87, with a background rejection of 17 for $\ptj \sim 1.5$ TeV (see Fig.~\ref{fig:ROC}), with similar values for $\ptj \sim 1$ TeV. 
\item For $W$ bosons (two-pronged), the autoencoder has an AUC of 0.74, and a background rejection  $\varepsilon_\text{bkg}^{-1} = 6.3$ for $\varepsilon_\text{sig} = 0.5$. For the MUST generic taggers the AUC ranges from 0.92 for $\ptj \sim 0.5$ TeV, to 0.95 for $\ptj \sim 1.5$ TeV. For $\ptj \sim 1$ TeV, as in Fig.~\ref{fig:ROC}, and a signal efficiency $\varepsilon_\text{sig} = 0.5$, the background rejection is $\varepsilon_\text{bkg}^{-1} = 41$.
\end{itemize}

\section{Mixed-sample estimator}
\label{sec:3}

For the detection of anomalous density distributions we use a mixed-sample estimator~\cite{mixedsample1,mixedsample2,Williams:2010vh}.\footnote{We have investigated the point-to-point dissimilarity method in Ref.~\cite{Williams:2010vh} with linear and Gaussian weighting functions and the sensitivity to jet mass bumps is smaller.} Let $A$ and $B$ be two (pseudo-)data sets, with numbers of events $n_A$ and $n_B$, respectively, and $N \equiv n_A + n_B$. In our application, these two sets will correspond to events before and after jet tagging. In the combined set $A \cup B$, let us define
\begin{equation}
I(i,j) = \left\{ \begin{array}{ll}
1 & x_i~\text{and}~x_j~\text{are in the same set}
\\[1mm]
0 & \text{otherwise}
\end{array} \right.
\end{equation}
The nearest neighbour of a point is defined in terms of the normalised euclidean distance
\begin{equation}
d(x_i,x_j)^2 = \sum_{l=1}^D \frac{|x_i^l-y_i^l |^2}{w_l^2} \,,
\end{equation}
with $l$ running over the different components of the $D$-dimensional vectors $x_i$ and $x_j$. A weight $w_l$ for each coordinate can be introduced, if desired. (We obtain good results by setting $w_l = 1$.) The estimator $T$ is then built by considering the $N_k$ nearest neighbours to a given point $x_i$, excluding $x_i$ itself, with the sum
\begin{equation}
T = \frac{1}{N\, N_k} \sum_{i=1}^{N} \sum_{k=1}^{N_k} I(i,k) \,.
\end{equation}
That is, one considers all the $N$ points $x_i$ in $A \cup B$, and for each point, its $N_k$ nearest neighbours. 

Provided the points in the sets $A$ and $B$ follow the same density distribution, for large $N$, $N_k$ and $D$ the p.d.f. of $T$ is a Gaussian, with mean and variance given respectively by~\cite{Williams:2010vh}
\begin{align}
& \mu_0 = \frac{n_A(n_A-1) + n_B(n_B-1)}{N(N-1)} \,, \notag \\
& \sigma_0^2 = \frac{n_A n_B}{N^3 \,N_k}  \left[ 1 + 4 \frac{n_A n_B}{N^2} \right] \,.
\label{ec:musigma}
\end{align}
In our examples in the following, the mean and standard deviation are quite close to the values obtained from the above expressions. In particular, we will always consider $n_A = n_B \equiv n$, in which case one has $\mu_0 \simeq 1/2$.

\begin{figure}[t]
\begin{center}
\begin{tabular}{c}
\includegraphics[width=9cm,clip=]{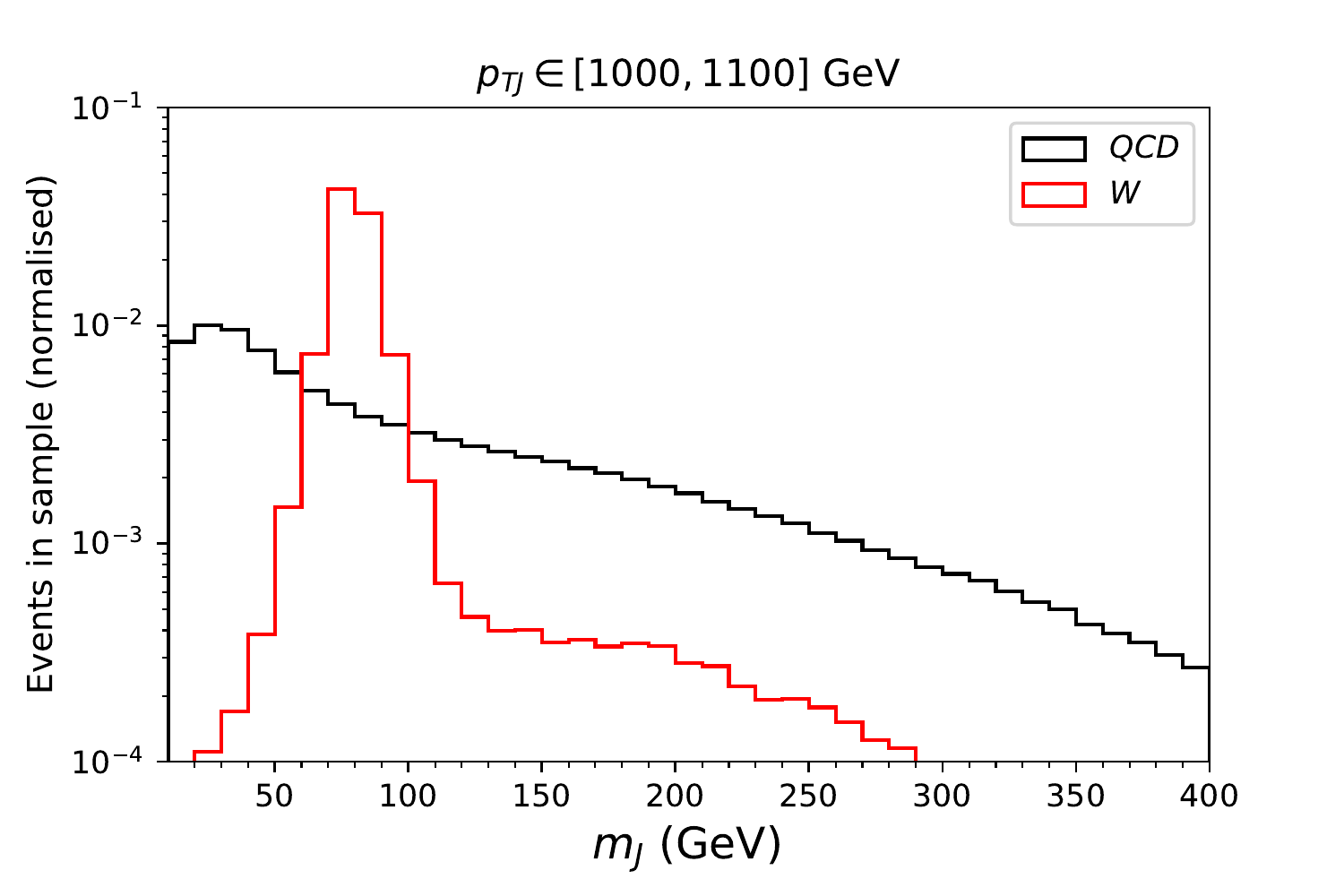}
\end{tabular}
\caption{Jet mass distribution for QCD and $W$ jets with $p_T \in [1,1,1]$ TeV.}
\label{fig:mJ-Tdemo}
\end{center}
\end{figure}

Let us study in a simplfied setup the sensitivity of this estimator to pinpoint, in a sample of QCD jets, a small `contamination' of $W$ jets. We take as single variable $x$ the jet mass $m_J$. For this exercise we take QCD jets from set JS1 with $p_T \in [1,1.1]$ TeV, and $W$ jets with $p_T \in [1,1.1]$ TeV  from $Z' \to WW$ with $M_{Z'} = 2.2$ TeV. The normalised jet mass distribution (i.e. the p.d.f. for our example) for each case is presented in Fig.~\ref{fig:mJ-Tdemo}.
With these events we build three jet `pools':
\begin{itemize}
\item Pool A ($\mathcal{P}_A$): half of the QCD jets, approximately $8 \times 10^5$ elements.
\item Pool B1 ($\mathcal{P}_{B1}$): the other half of the QCD jets.
\item Pool B2 ($\mathcal{P}_{B2}$): jets from $Z' \to WW$, around $10^5$ elements.
\end{itemize}

\begin{figure}[t]
\begin{center}
\begin{tabular}{c}
\includegraphics[width=9cm,clip=]{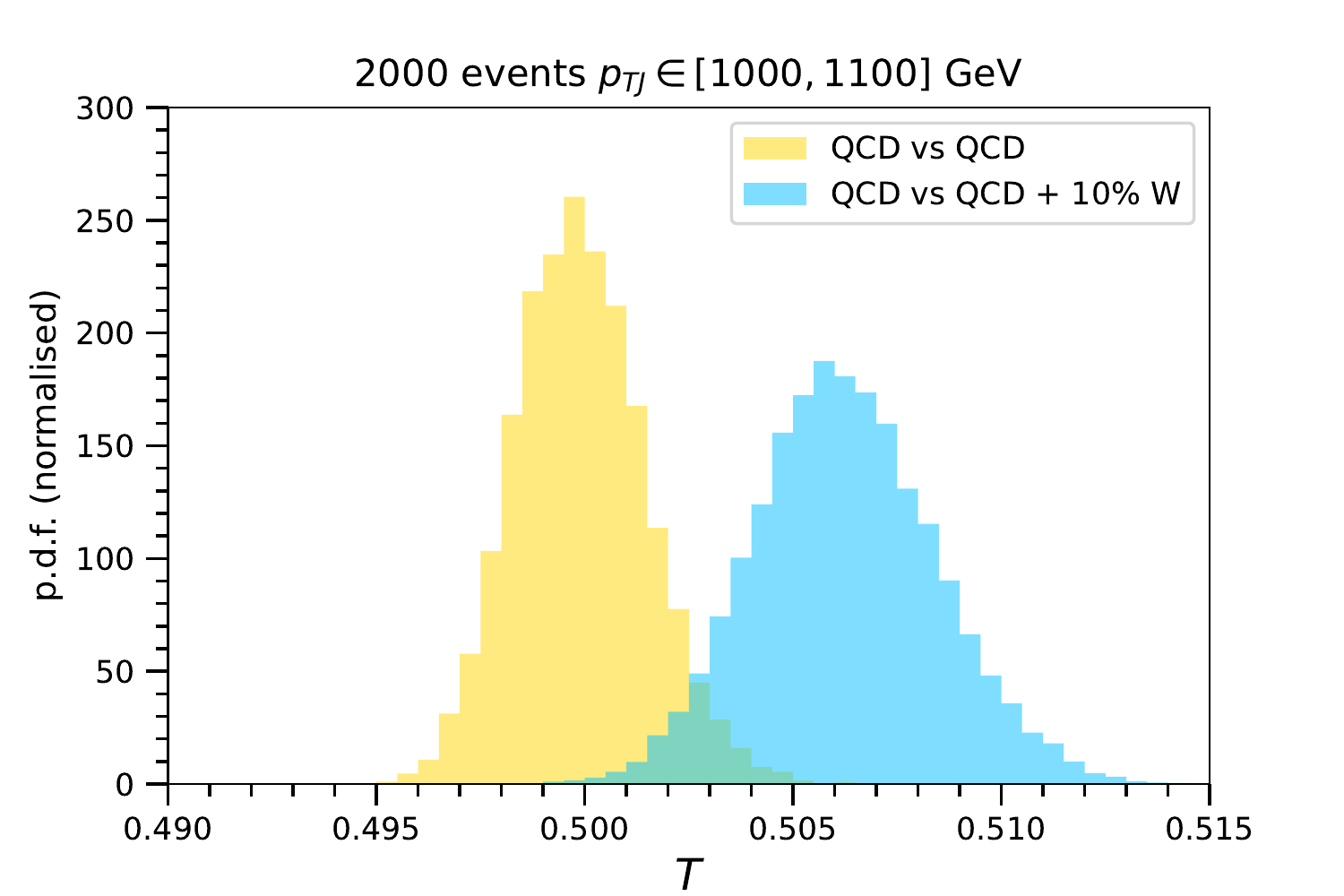}
\end{tabular}
\caption{Probability density functions for the estimator $T$, for two equivalent QCD jet samples (yellow) or with a small contamination of $W$ jets (blue). See the text for details.}
\label{fig:T-Tdemo}
\end{center}
\end{figure}

For the comparison of equivalent sets of QCD jets we fix a value of $N_k$ and perform a series of pseudo-experiments: in each pseudo-experiment we take $n$ random elements from $\mathcal{P}_A$ and $n$ random elements from $\mathcal{P}_{B1}$, computing the value of $T$. This procedure is repeated $10^4$ times in order to get the distribution of $T$ for equivalent sets. An example for $N_k = 50$, $n = 2000$ is shown in Fig.~\ref{fig:T-Tdemo} (yellow distribution). The mean $\mu_0 = 0.49990$ and standard deviation $\sigma_0 = 0.00156$ obtained from the numerical distribution are quite close to the expectation $\mu_0 = 0.5$, $\sigma_0 = 0.00158$ from Eqs.~(\ref{ec:musigma}). 

For the comparison of a set of QCD jets with a set containing a small fraction $f$ of $W$ jets, we randomly take $n$ elements from $\mathcal{P}_A$, $n(1-f)$ elements from $\mathcal{P}_{B1}$ and $n f$ elements from $\mathcal{P}_{B2}$. This procedure is repeated $10^4$ times. An example for $N_k = 50$, $n = 2000$, $f=0.1$ is shown in Fig.~\ref{fig:T-Tdemo} (blue distribution). The mean and standard deviation of this distribution, which is not Gaussian in general, are $\mu = 0.5062$, $\sigma = 0.00217$. For a given pseudo-experiment the resulting value of $T$ can be used to quantify the likelihood that both sets have the same density distribution, with the pull $p = (T - \mu_0)/\sigma_0$. On the other hand, the expected sensitivity to detect deviations for given values of $N_k$, $n$ and $f$ can be quantified by the same expression but replacing $T$ by the mean of its distribution $\mu$, i.e. $p = (\mu - \mu_0)/\sigma_0$. It amounts to $4.1\sigma$ in the example shown in Fig.~\ref{fig:T-Tdemo}.
Note that the distribution for the comparison of two equivalent sets (yellow in Fig.~\ref{fig:T-Tdemo}) may exhibit some skewness when the event samples are very small. In such case, the effect is taken into account in the computation of the pulls.

The sensitivity of the $T$ estimator to pinpoint sample contaminations grows with $f$ (the size of the contamination) and also with $n$, the sample size. It also grows with $N_k$ but only up to a certain point: if $N_k = N$, then $T = 1/2 - 1/N$ irrespectively of whether the two samples follow the same density distribution or not. In order to get a better insight on the numerical dependence for small $f$ and $N_k \ll N = 2n$ we have calculated the expected sensitivity for different values of the parameters. We collect in Table~\ref{tab:pull-Tdemo} the expected pulls obtained computing $T$ in sets with $n = 1000,2000,4000$ events each, one of them with QCD jets and the other one containing a fraction $f = 0.05,0.1,0.15$ of $W$ jets. We use $N_k = 25,50,100$ nearest neighbours.

\begin{table}[htb]
\begin{center}
\begin{tabular}{cccc}
$N_k = 25$ \\
\hline
$n\backslash f$ & 0.05 & 0.10 & 0.15 \\
1000 & 0.60 & 2.06 & 4.10 \\
2000 & 0.86 & 2.92 & 5.76 \\
4000 & 1.23 & 4.17 & 8.19
\end{tabular}
\quad
\begin{tabular}{cccc}
$N_k = 50$ \\
\hline
$n\backslash f$ & 0.05 & 0.10 & 0.15 \\
1000 & 0.87 & 2.93 & 5.82 \\
2000 & 1.19 & 4.06 & 8.04 \\
4000 & 1.72 & 5.79 & 11.5
\end{tabular}
\vspace{5mm}

\begin{tabular}{cccc}
$N_k = 100$ \\
\hline
$n\backslash f$ & 0.05 & 0.10 & 0.15 \\
1000 & 1.17 & 4.09 & 8.20 \\
2000 & 1.68 & 5.77 & 11.4 \\
4000 & 2.37 & 8.11 & 16.0
\end{tabular}
\caption{Pulls obtained for the comparison of two sets with different values of $N_k$, $n$ and $f$. }
\label{tab:pull-Tdemo}
\end{center}
\end{table}

For the examples considered, the pull can be well approximated by the expressions
\begin{align}
& N_k = 25 : &&  p \sim 3.2 \, n^{1/2} f^{1.7} \,, \notag \\
& N_k = 50 : && p \sim 4.8 \, n^{1/2} f^{1.7} \,, \notag \\
& N_k = 100: &&  p \sim 7.3  \, n^{1/2} f^{1.7} \,,
\end{align}
in which the dependence on $n$ and $f$ is almost the same, with different prefactors that depend on $N_k$. Writing $n$ and $f$ in terms of the signal $n_\text{sig} = nf$ and background $n_\text{bkg} = n(1-f)$, the common dependence on $n$ and $f$ can be rewritten as
\begin{equation}
n^{1/2} f^{1.7} = \frac{n_\text{sig} ^{1.7}}{(n_\text{sig} +n_\text{bkg})^{1.2}}  \,.
\end{equation}
It is beyond our scope to determine the dependence of $p$ on $n$, $f$ and $N_k$ in full generality; however, the examples shown provide insight on the expected behaviour in the situations of interest, and may guide the optimisation of the event selection to search for small signals.

\begin{figure}[t]
\begin{center}
\begin{tabular}{c}
\includegraphics[width=9cm,clip=]{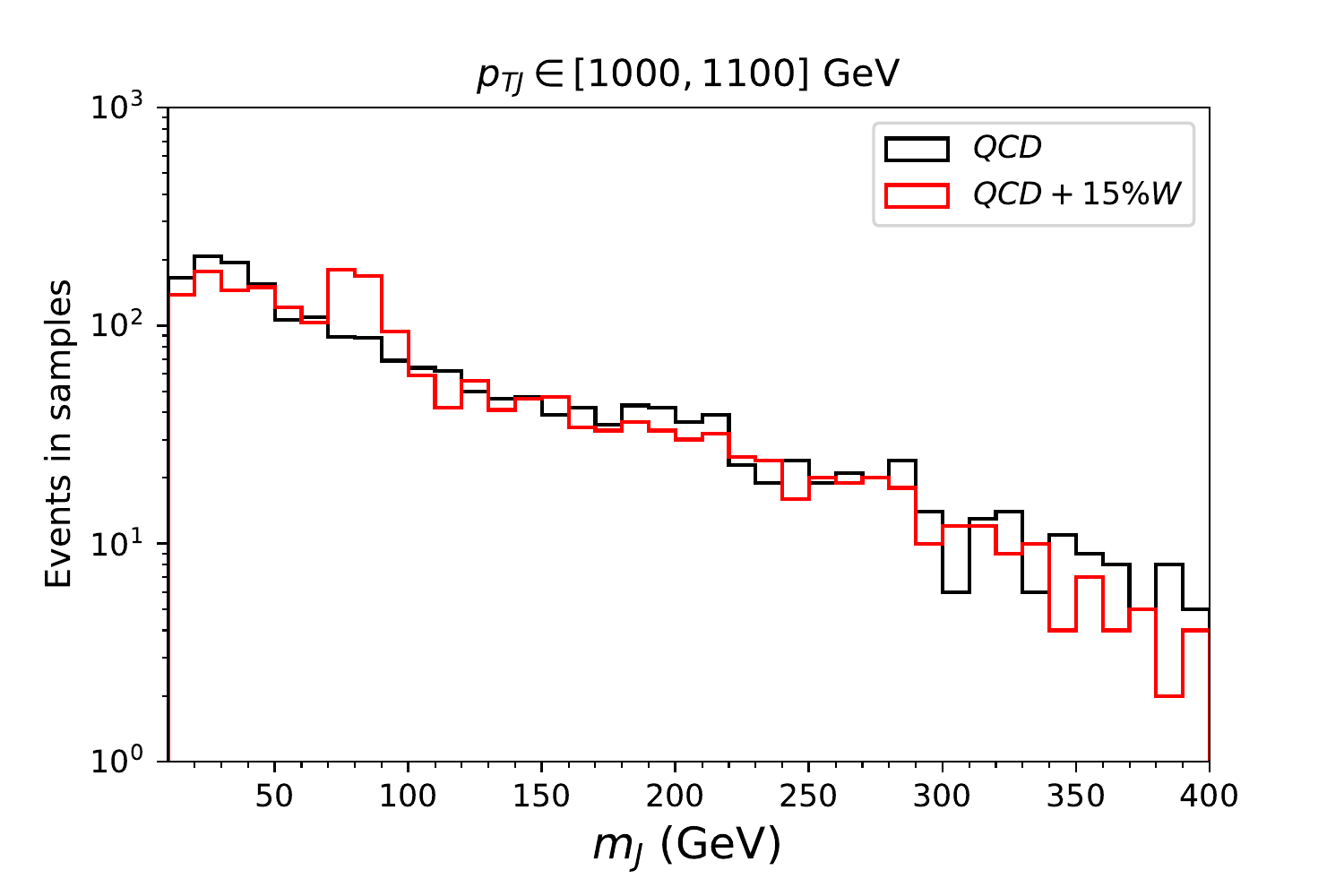}
\end{tabular}
\caption{Jet mass distributions for two pseudo-experiments with $n = 2000$. The first sample (black line) only contains QCD jets and the second sample (red line) contains a fraction $f=0.15$ of $W$ jets.}
\label{fig:Tdemo-PE}
\end{center}
\end{figure}

Finally, we stress again that the pull $p = (T - \mu_0)/\sigma_0$ as well as the expected sensitivity $p = (\mu - \mu_0)/\sigma_0$ measure the {\em global} agreement between the two distributions, and may be smaller than the local significance of a bump. As an example, we show in Fig.~\ref{fig:Tdemo-PE} the jet mass distribution for two pseudo-experiments with $n = 2000$, one with QCD events only, and the other one with $f=0.15$. The estimator for these sets with $N_k = 100$ yields $T = 0.51$, which amounts to a $9.2\sigma$ deviation. On the other hand, the local excess between 70 and 90 GeV amounts to $12.9\sigma$. 

\section{Mass decorrelation}
\label{sec:4}

The mass decorrelation of the tagger, which is essential for the anomaly search method discussed, is done by the method of the varying threshold~\cite{CMS:2017dcz}, with several differences with respect to the original proposal. We consider two methods for mass decorrelation:
\begin{enumerate}
\item Exclusive: the decorrelation is performed separately for quark and gluon jets, which requires the use of a three-class tagger such as \Gent\ and the variation of two thresholds.
\item In situ: the decorrelation is done for a given sample with a specific (although $p_T$-dependent) mixture of quark and gluon jets, by varying a single threshold. This can be performed with any two-class tagger such as \gentx.
\end{enumerate}
In both cases, the threshold adjustment could directly be done in data. For the in situ decorrelation a collection of jets obtained from a subset of the target event sample, or an equivalent one, could be used. For the exclusive decorrelation, quark- and gluon-enriched datasets could in principle be used, as already done by the CMS Collaboration to for the comparison of the jet substructure with Monte Carlo predictions~\cite{CMS:2021iwu}. We also point out that the jet tagger does not need to be supervised. For in situ decorrelation, one could as well use an unsupervised tagger such as those in Refs.\cite{Heimel:2018mkt,Farina:2018fyg,Dillon:2019cqt, Amram:2020ykb,Cheng:2020dal,Dillon:2021nxw}. We now discuss in turn both decorrelation methods, using the supervised taggers \Gent\ and \gentx\ introduced in section~\ref{sec:2}.

\begin{figure*}[p]
\begin{center}
\begin{tabular}{ccc}
\includegraphics[width=5.5cm,clip=]{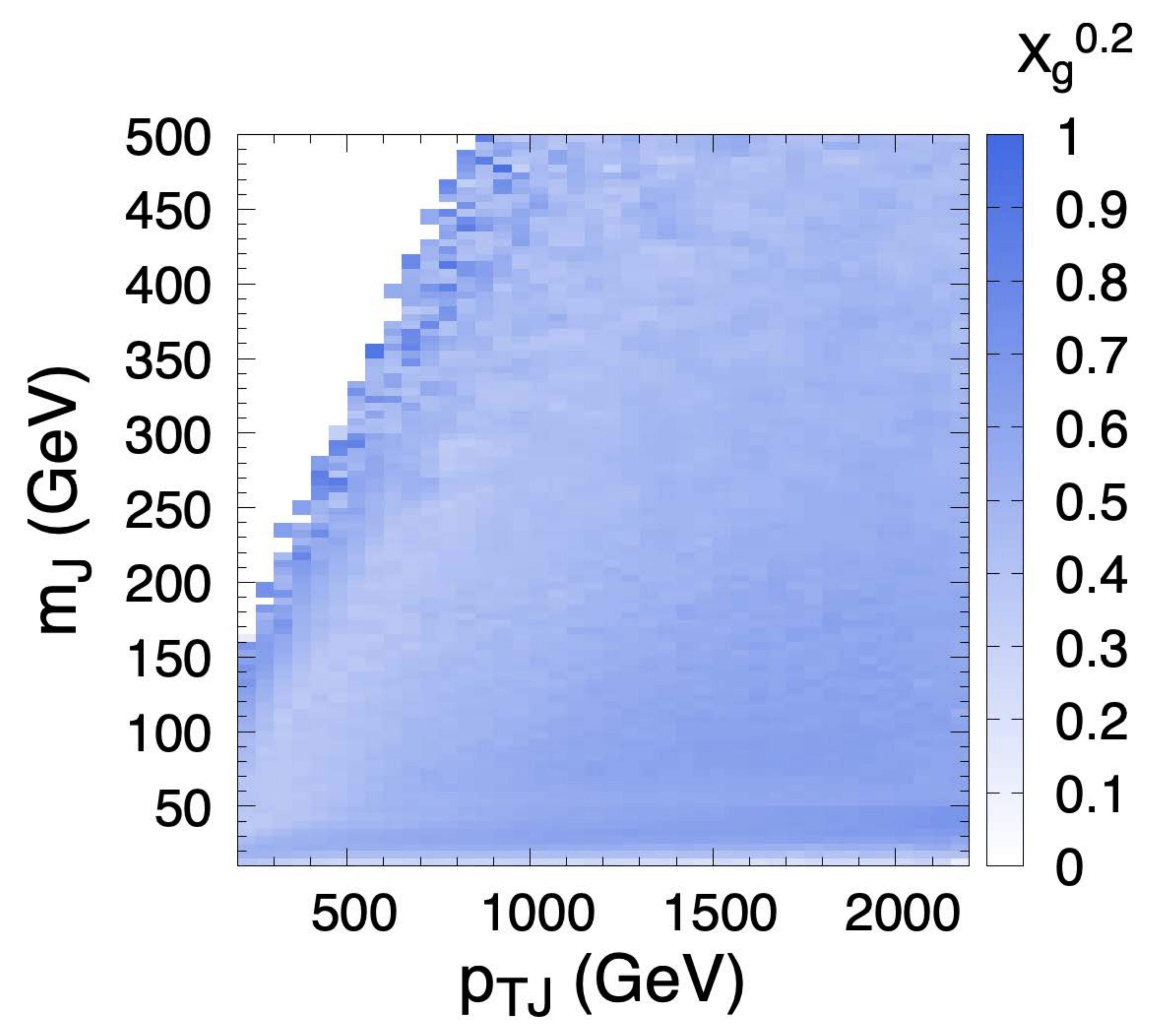} &
\includegraphics[width=5.5cm,clip=]{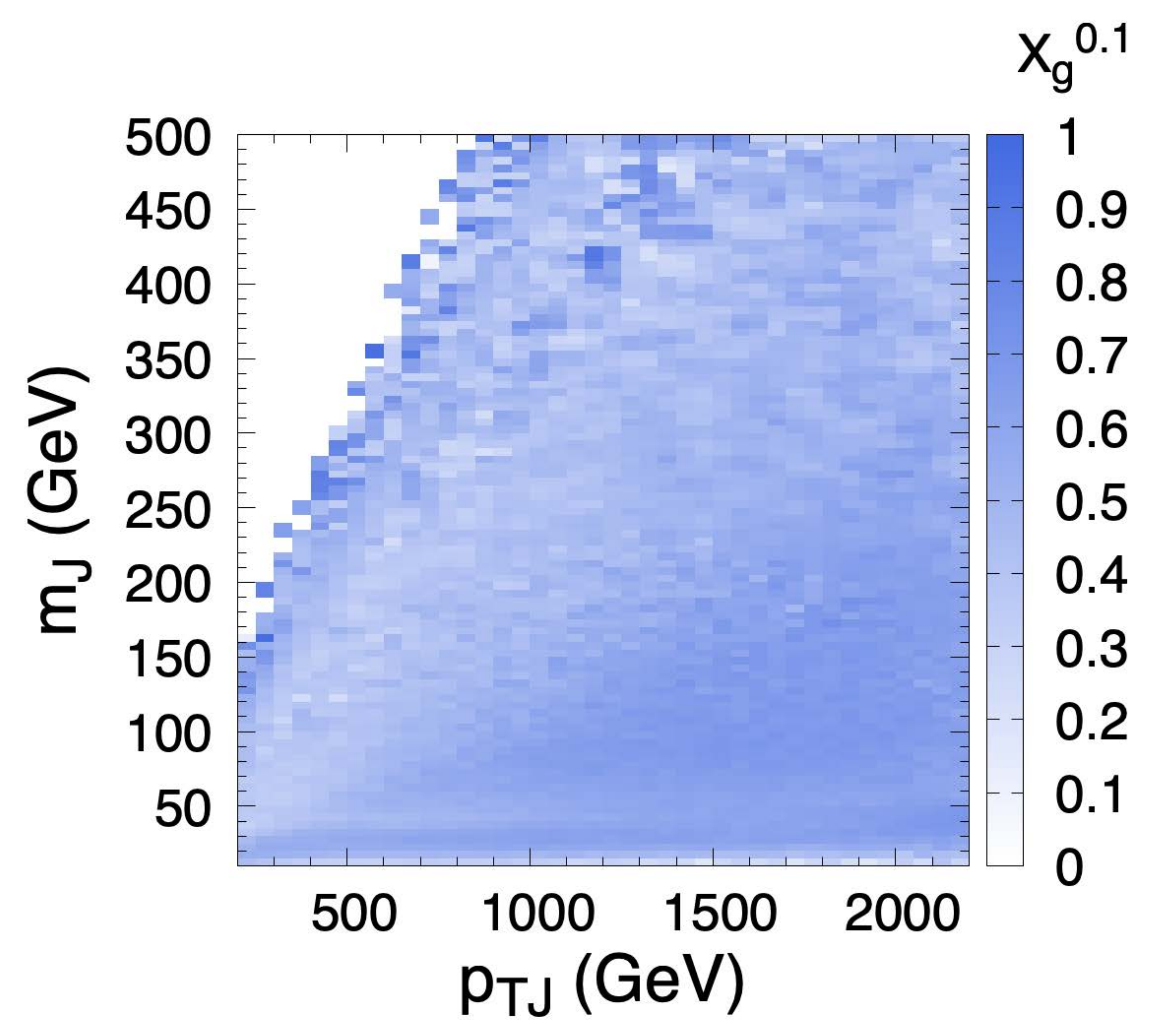} &
\includegraphics[width=5.5cm,clip=]{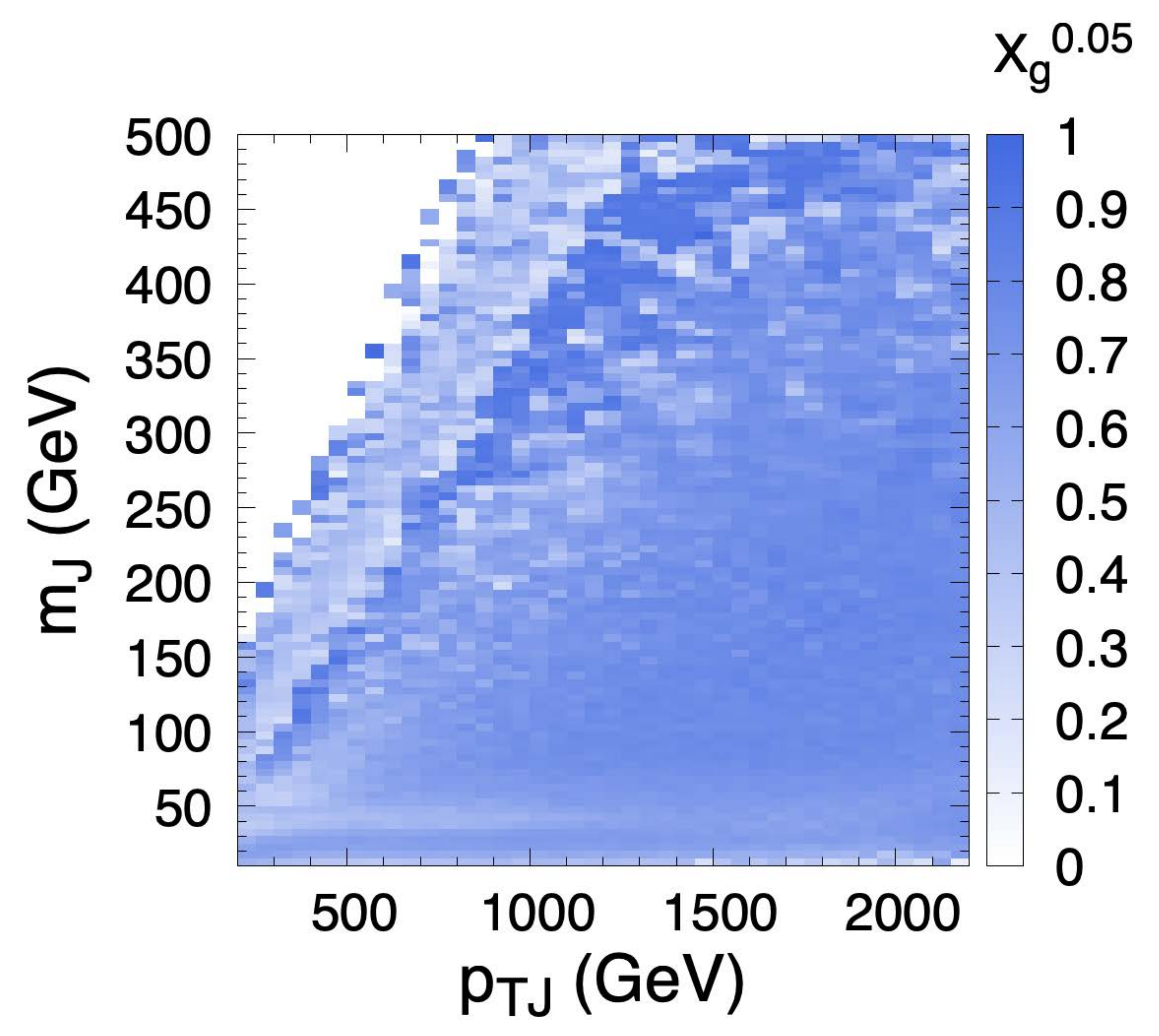} \\
\includegraphics[width=5.5cm,clip=]{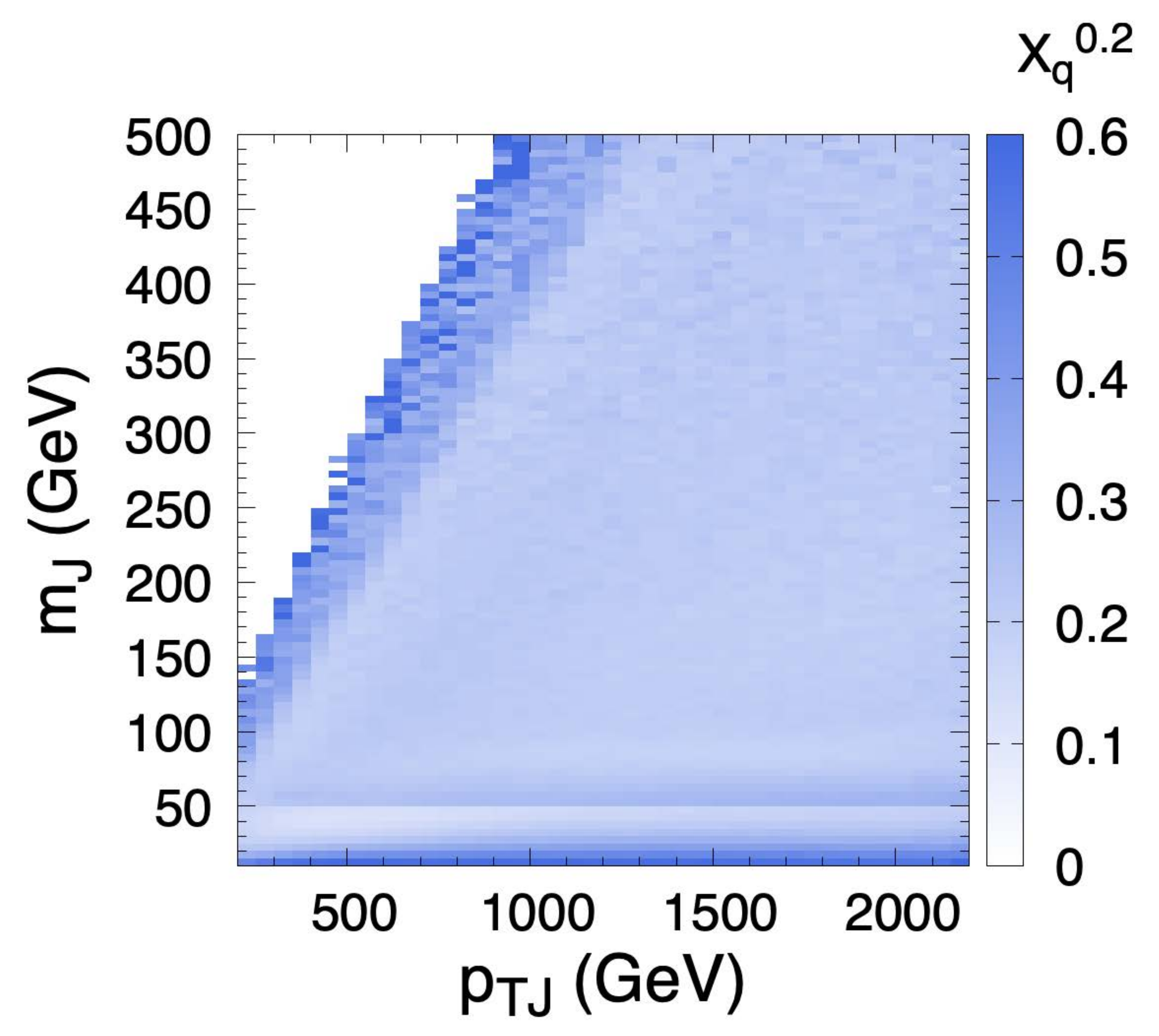} &
\includegraphics[width=5.5cm,clip=]{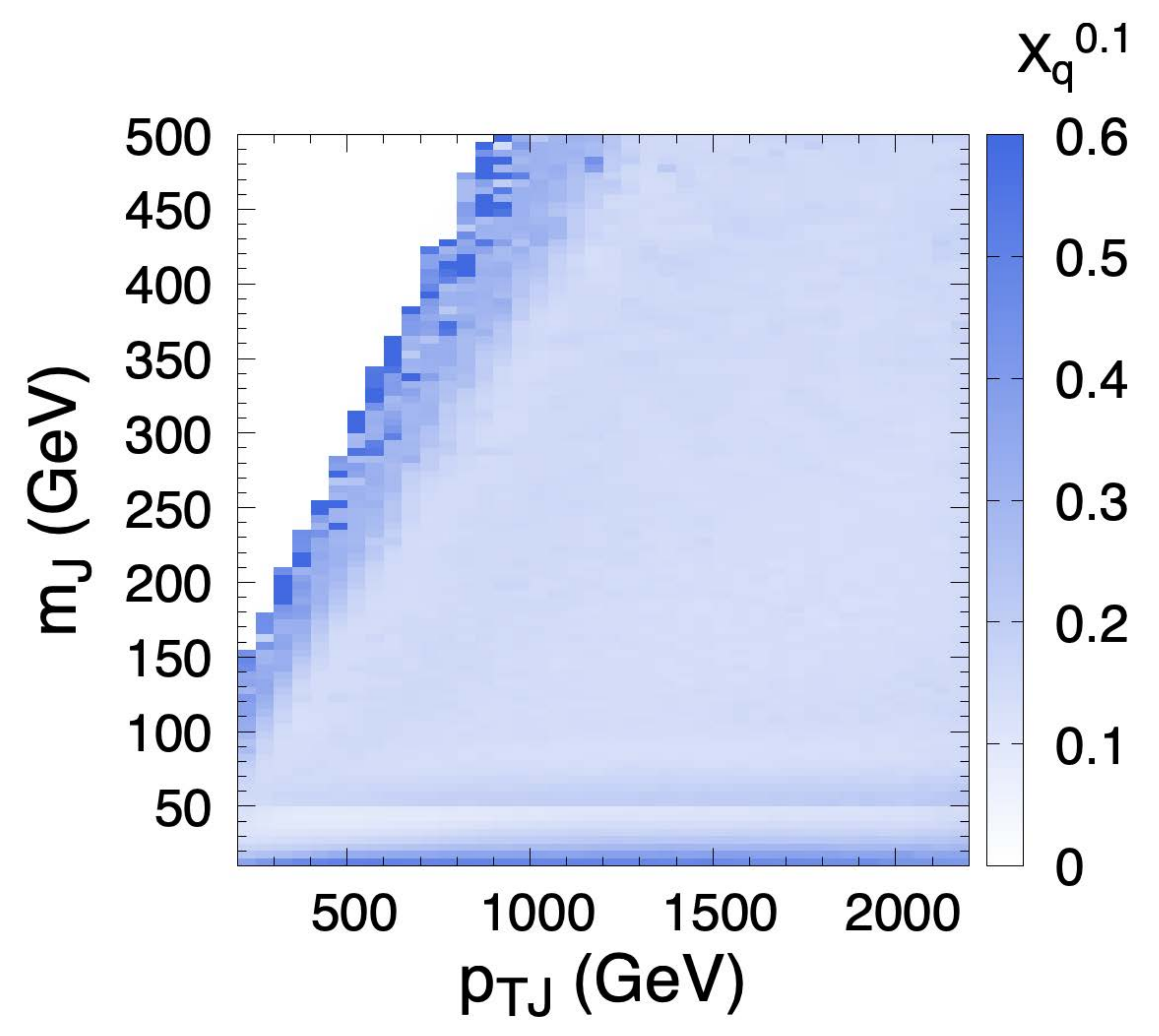} &
\includegraphics[width=5.5cm,clip=]{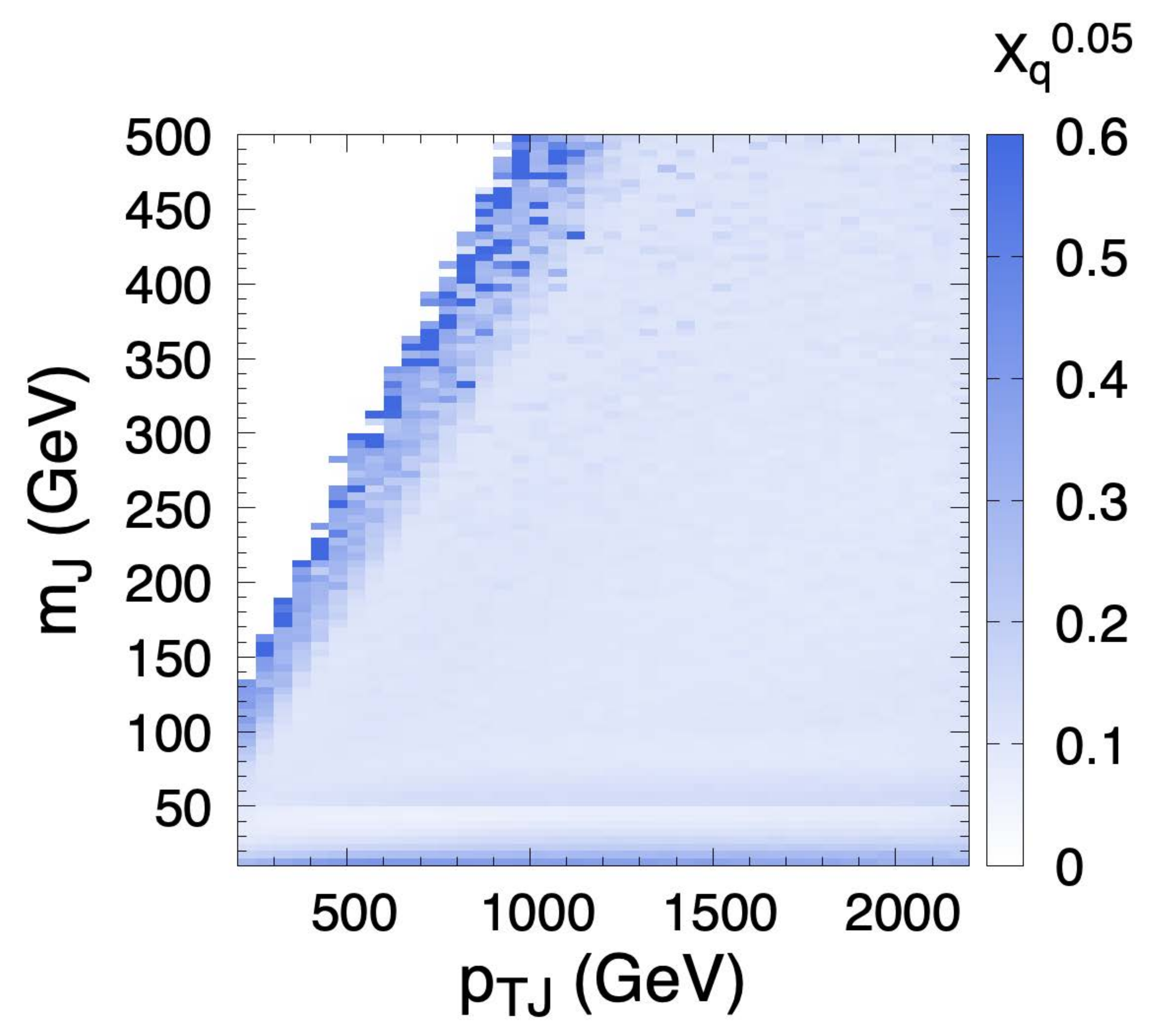}
\end{tabular}
\caption{Thresholds for exclusive mass decorrelation computed in bins of jet mass and $p_T$ for three fixed efficiencies $\varepsilon_b = 0.2$, 0.1, 0.05 from left to right. Empty bins on the upper left side of the plots are shown in white.}
\label{fig:thr3}
\end{center}
\end{figure*}

\begin{figure*}[p]
\begin{center}
\begin{tabular}{cc}
\includegraphics[width=5.5cm,clip=]{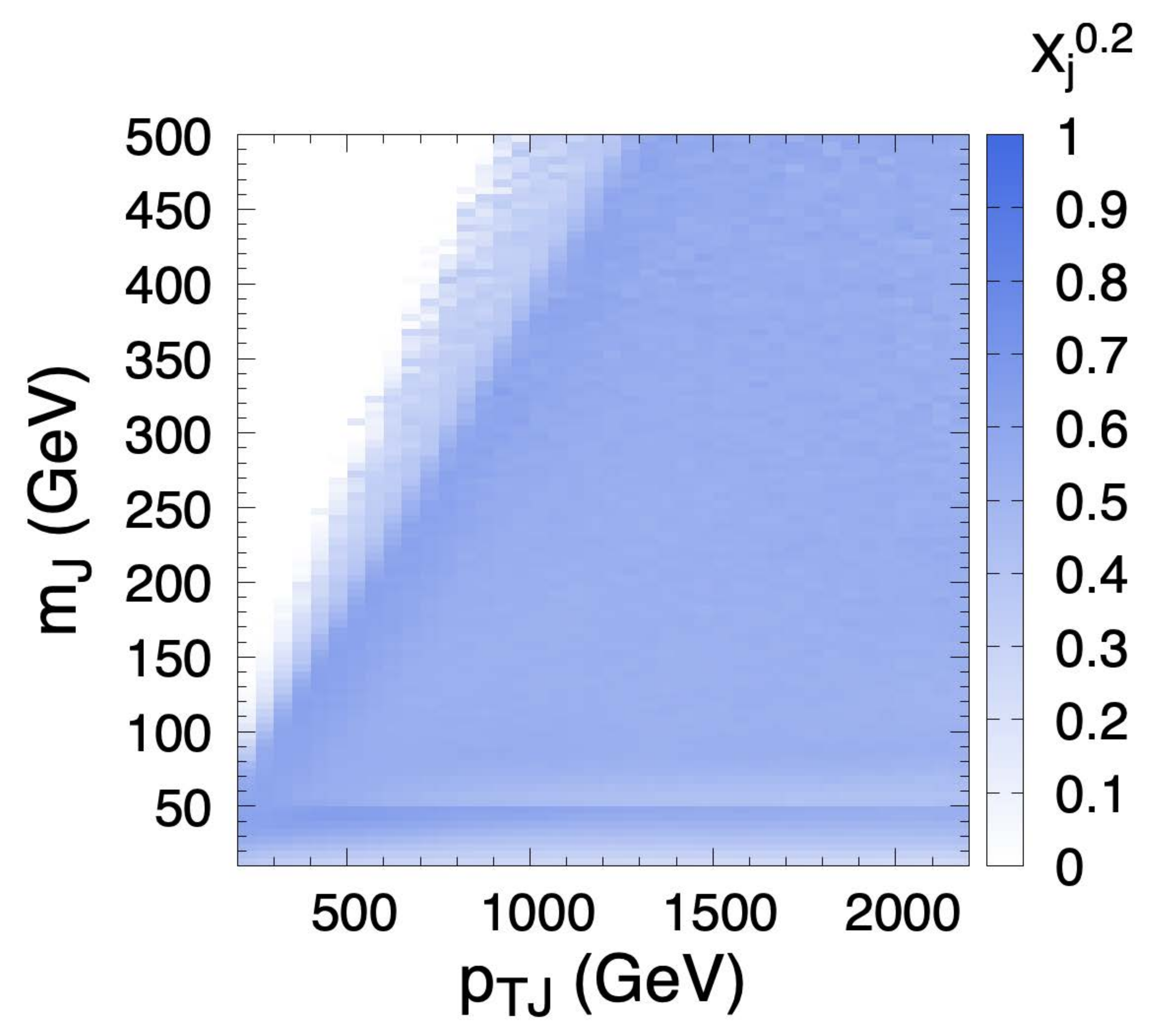} &
\includegraphics[width=5.5cm,clip=]{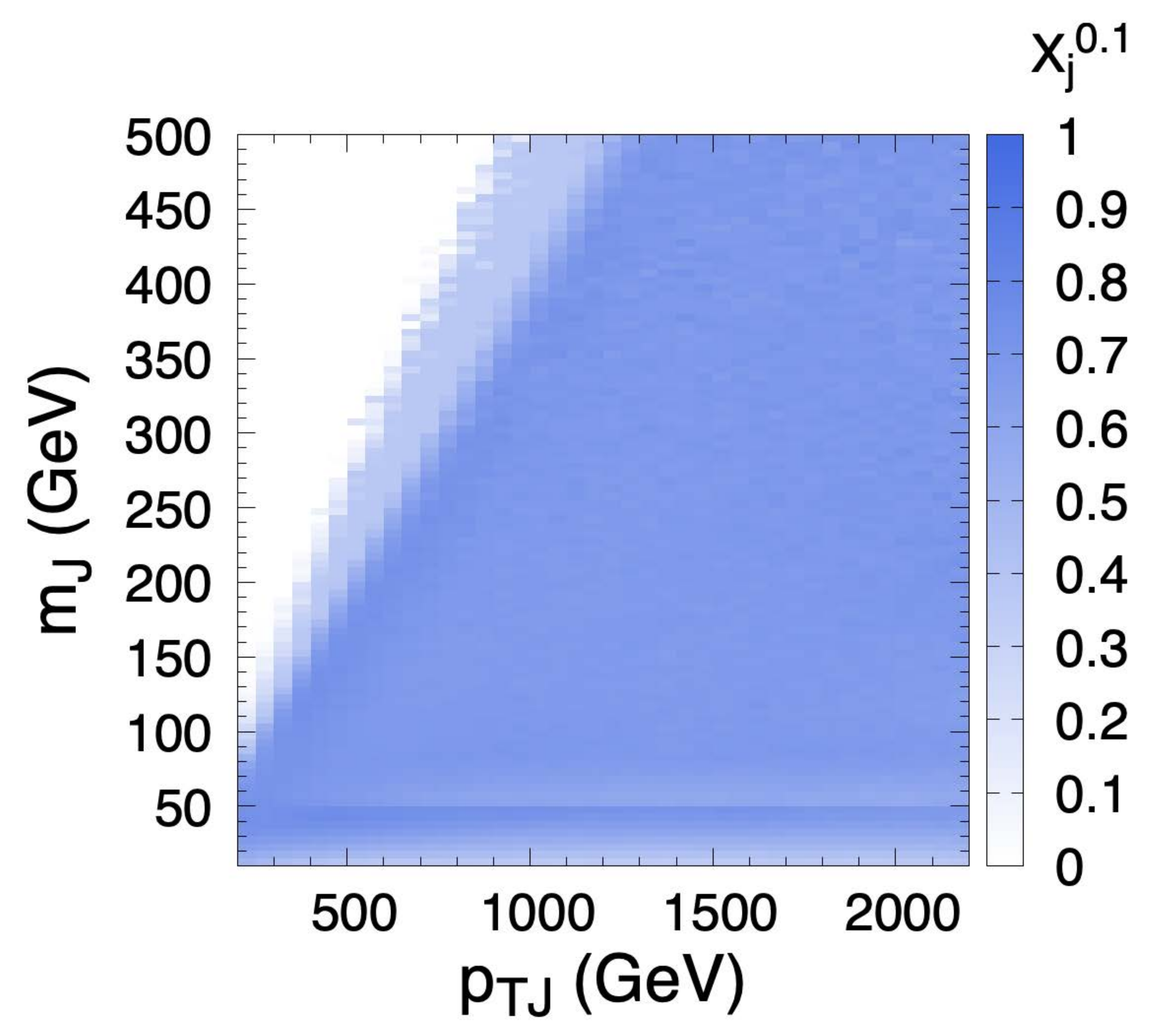} \\
\includegraphics[width=5.5cm,clip=]{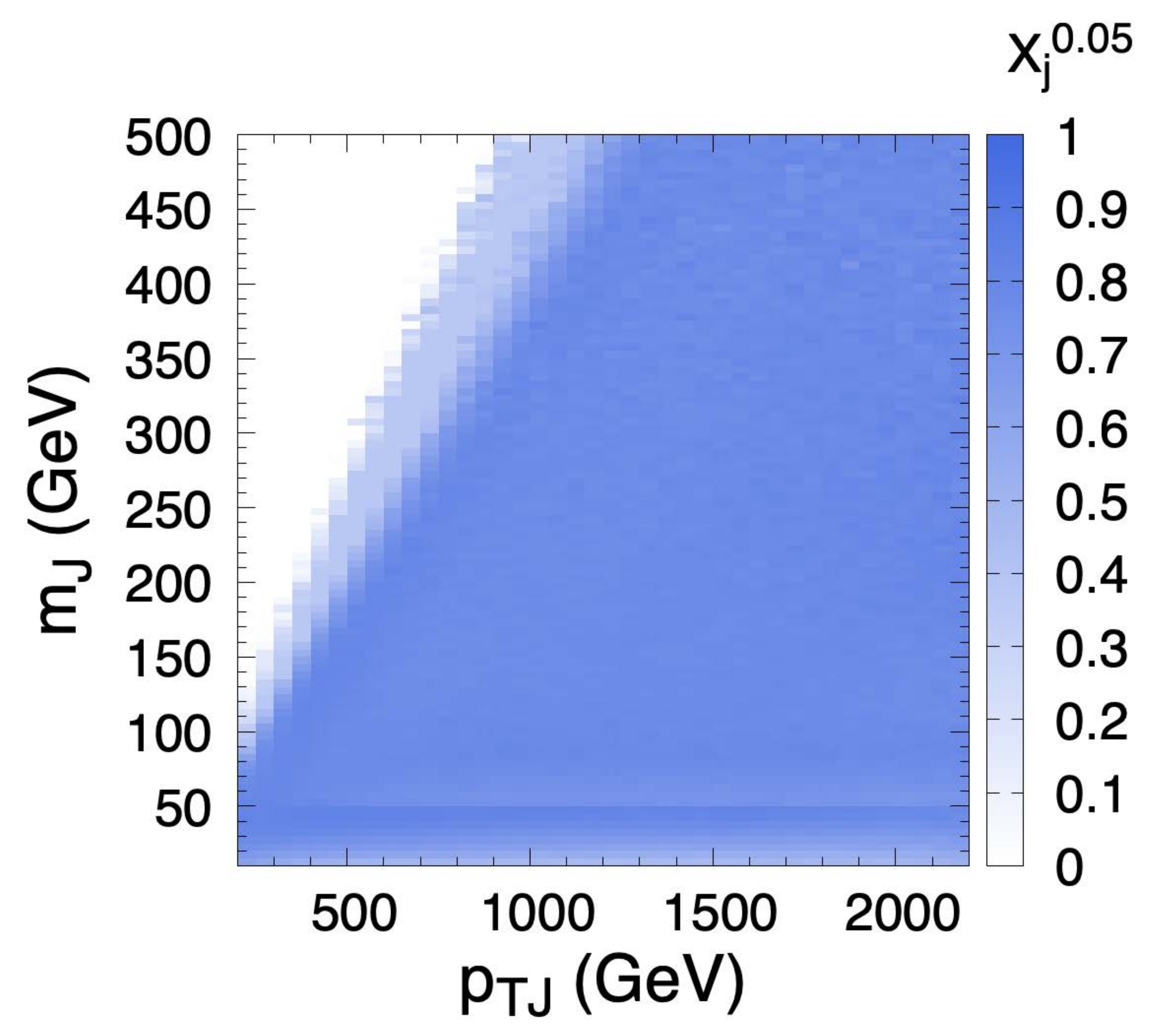} &
\includegraphics[width=5.5cm,clip=]{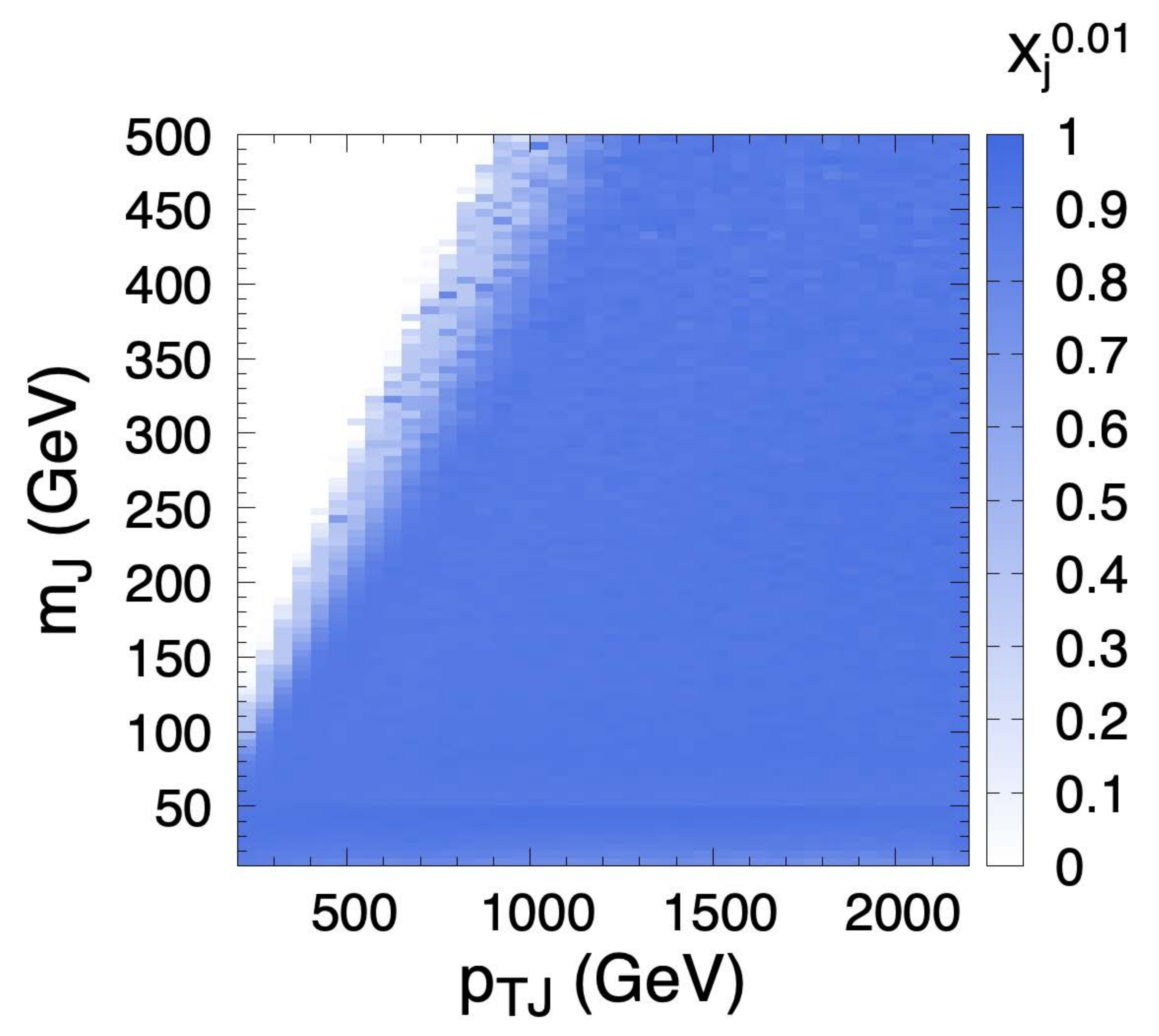}
\end{tabular}
\caption{Thresholds for in-situ mass decorrelation computed in bins of jet mass and $p_T$ for four fixed efficiencies $\varepsilon_b = 0.2$, 0.1, 0.05, 0.01 from left to right and top to bottom. Empty bins on the upper left side of the plots are shown in white.}
\label{fig:thr1}
\end{center}
\end{figure*} 

\subsection{Exclusive mass decorrelation}
\label{sec:4.1}

For the exclusive mass decorrelation the tagger \Gent\, provides the respective probabilities $(P_g,P_q,P_s)$ that a jet is initiated by a gluon, quark, or a signal particle. For a target background efficiency $\varepsilon_b$ (which is considered fixed), and in a given bin of jet mass and $p_T$, one determines the values $X_g^{\varepsilon_b}$, $X_q^{\varepsilon_b}$ such that simultaneously
\begin{itemize}
\item[(i)] the fraction of gluon jets fulfiling $P_g \leq X_g^{\varepsilon_b}$ and $P_q \leq X_q^{\varepsilon_b}$ equals $\varepsilon_b$;
\item[(ii)] the fraction of quark jets fulfiling $P_g \leq X_g^{\varepsilon_b}$ and $P_q \leq X_q^{\varepsilon_b}$ also equals $\varepsilon_b$.
\end{itemize}
An important point here is that the quark versus gluon discrimination does not need to be very good: it is sufficient that the distributions of $P_g$ and $P_q$ are different, so that the solution to the above conditions is unique. This is a posteriori confirmed by the explicit calculation, where we find a quite stable behaviour up to statistical fluctuations.

\begin{figure*}[p]
\begin{center}
\begin{tabular}{cc}
\includegraphics[width=9cm,clip=]{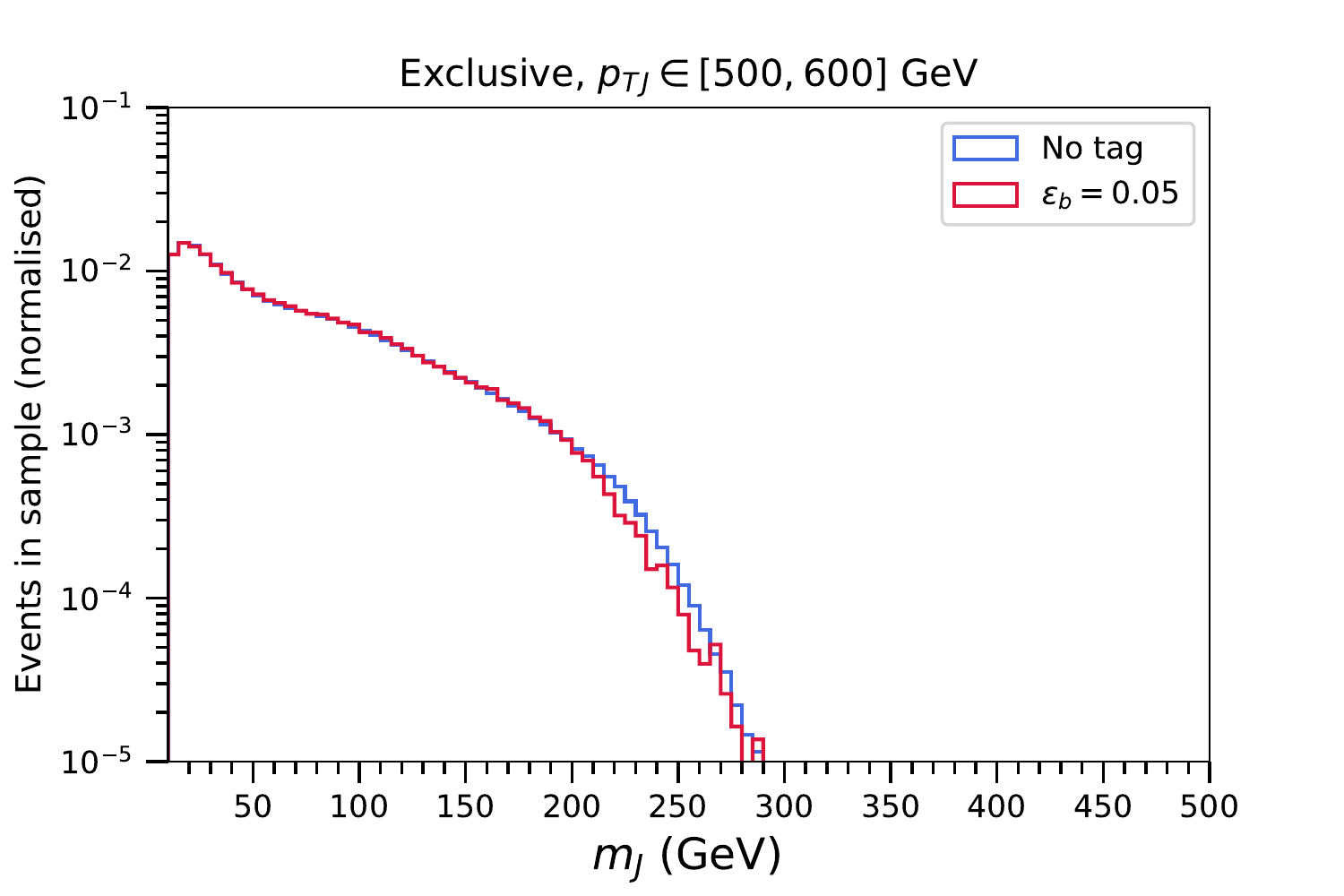} &
\includegraphics[width=9cm,clip=]{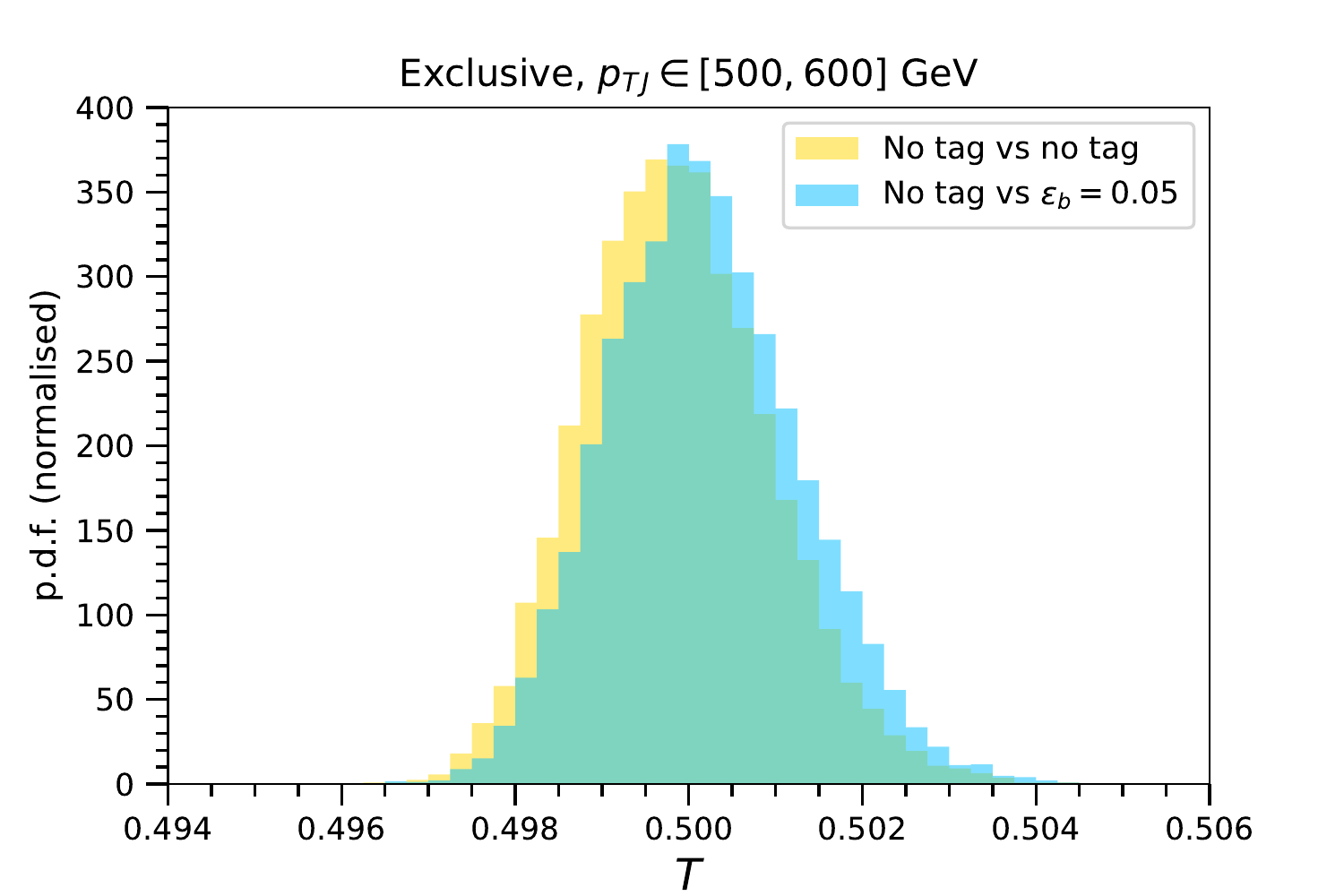} \\
\includegraphics[width=9cm,clip=]{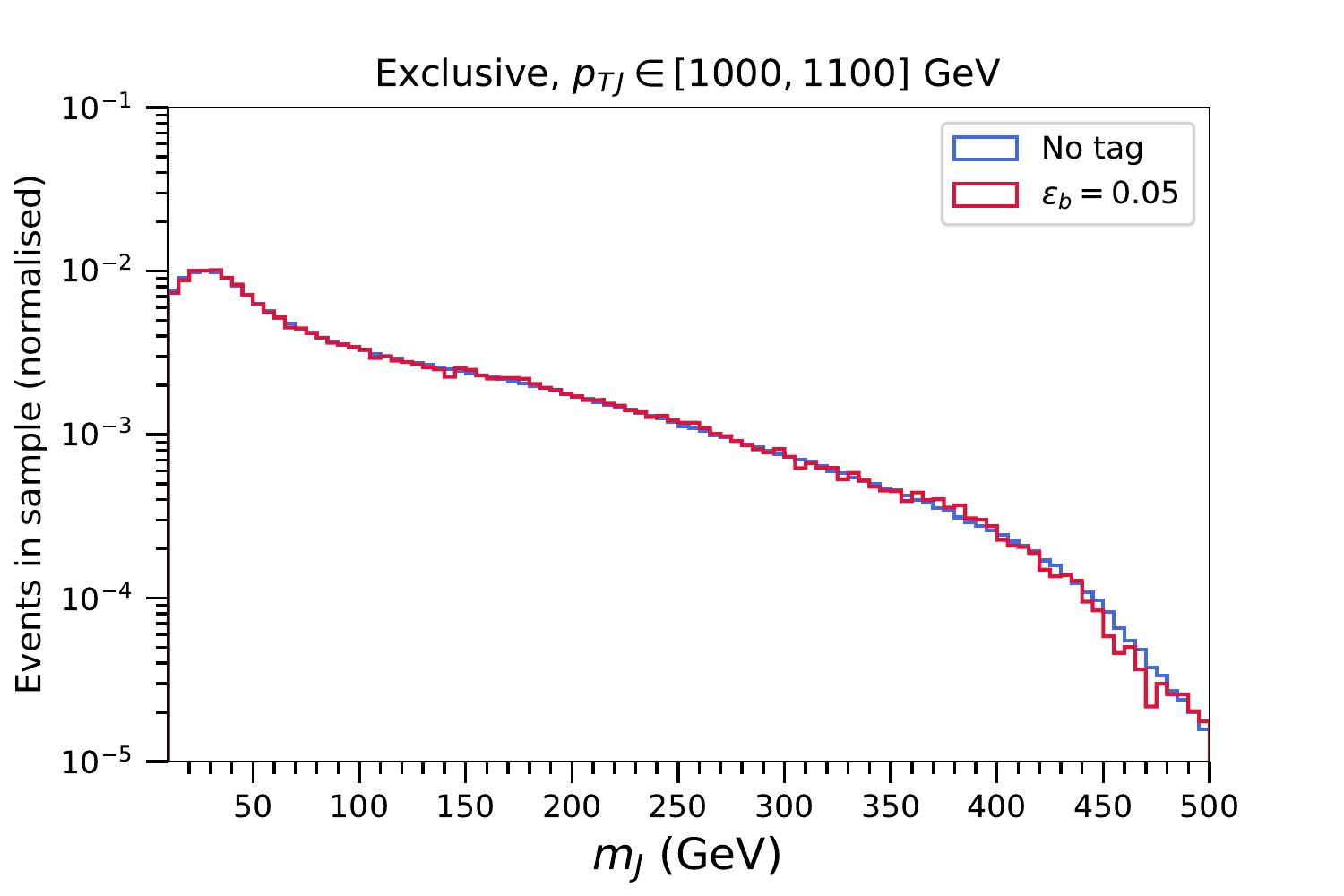} &
\includegraphics[width=9cm,clip=]{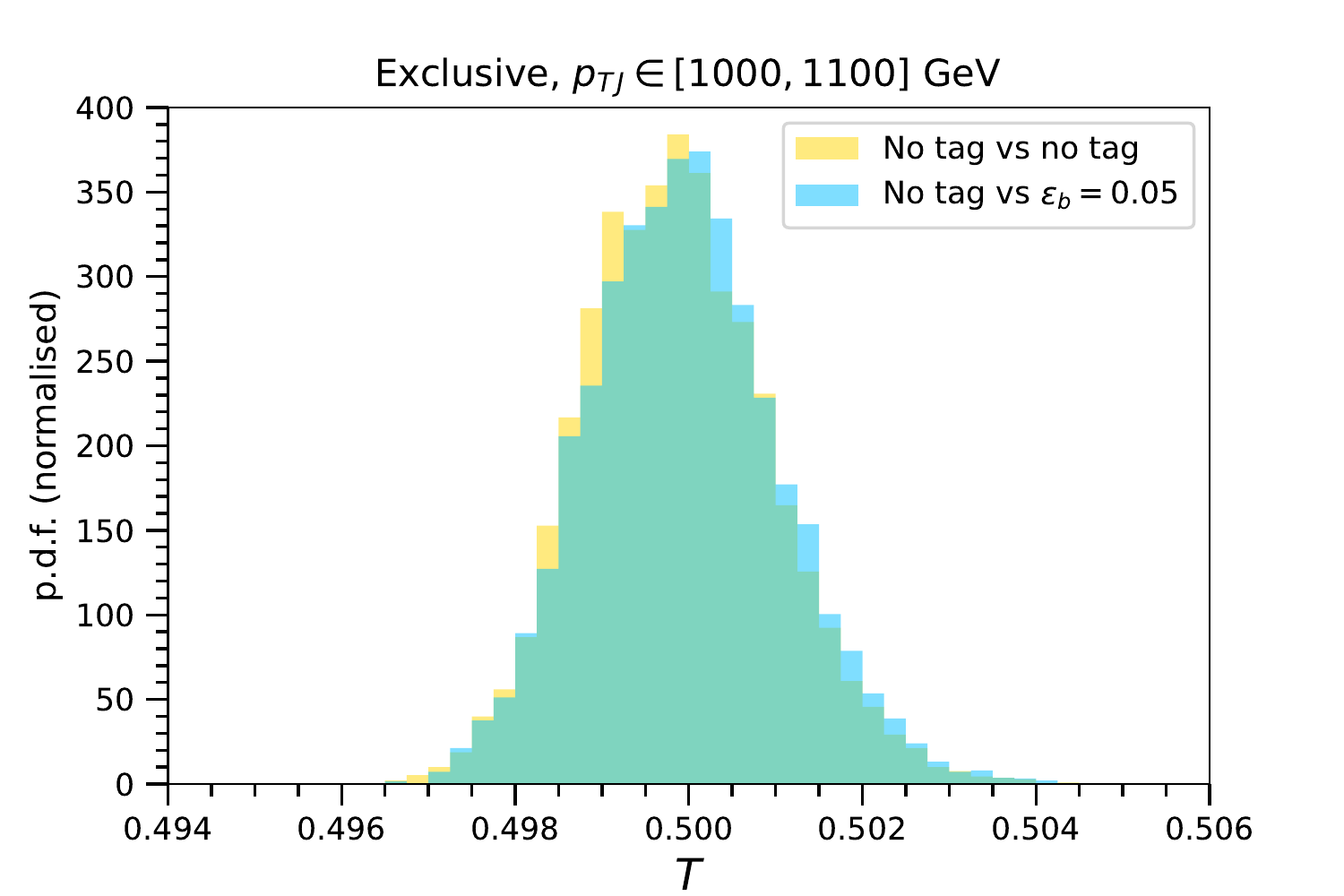} \\
\includegraphics[width=9cm,clip=]{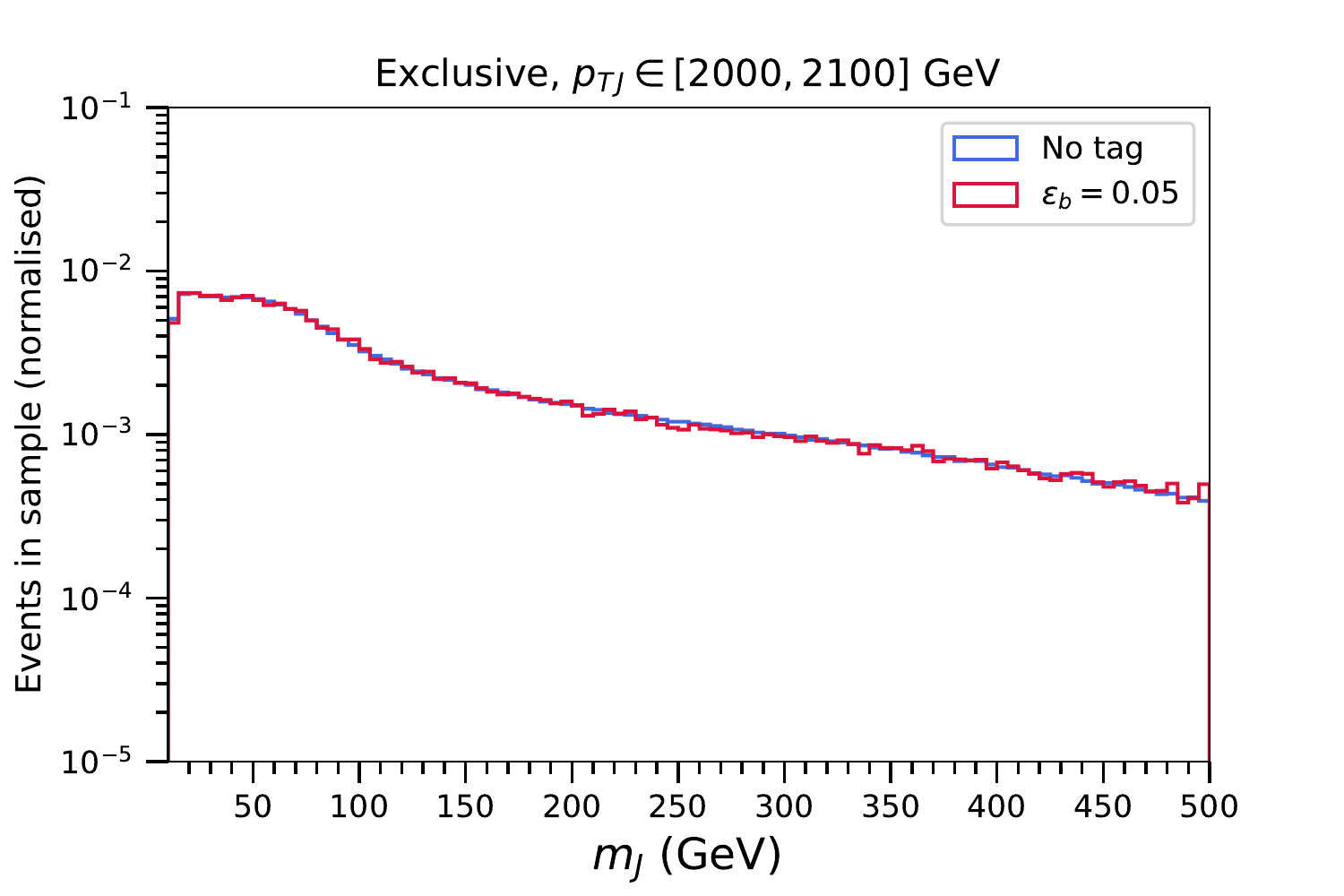} &
\includegraphics[width=9cm,clip=]{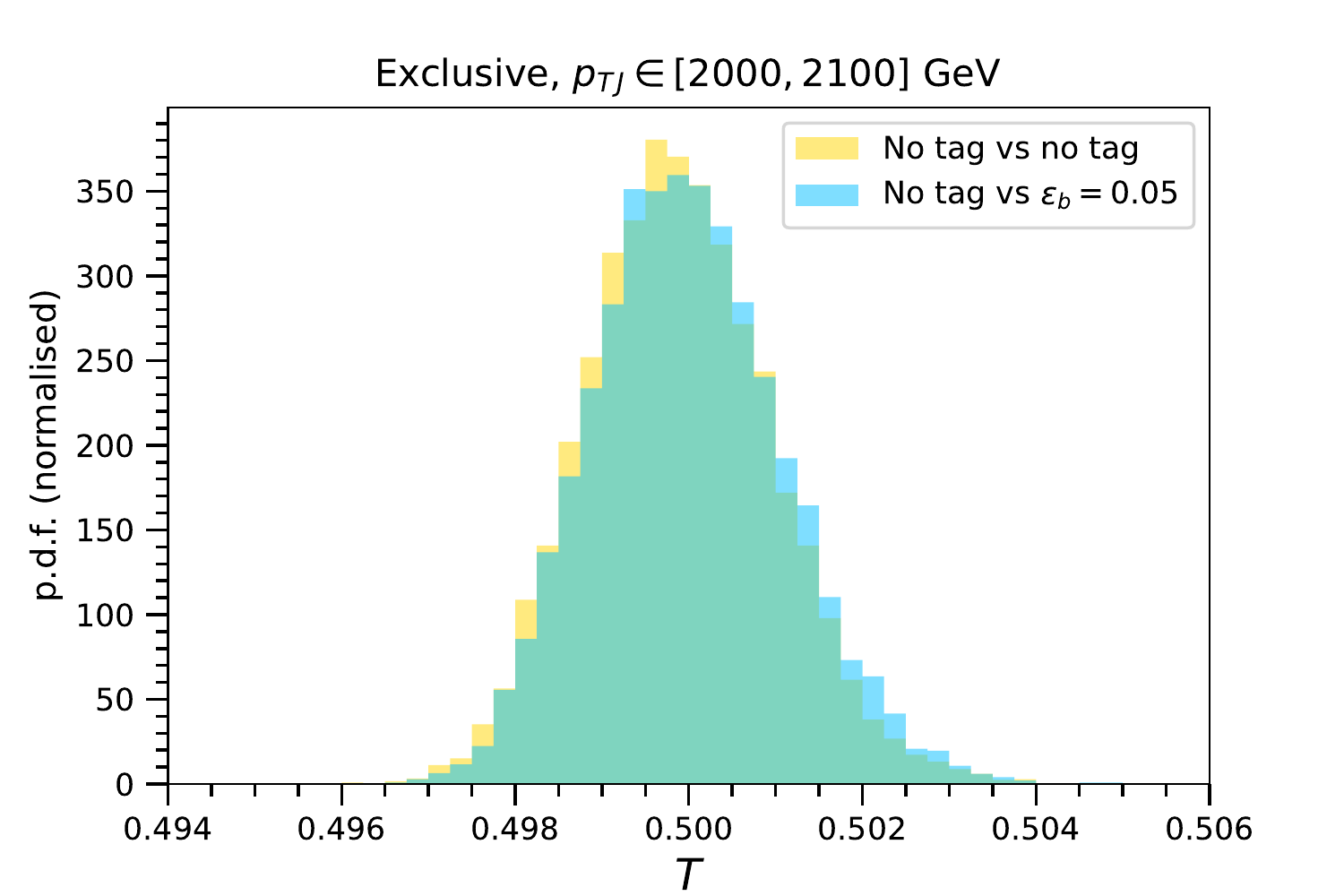}
\end{tabular}
\caption{Left: jet mass distribution for QCD jets in three narrow $p_T$ samples, before tagging and after the application of the \Gent\ tagger and exclusive mass decorrelation with $\varepsilon_b = 0.05$. Right: probability density functions for the estimator $T$, obtained when the two samples are drawn from the untagged jet pool (yellow) or from the untagged and tagged jet pools (blue).}
\label{fig:dec3}
\end{center}
\end{figure*} 

In the computation of the the thresholds we use the sets GS1 and QS1, excluding those jets used in the training of the tagger (but including the ones used for validation), plus the full sets GS2 and QS2 (see appendix~\ref{sec:a}). We divide the $m_J$ range $[10,500]$ GeV in 5 GeV bins, and the $\ptj$ range $[200,2200]$ GeV in 50 GeV bins. In order to decrease statistical fluctuations, in the bins with $m_J \geq 50$ GeV that have fewer than 4000 events either of quark or gluon jets, we increase the bin size by extending 2.5 GeV the $m_J$ interval and 25 GeV the $\ptj$ interval on each side. In such case, overlapping bins have some common elements but this is not an issue since the dependence on mass and $p_T$ of the thresholds is sufficiently smooth. The values of $(X_g^{\varepsilon_b},X_q^{\varepsilon_b})$ are found using Powell's conjugate direction method~\cite{powell} and verifying that the fractions of quarks and gluon jets fulfiling the conditions (i) and (ii) above both correspond to $\varepsilon_b$ within a 10\% relative uncertainty. If the solution provided does not match the required precision, a brute force scan is performed.

Fixed efficiencies $\varepsilon_b = 0.2$, 0.1 and 0.05 are considered, and the computed thresholds are presented in Fig.~\ref{fig:thr3}. Their smooth variation with mass and $p_T$ in most regions shows that the conditions (i), (ii) above lead to a unique solution. For bins at high $m_J$ with lower statistics there are visible fluctuations in the case $\varepsilon_b= 0.05$. These fluctuations lead to small departures from the target efficiency $\varepsilon_b= 0.05$, but they have a minor impact for the examples discussed: they would only become a problem for datasets where the number of events is large even after the application of the tagger.

The mass-decorrelated tagger is tested on the inclusive jet samples JS1$+$JS2. The jet mass distributions without tagging and with $\varepsilon_b = 0.05$ are presented in the left column of Fig.~\ref{fig:dec3} for jets in three narrow $p_T$ intervals $[500,600]$, $[1000,1100]$ and $[2000,2100]$ GeV. As it can be seen, the $m_J$ distribution is well preserved across all the $p_T$ range, for jets that are initiated by quarks and gluons in different proportions (depending on $p_T$). The limited statistics of high-mass jets in the decorrelation samples makes some features visible at high $m_J$, especially for the sample with $\ptj \in [500,600]$ GeV.

We test whether the jet tagging alters the $(m,\ptj,\eta)$ distribution of the jets by using the mixed-sample estimator $T$. We consider the same $p_T$ intervals as in the left column of Fig.~\ref{fig:dec3}. In each interval we consider three pools of jets:
\begin{itemize}
\item $\mathcal{P}_A$ with one tenth of the jets.
\item $\mathcal{P}_B$ with the remaining $9/10$.
\item The subset $\mathcal{P}_{B'}$ with those jets in $\mathcal{P}_B$ that pass the tagger with $\varepsilon_b= 0.05$, which has around $1/20$ of the size of $\mathcal{P}_B$.
\end{itemize}
The pools $\mathcal{P}_A$ and $\mathcal{P}_B$, as well as $\mathcal{P}_A$ and $\mathcal{P}_{B'}$, are statistically independent. For the $p_T$ intervals under consideration, the pools $\mathcal{P}_A$  contain around $290$K, $290$K and $170$K events, respectively, and the pools $\mathcal{P}_{B'}$ contain around $130$K, $130$K and $80$K.

We perform (a) $N=10000$ pseudo-experiments, drawing $n_A = n_B = 2000$ random events from $\mathcal{P}_A$ and $\mathcal{P}_B$; and (b) the same, but drawing the second set of events from $\mathcal{P}_{B'}$. In each pseudo-experiment we calculate $T$ with $N_k = 100$. The resulting p.d.f.'s of $T$ obtained are shown in the right column of Fig~\ref{fig:dec3}. 
The distributions of $T$ obtained when events are all drawn from the untagged jet pools agree very well with the expectation of $\mu_0 = 1/2$ for events following the same distribution, c.f. (\ref{ec:musigma}) with $n_A = n_B$, and the standard deviation also agrees very well with the limit value.
In the three $\ptj$ bins considered, the p.d.f. of $T$ when comparing tagged ($\mathcal{P}_{B'}$) versus untagged ($\mathcal{P}_A$) samples exhibits deviations of $0.29\sigma$, $0.09\sigma$ and $0.10\sigma$, with respect to the p.d.f. for $\mathcal{P}_{B}$ versus $\mathcal{P}_{A}$. In neither of these cases the deviation is significant, and for the lower $\ptj$ bin it could be reduced with more statistics to compute the thresholds for mass decorrelation. For comparison, we show in appendix~\ref{sec:b} how applying a fixed cut on $P_s$ severely distorts the mass distribution, leading to shifts of several standard deviations in the mean value of $T$.

In conclusion, the exclusive mass decorrelation with \Gent\ allows to use the mixed sample estimator $T$ and the jet mass as variable, in order to search for anomalies in events with tagged jets. The extra difficulty of having to deal with a three-class tagger and performing the decorrelation for quark and gluon jets separately, is compensated by the fact that the decorrelation serves for any jet sample, by construction.

\subsection{In-situ mass decorrelation}
\label{sec:4.2}

With a tagger such as \gentx\ that does not attempt to discriminate quarks from gluons, an in situ mass decorrelation must be performed. The tagger \gentx\ provides as output the probability $P_s$ that a jet corresponds to signal. Then, for a target background efficiency $\varepsilon_b$, and in a given bin of jet mass and $p_T$, one determines the value $X_j^{\varepsilon_b}$ such that the fraction of jets with $P_s \geq X_j^{\varepsilon_b}$ equals $\varepsilon_b$.

\begin{figure*}[p]
\begin{center}
\begin{tabular}{cc}
\includegraphics[width=9cm,clip=]{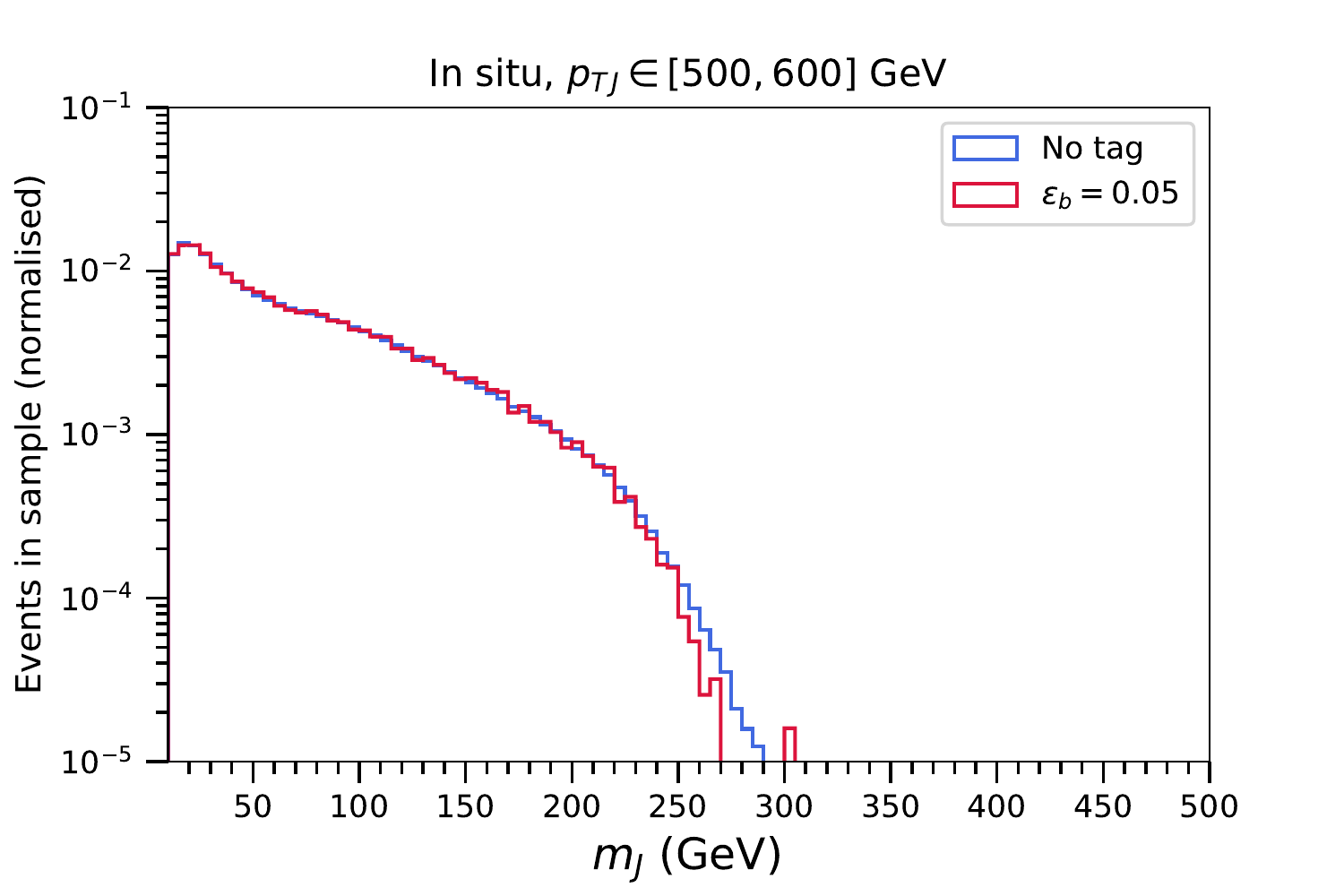} &
\includegraphics[width=9cm,clip=]{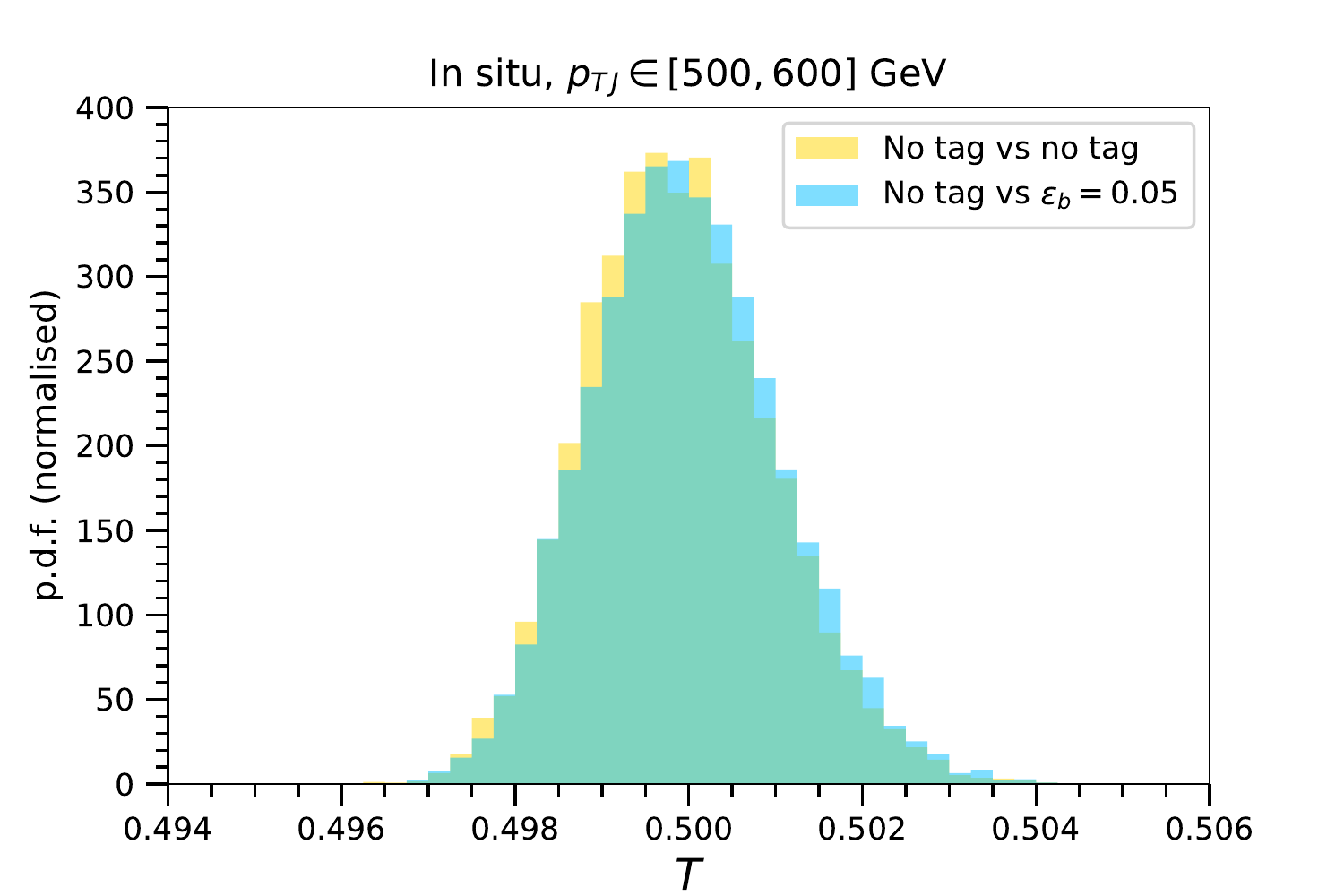} \\
\includegraphics[width=9cm,clip=]{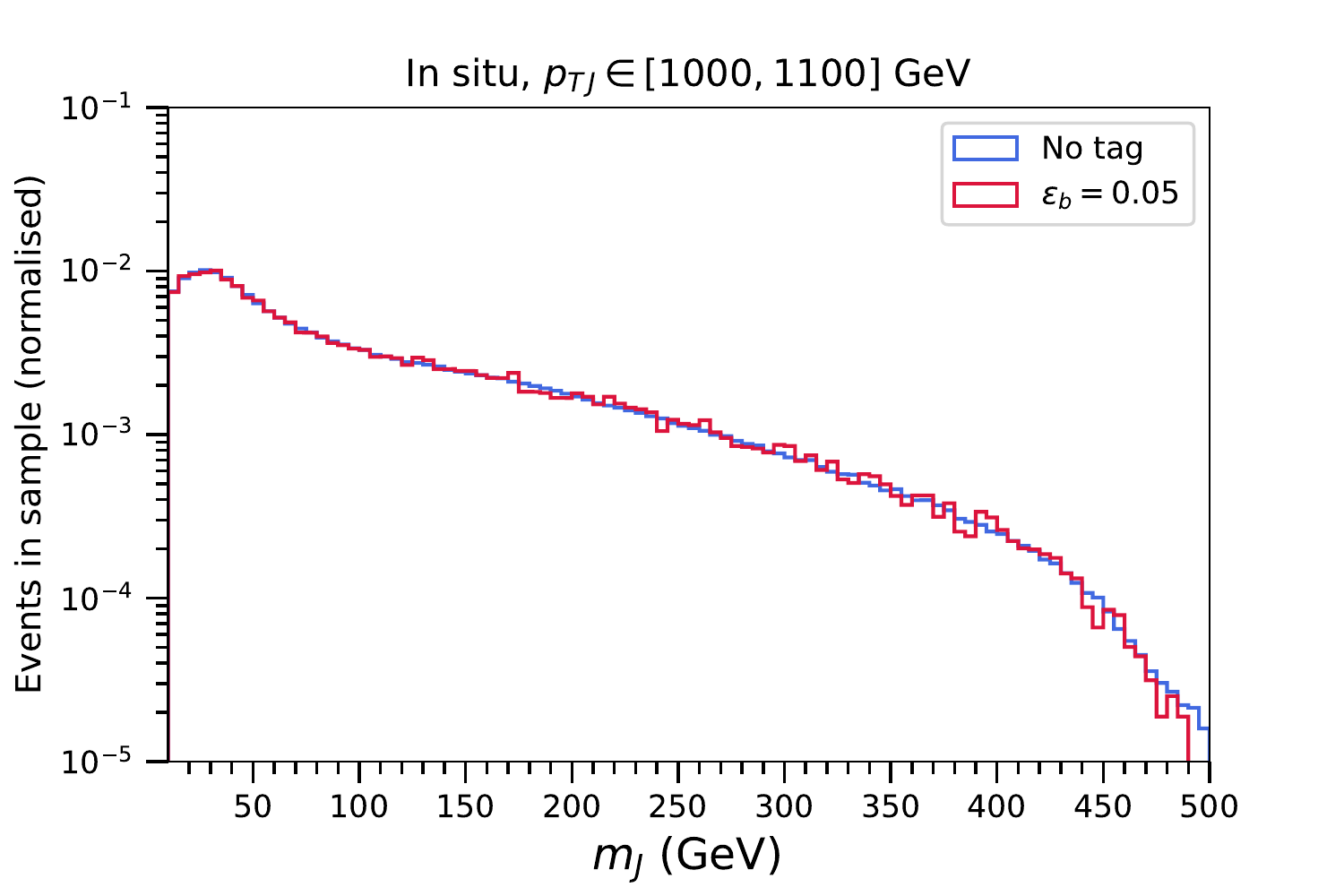} &
\includegraphics[width=9cm,clip=]{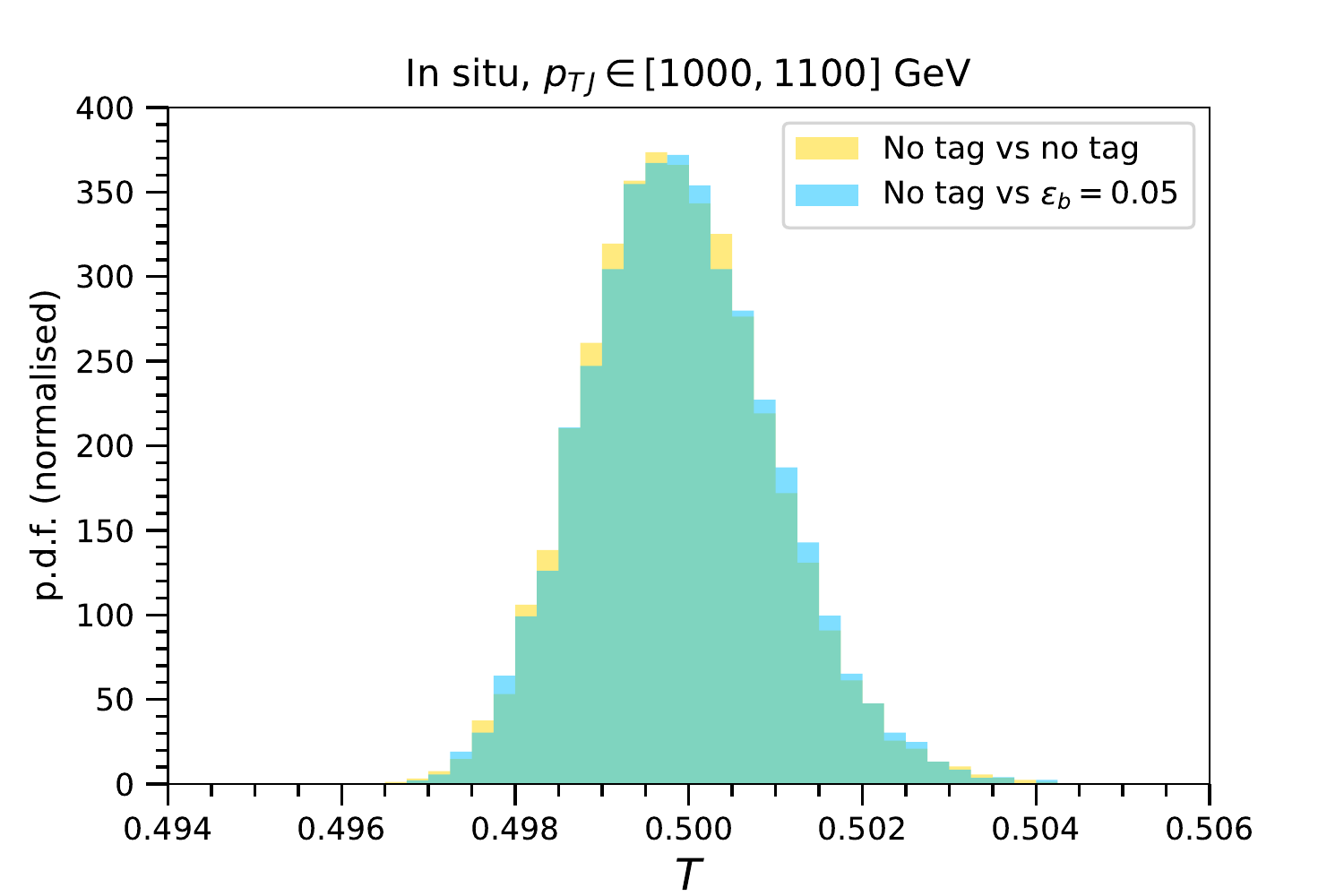} \\
\includegraphics[width=9cm,clip=]{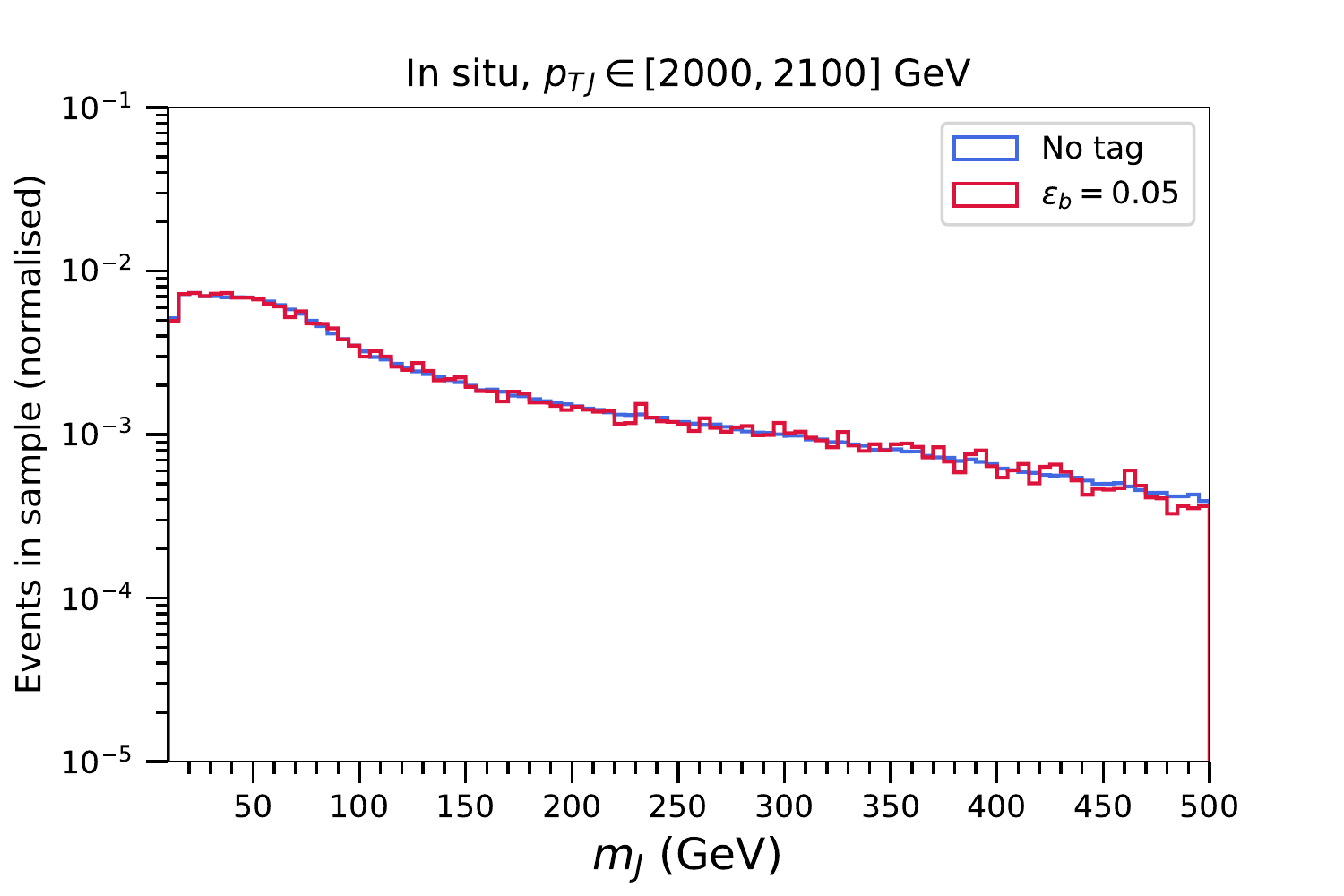} &
\includegraphics[width=9cm,clip=]{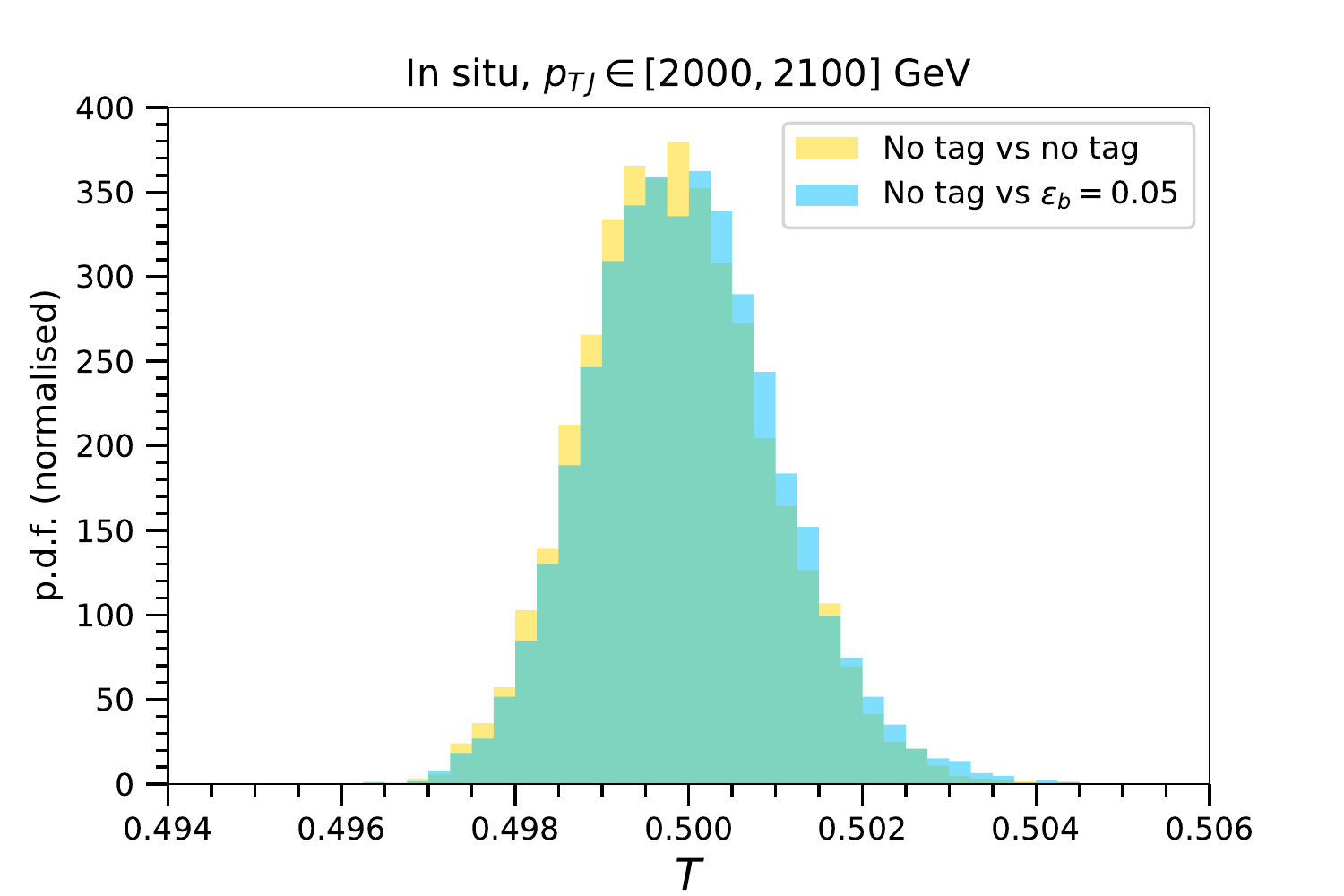}
\end{tabular}
\caption{Left: jet mass distribution for QCD jets in three narrow $p_T$ samples, before tagging and after the application of the \gentx\ tagger and in-situ mass decorrelation with $\varepsilon_b = 0.05$. Right: probability density functions for the estimator $T$, obtained when the two samples are drawn from the untagged jet pool (yellow) or from the untagged and tagged jet pools (blue).}
\label{fig:dec1}
\end{center}
\end{figure*}

As in the previous subsection, we divide the $m_J$ range $[10,500]$ GeV in 5 GeV bins, and the $p_T$ range $[200,2200]$ GeV in 50 GeV bins.\footnote{By using $\ptj$ and $\rho = 2 \log m_J / \ptj$, as in Refs.~\cite{CMS:2017dcz,Aguilar-Saavedra:2020uhm}, the decorrelation is not good for large jet momenta $\ptj \gtrsim 1.5$ TeV.} In each bin, we calculate the thresholds for fixed efficiencies $\varepsilon_b = 0.2$, $0.1$, $0.05$ and also $0.01$. In the computation of the efficiency percentiles we use the inclusive set JS1 excluding those jets used in the tagger training (but not those used for validation). The obtained thresholds are presented in Fig.~\ref{fig:thr1}. Their variation is quite smooth with mass and $p_T$ across all the range, even for $\varepsilon_b = 0.01$.

The mass-decorrelated tagger is tested on the inclusive jet sample JS2. The jet mass distributions without tagging and for $\varepsilon_b = 0.05$ are presented in the left column of Fig.~\ref{fig:dec1} for jets in the same $p_T$ intervals $[500,600]$, $[1000,1100]$ and $[2000,2100]$ GeV. The $m_J$ distribution is well preserved across all the $p_T$ range by construction, and the deviations for large $m_J$ in the sample with $\ptj \in [500,600]$ are smaller than with the exclusive decorrelation method, as expected. 

The effect in the $(m,\ptj,\eta)$ distribution of the jets is tested as in the previous subsection, using the inclusive jet sample JS2 to obtain pools of untagged jets and tagged with $\varepsilon_b = 0.05$. The pools of untagged jets contain around $120$K, $120$K and $80$K events, respectively, and the pools of tagged jets contain around $60$K, $60$K and $40$K. The p.d.f. distributions of $T$ are shown in the right column of Fig.~\ref{fig:dec1}. The shifts in the mean value are quite small, $0.09\sigma$, $0.03\sigma$ and $0.10\sigma$. Additional tests are performed for $\varepsilon_b = 0.01$, showing no significant deviations in $T$ when comparing tagged and untagged samples. Therefore, an in situ mass decorrelation can also be implemented to use the mixed sample estimator $T$ to search for anomalies in events with tagged jets.

\section{Application to $JJ$ final states}
\label{sec:5}

Let us consider a final state with two boosted large-radius jets $J$. We address the discovery potential for a heavy $Z'$ resonance with cascade decay into new scalars, yielding two massive four-pronged jets. Explicit models where this cascade decay can happen at observable rates, given the current limits on $Z'$ bosons from other decay modes, range from the minimal implementation~\cite{Aguilar-Saavedra:2019adu} to a full-fledged supersymmetric model~\cite{Aguilar-Saavedra:2021smc}. We consider four benchmark points for $M_{Z'} = 2.2$, 3.3 TeV in two decay channels:
\begin{itemize}
\item[(i)] $Z' \to S_1 S_2$, $S_{1,2} \to AA \to 4b$, with $M_{S_1} = 100$ GeV, $M_{S_2} = 80$ GeV, $M_A = 30$ GeV;
\item[(ii)] $Z' \to S_1 S_2$, $S_{1,2} \to WW \to 4q$, with $M_{S_1} = 300$ GeV, $M_{S_2} = 200$ GeV.
\end{itemize}
These two signals are denoted as $Z' \to 8b$, $Z' \to 4W$ in the following.
The first one is particularly elusive --- it has been dubbed as `stealth boson' in Ref.~\cite{Aguilar-Saavedra:2017zuc} --- and constitutes a very demanding test of the anomaly search method proposed here. We only test four-pronged signals in this section. The tagger performance for two-pronged signals is similar (see Fig.~\ref{fig:ROC}) and the ability of the mixed-sample estimator to spot mass bumps is obviously the same.

We present in turn results for exclusive and in-situ mass decorrelation. In both cases the significance improvement obtained is about the same size, but slightly better in the latter due to a larger signal tagging efficiency for the same background rejection.

\subsection{Exclusive mass decorrelation}
\label{sec:5.1}

For the jet tagging we select the tagger \Gent\ with exclusive mass decorrelation and $\varepsilon_b = 0.05$. From the ROC curves for this type of signals in Fig.~\ref{fig:ROC} one can observe that a tighter background rejection would increase the sensitivity. Even if the optimal point is signal-dependent, for a wide variety of signals (see Ref.~\cite{Aguilar-Saavedra:2020uhm}), and in particular for those studied here, a better significance improvement $s_T$ is achieved for larger background rejections. However, the available Monte Carlo statistics do not permit a reliable calculation of thresholds for exclusive mass decorrelation with $\varepsilon_b = 0.01$. Therefore, in this respect the results presented here are conservative and not optimised. Results with a tighter tagger selection are given in the next subsection, with in-situ decorrelation.

\begin{figure}[t]
\begin{center}
\begin{tabular}{c}
\includegraphics[width=9cm,clip=]{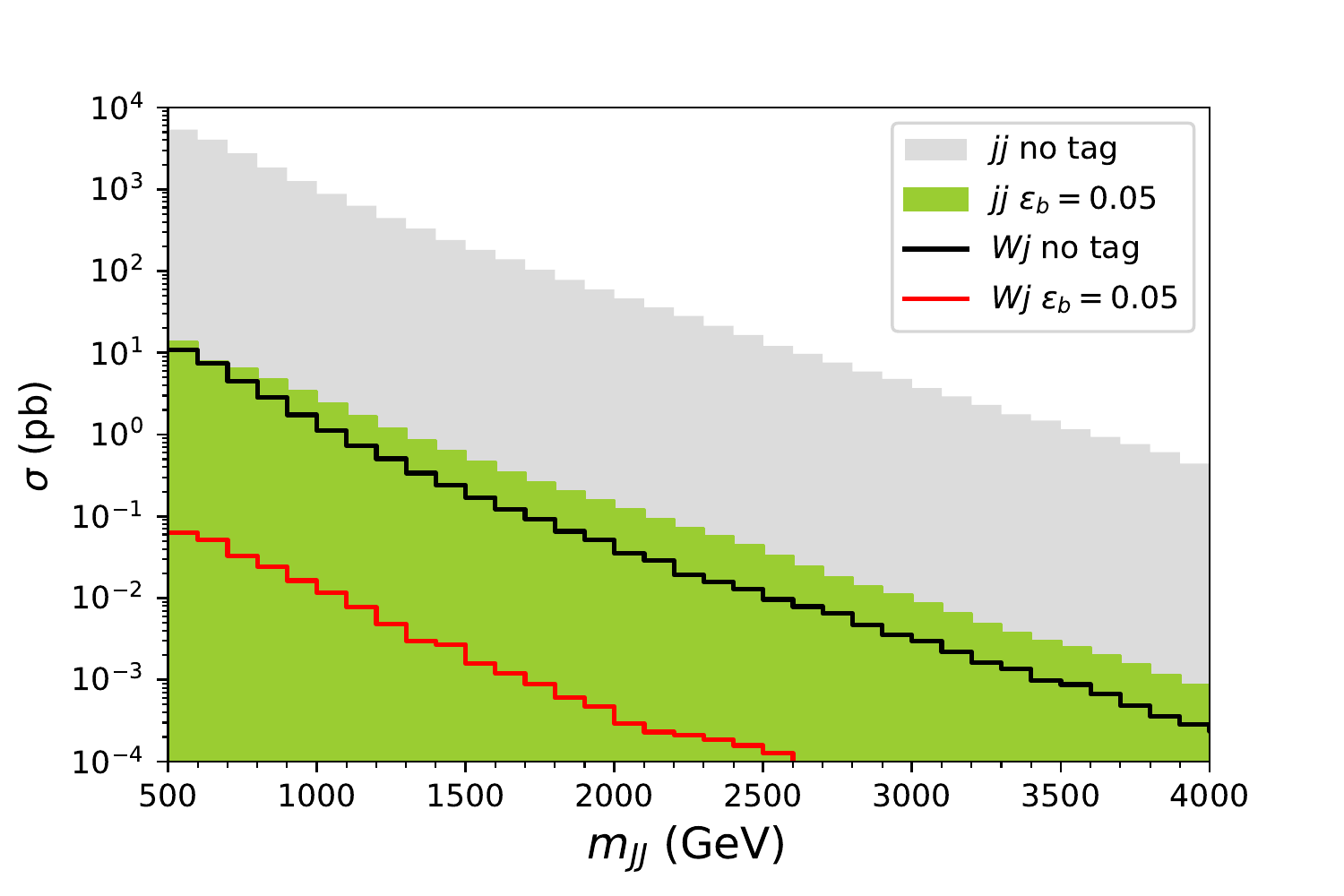} \\
\includegraphics[width=9cm,clip=]{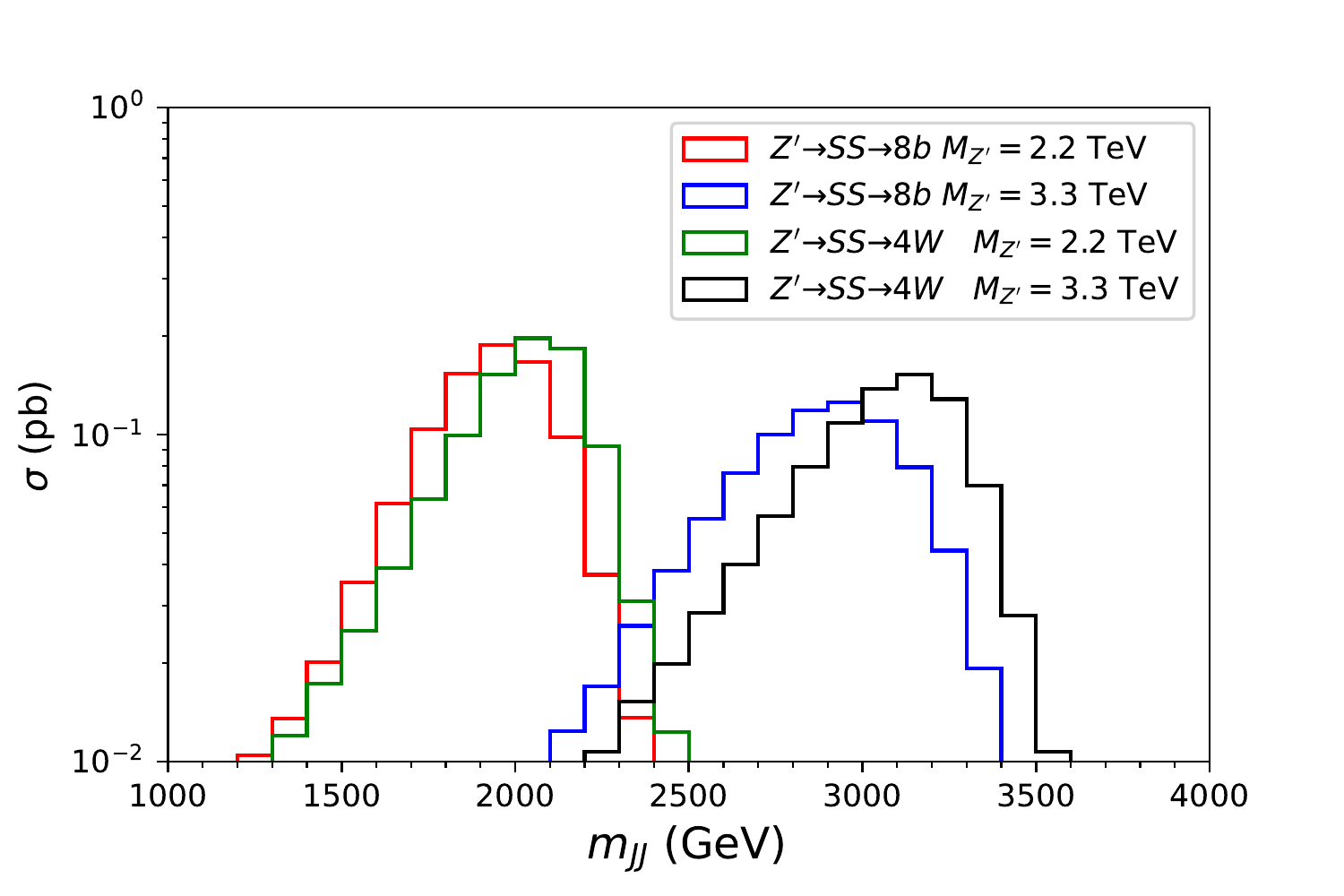}
\end{tabular}
\caption{Top: $JJ$ invariant mass distribution for the QCD dijet ($jj$) and $Wj$ backgrounds before and after jet tagging. Bottom: the same, for the signals before jet tagging, with exclusive mass decorrelation.}
\label{fig:mJJ-2body}
\end{center}
\end{figure}

As event selection we require that events contain two large-radius $R=0.8$ jets (as used throughout the paper) with mass $m_J \geq 10$ GeV, transverse momentum $\ptj \geq 200$ GeV and pseudo-rapidity $|\eta| \leq 2.5$. 
Figure~\ref{fig:mJJ-2body} (top) shows the distribution of the $JJ$ invariant mass $m_{JJ}$ for the QCD background, before and after the tagging.\footnote{At low $m_{JJ}$ the efficiency of the triggers is small, see e.g. Ref.~\cite{CMS:2019qem}, but for $m_{JJ} \geq 1$ TeV it is nearly 100\%. We prefer to include the $500-1000$ GeV mass range to better observe how the varying-threshold mass decorrelation preserves the shape of the $m_{JJ}$ distribution.} We use the inclusive samples JS1$+$JS2. We also include the electroweak background from $Wj$ production with $W \to q \bar q$, generated in 100 GeV bins of $p_T$. It turns out that its contribution is negligible and therefore it is neglected in the following, as well as $Zj$ with $Z \to q \bar q$, whose cross section is smaller.

The signal normalisation is fixed so that in the untagged sample the excess $n_\text{sig}/\sqrt n_\text{bkg}$ (being $n_\text{sig}$ and $n_\text{bkg}$ the number of signal and background events, respectively) summing the three 100 GeV bins with largest signal, amounts to $1\sigma$ for a reference luminosity $L = 20$ fb$^{-1}$. This luminosity is chosen so as to have a moderate number of background events of the order of $10^2$-$10^3$ after tagging with $\varepsilon_b = 0.05$ in the region of interest, so that our Monte Carlo generation has more statistics than the number of expected events.
%\footnote{For the 3.3 TeV $Z' \to 4W$ signal we have verified that with $L = 40$ fb$^{-1}$ and the corresponding signal normalisation, the significance improvement brought by the SOFIE anomaly detection method is quite similar.}
The signal to background ratio in these bins is around $6 \times 10^{-4}$ for the signals with $M_{Z'} = 2.2$ TeV and $2 \times 10^{-3}$ for the signals with $M_{Z'} = 3.3$ TeV. The $m_{JJ}$ distribution for the signals with this normalisation is presented in the bottom panel of Fig.~\ref{fig:mJJ-2body}.

The tagging of both jets already increases the signal significance $n_\text{sig}/\sqrt n_\text{bkg}$ by factors of $4-8$. But, by using the $T$ estimator on the tagged jet samples, a much larger significance can be achieved. The anomaly search is performed with a scan on narrow $m_{JJ}$ bins, comparing the features of the sample with both jets tagged with the features of the untagged sample. The width of the $m_{JJ}$ bins should be of the order of the experimental resolution. Too narrow bins suffer from smaller statistics and the mixed sample method performs worse (see section~\ref{sec:3}). On the other hand, in too wide bins the $m_{JJ}$ peak of the signal is more diluted, and additionally, the variation of the $m_J$ distribution with $\ptj$ may make it more difficult to spot anomalies. We therefore choose to analyse the sensitivity by dividing the $m_{JJ}$ spectrum in 100 GeV bins. In each bin, independently, we calculate the expected sensitivity by the following procedure that imitates, with Monte Carlo pseudo-data, what would be done in an experiment.
\begin{enumerate}
\item The number of `observed' events $n$ equals the expected background plus the injected signal, $n = n_\text{bkg} + n_\text{sig}$, with the assumed luminosity $L = 20$ fb$^{-1}$. Notice that this number may already exhibit an excess, characterised by the pull $p_1 = n_\text{sig} / \sqrt{n_\text{bkg}}$ (in all relevant cases the number of events is sufficiently large so as to assume Gaussian statistics).
\item The distribution of the $T$ estimator for the `null hypothesis' is obtained with 2000 pseudo-experiments where $n$ random events are drawn from a pool $\mathcal{P}_A$ of untagged background events, and $n$ events from a pool $\mathcal{P}_{B1}$ of tagged background events.\footnote{For the test of the 2.2 TeV signals we use the tagged samples with $\varepsilon_b = 0.1$, in order to have sufficient Monte Carlo statistics in the event pools. Results in section~\ref{sec:4} validate proceeding in this fashion: the mixed-sample comparison of jets before and after tagging in section~\ref{sec:4} shows no significant deviations for either background efficiency. We have tested with the 3.3 TeV signals that results are the same whether using the tagged jet pools with $\varepsilon_b = 0.1$ or $\varepsilon_b = 0.05$.} The procedure is analogous to the examples of sections~\ref{sec:3} and \ref{sec:4}, but now we use as variables the masses of the two jets, ordered from higher to lower. Using the masses of the leading and subleading jet as variables yields a double peak structure in the signal, and a smaller significance. The obtained p.d.f. of $T$ has $\mu_0 \simeq 1/2$, as expected. An example is shown in Fig.~\ref{fig:T-2body} for the bin $m_{JJ} \in [2100,2200]$ GeV (yellow distribution).

\begin{figure}[t]
\begin{center}
\includegraphics[width=9cm,clip=]{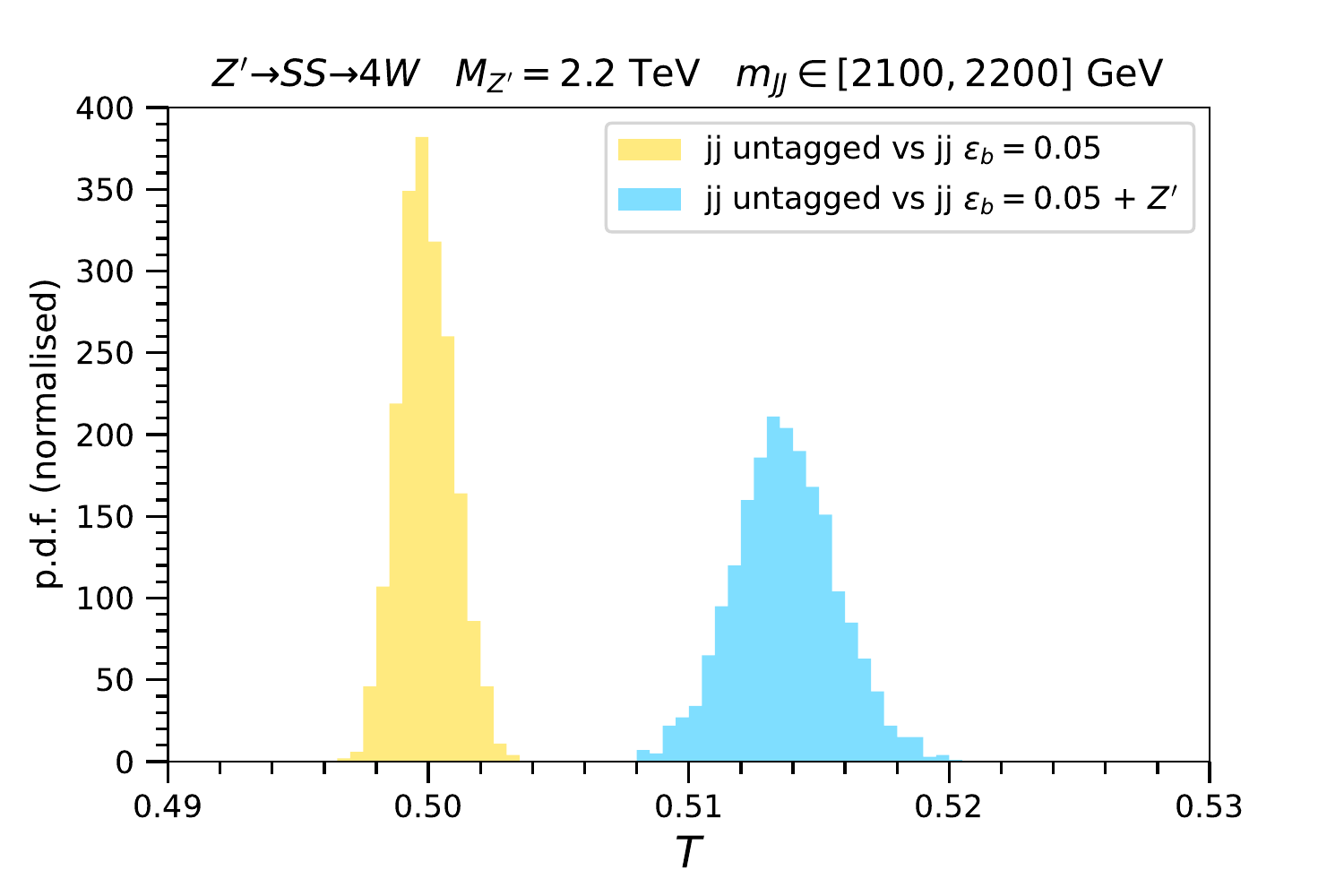}
\caption{Example of probability density functions for the estimator $T$, obtained from the comparison of untagged and tagged background events (yellow) or the same, with a signal injected (blue).}
\label{fig:T-2body}
\end{center}
\end{figure}
\item The distribution of the $T$ estimator in the presence of signal is obtained with 2000 pseudo-experiments where, as before, $n$ random events are drawn from $\mathcal{P}_A$, but now $n_\text{bkg}$ events are drawn from $\mathcal{P}_{B1}$ and $n_\text{sig}$ events are drawn from a pool $\mathcal{P}_{B2}$ of tagged signal events. An example is shown in Fig.~\ref{fig:T-2body} for the $Z' \to 4W$ signal with $M_{Z'} = 2.2$ TeV, in the bin $m_{JJ} \in [2100,2200]$ GeV (blue distribution).
\item The significance of the difference is computed, as in section~\ref{sec:3}, as $p_2 = (\mu - \mu_0)/\sigma_0$. For large values of $p_2$, these must be regarded as an estimate. Although the p.d.f. of $T$ quickly converges to a Gaussian~\cite{Williams:2010vh}, the accuracy of the approximation cannot be guaranteed at the distance of several standard deviations from the mean value.
\begin{figure*}[p]
\begin{center}
\begin{tabular}{cc}
\includegraphics[width=8.3cm,clip=]{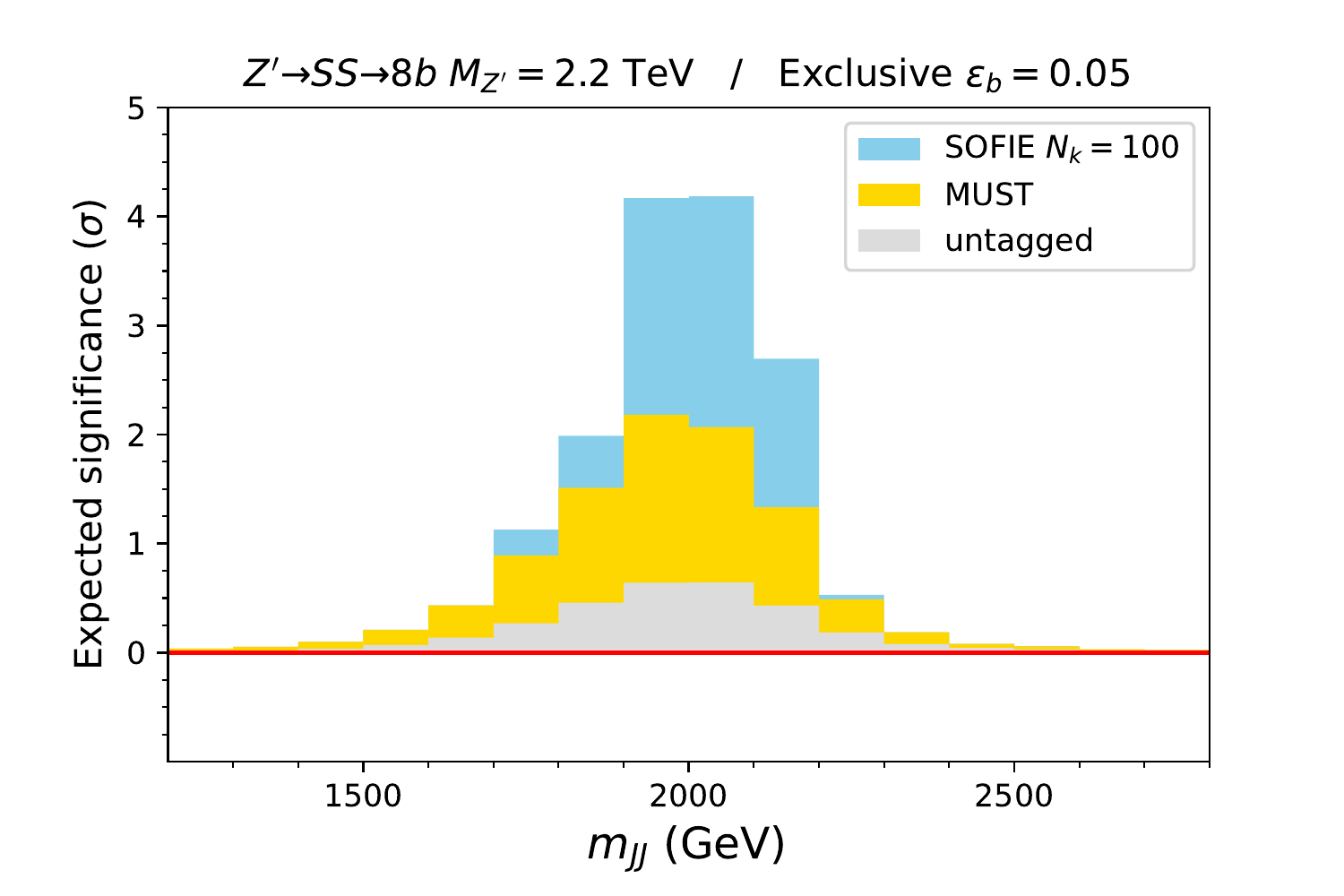} &
\includegraphics[width=8.3cm,clip=]{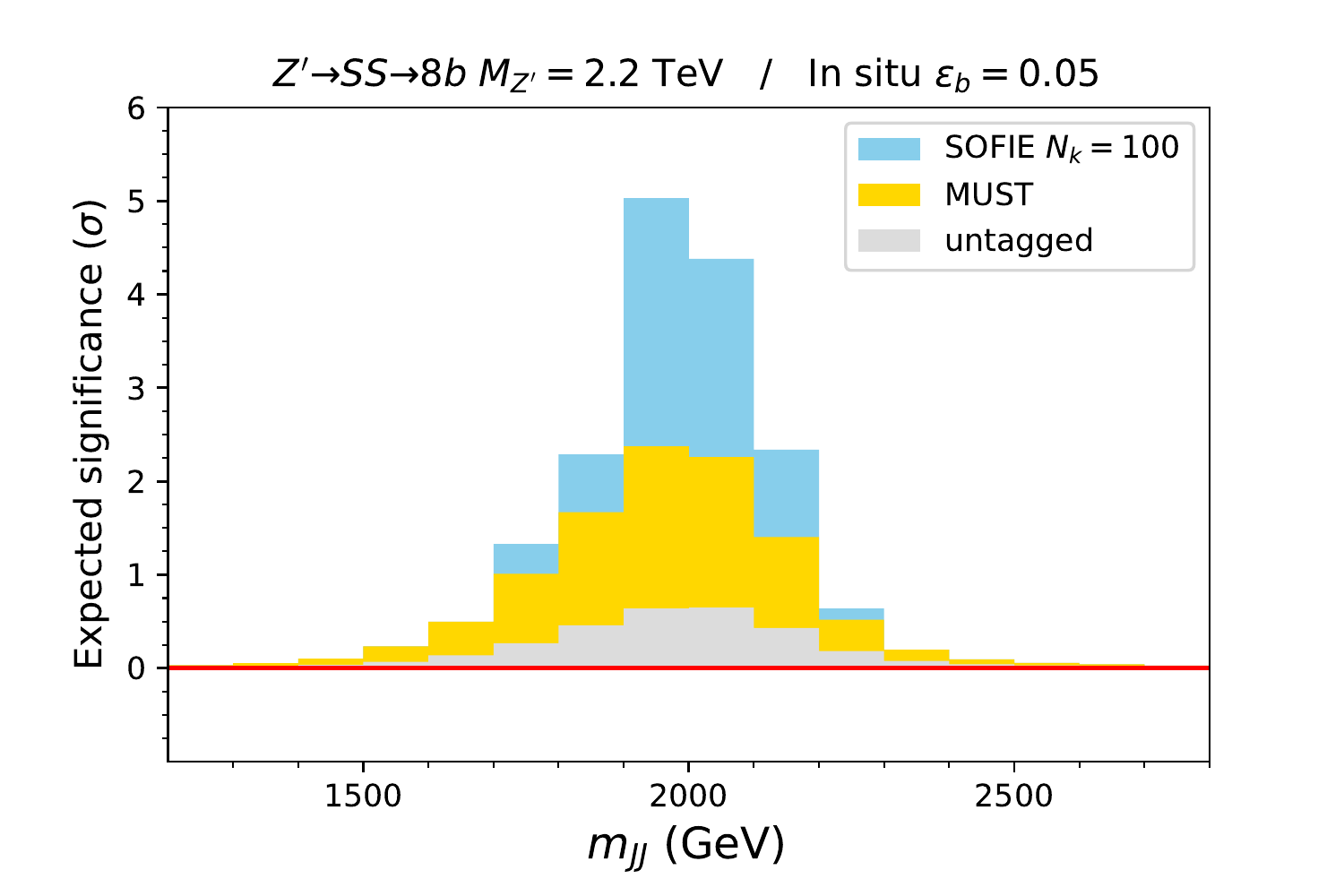} \\
\includegraphics[width=8.3cm,clip=]{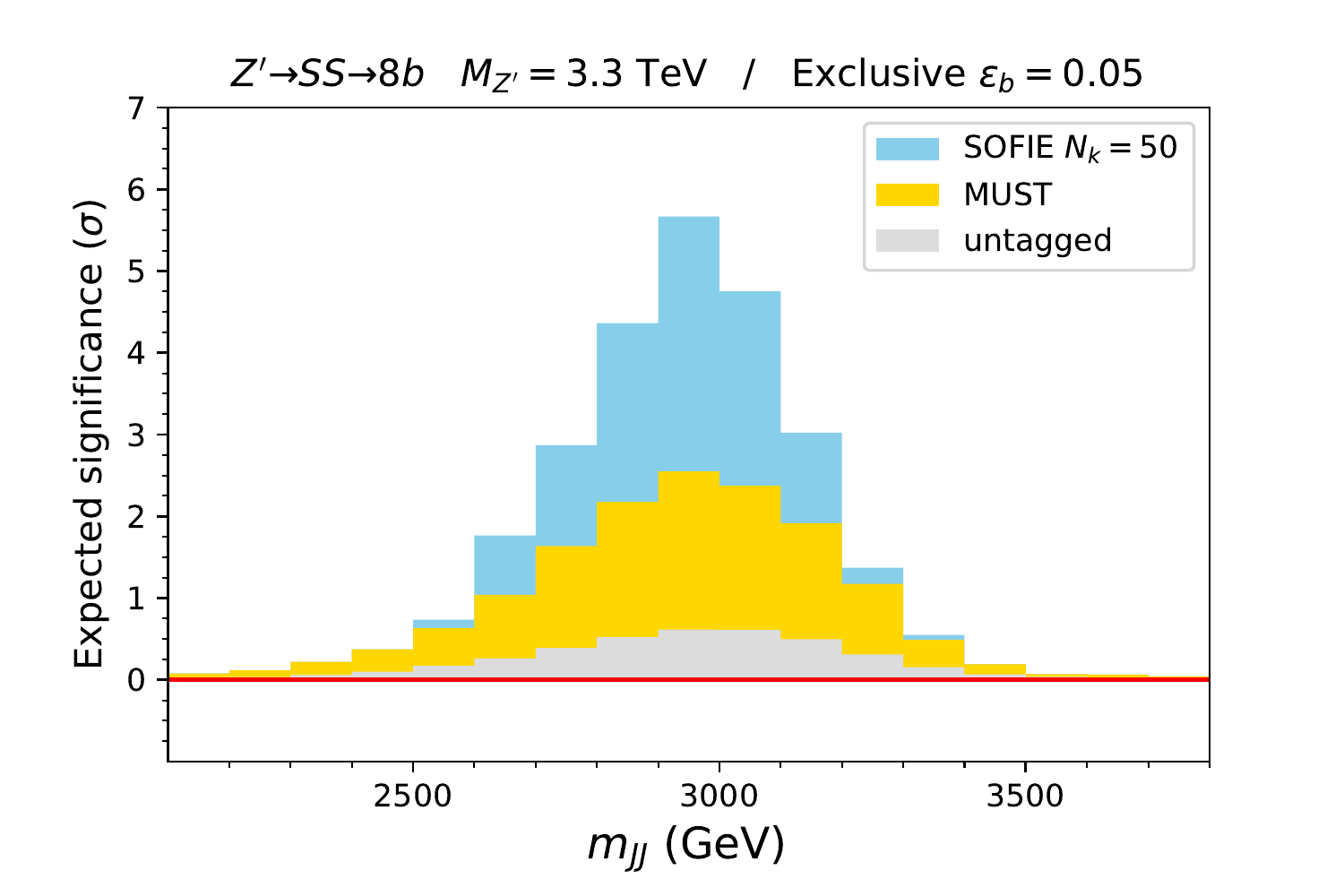} &
\includegraphics[width=8.3cm,clip=]{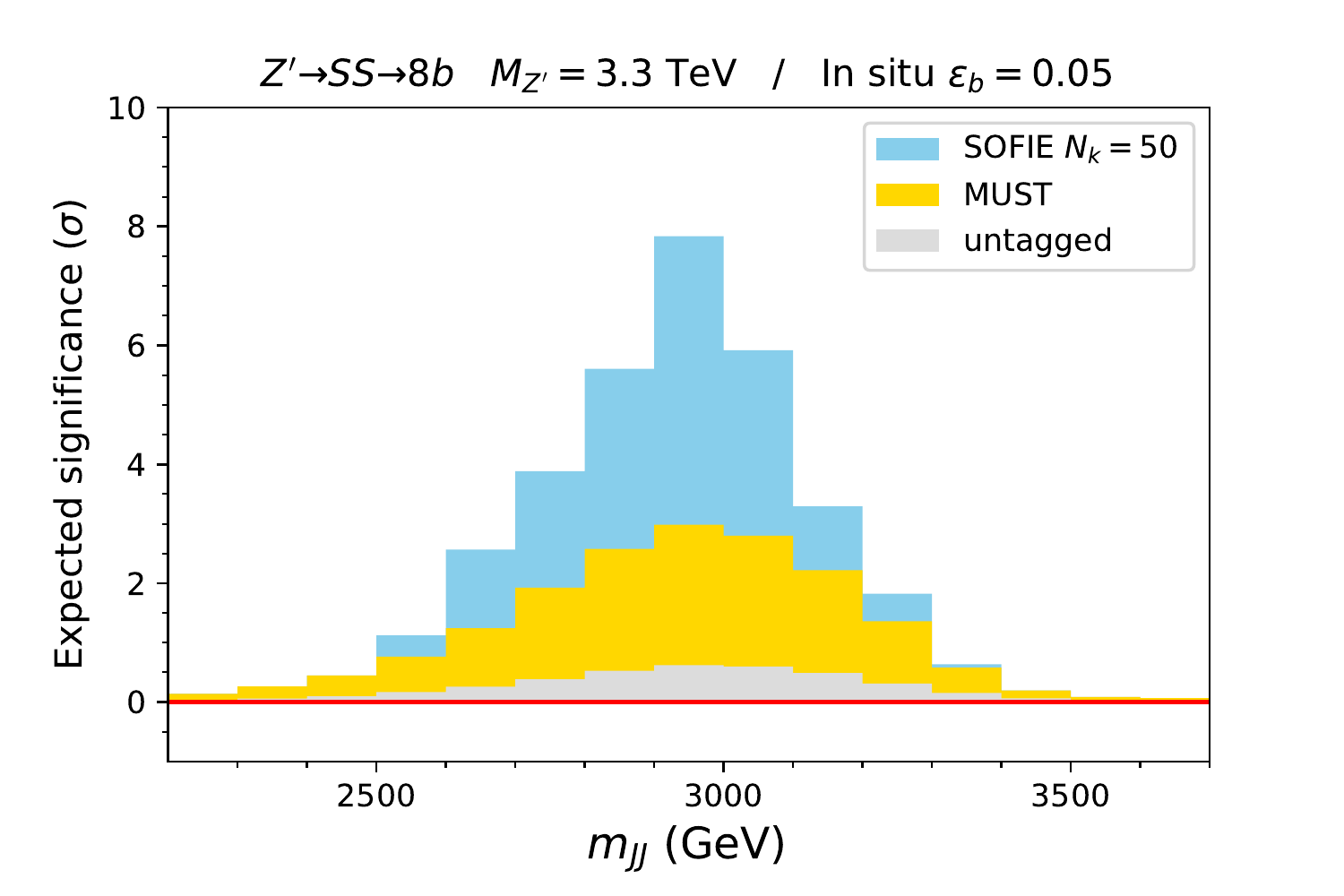} \\
\includegraphics[width=8.3cm,clip=]{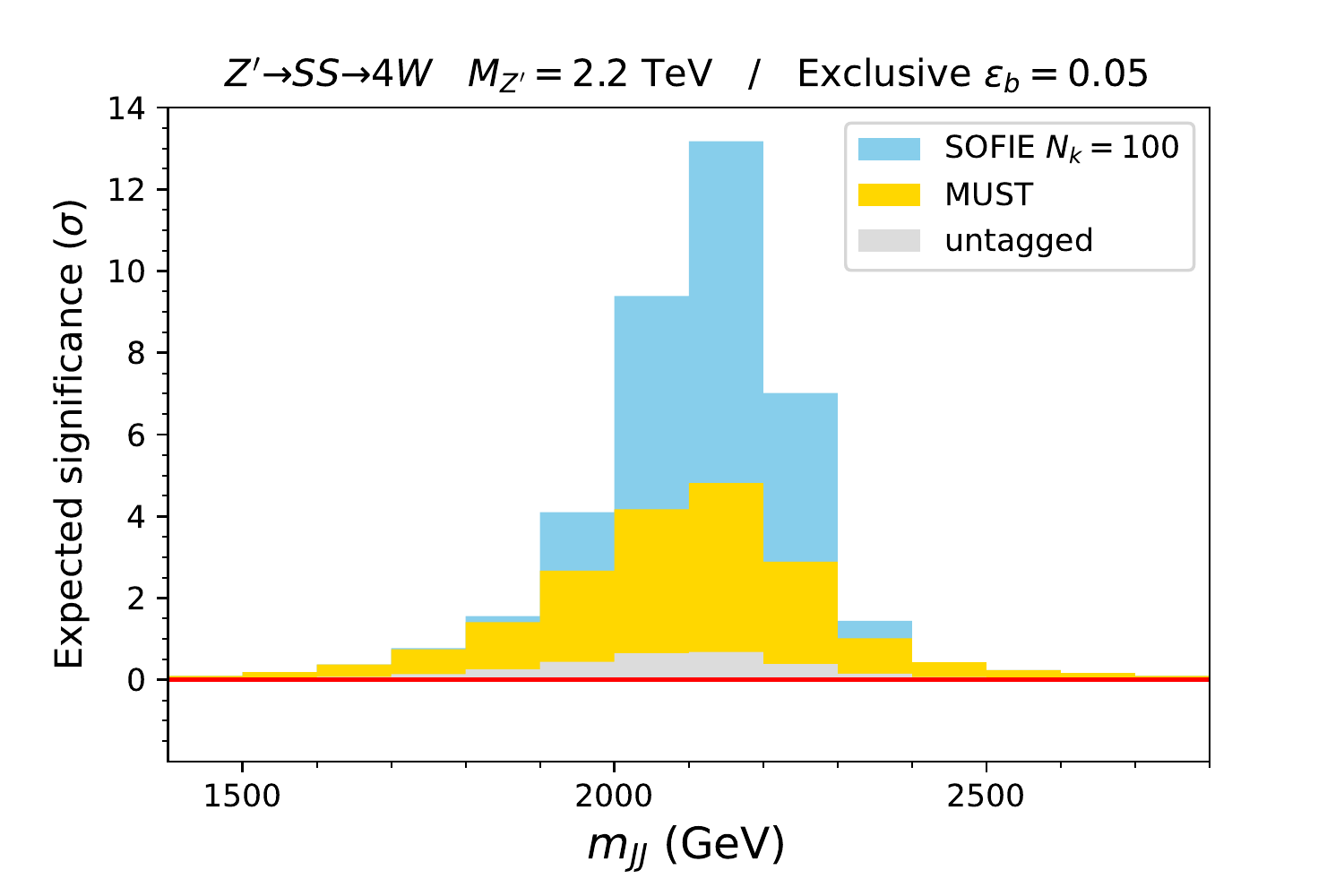} &
\includegraphics[width=8.3cm,clip=]{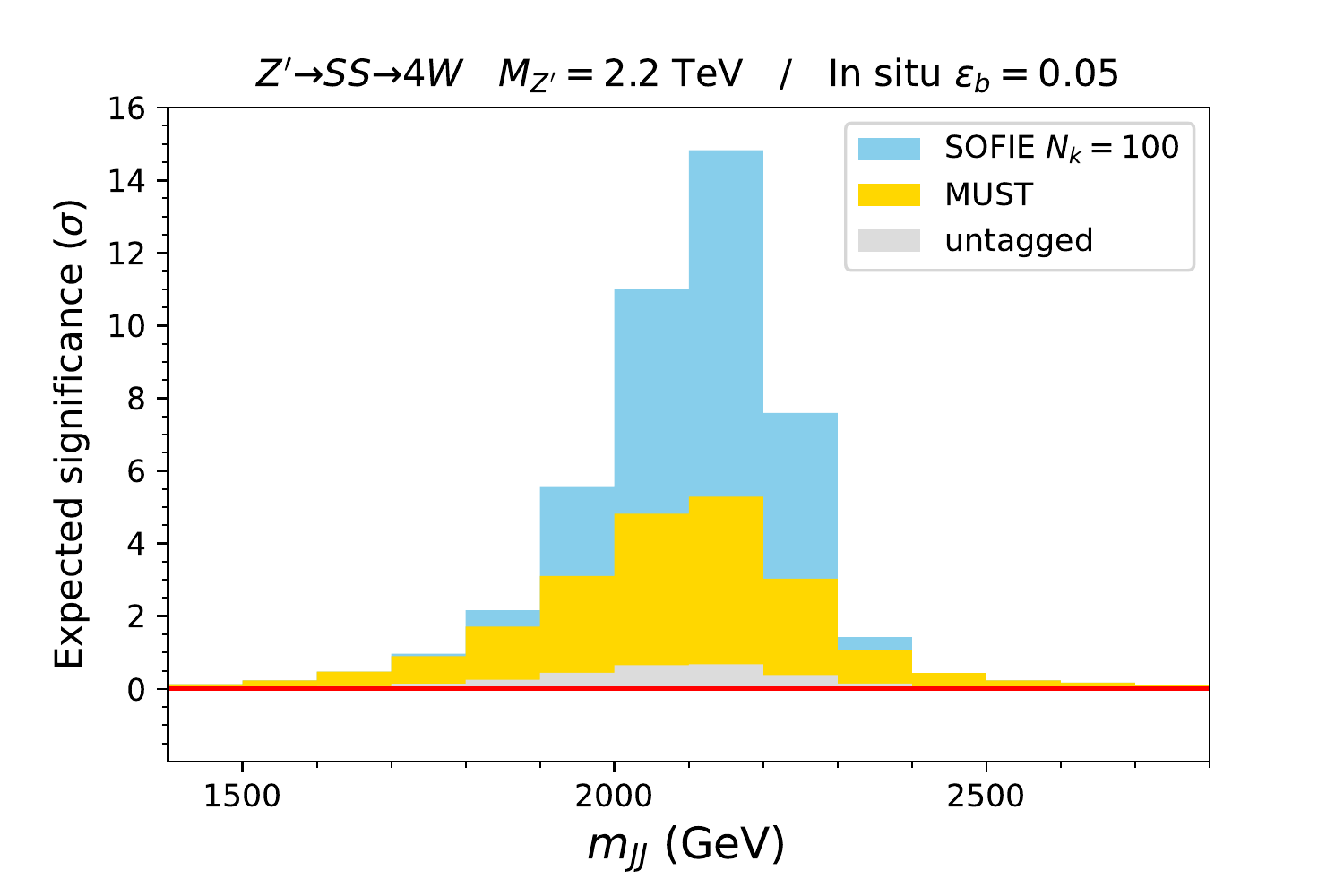} \\
\includegraphics[width=8.3cm,clip=]{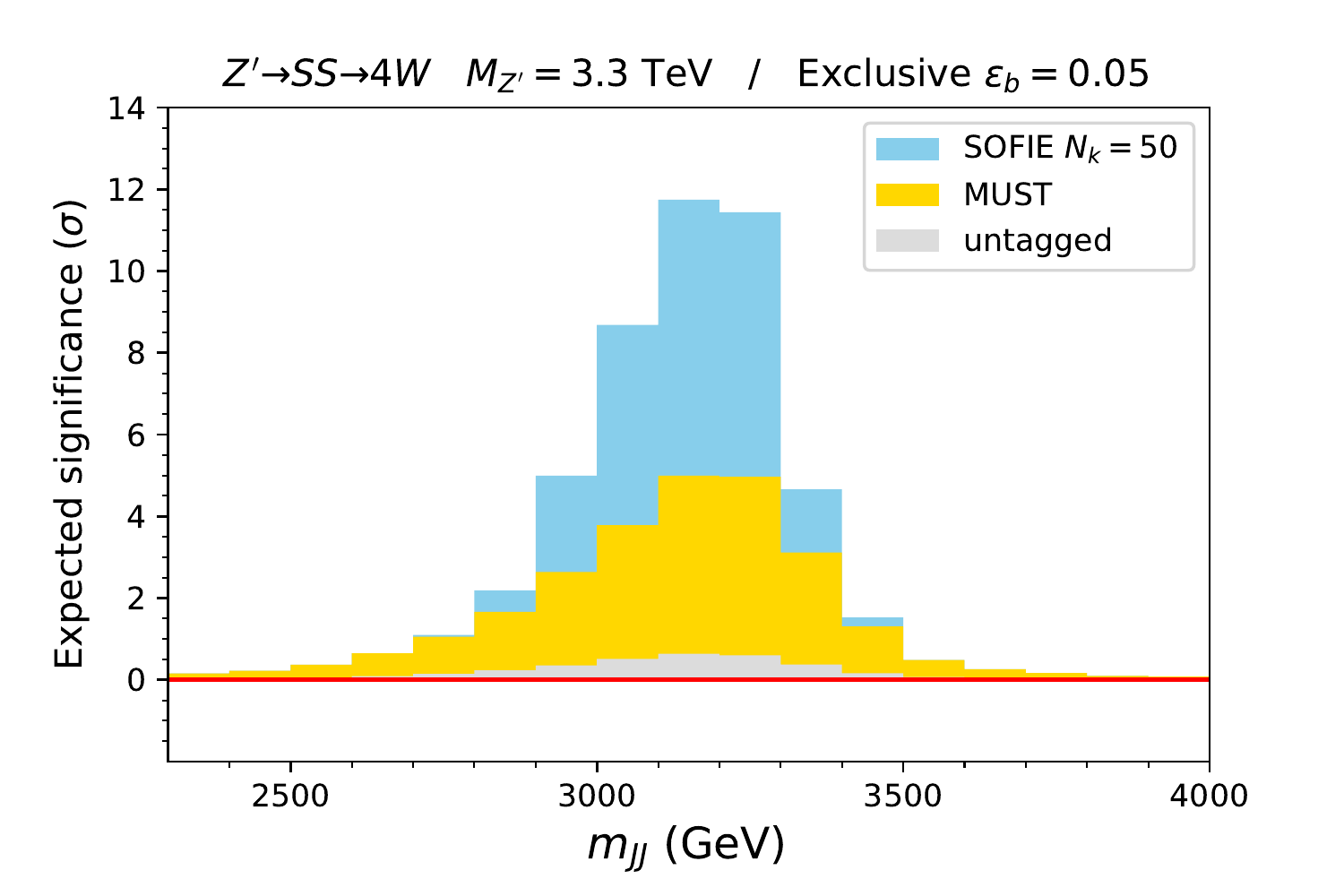} &
\includegraphics[width=8.3cm,clip=]{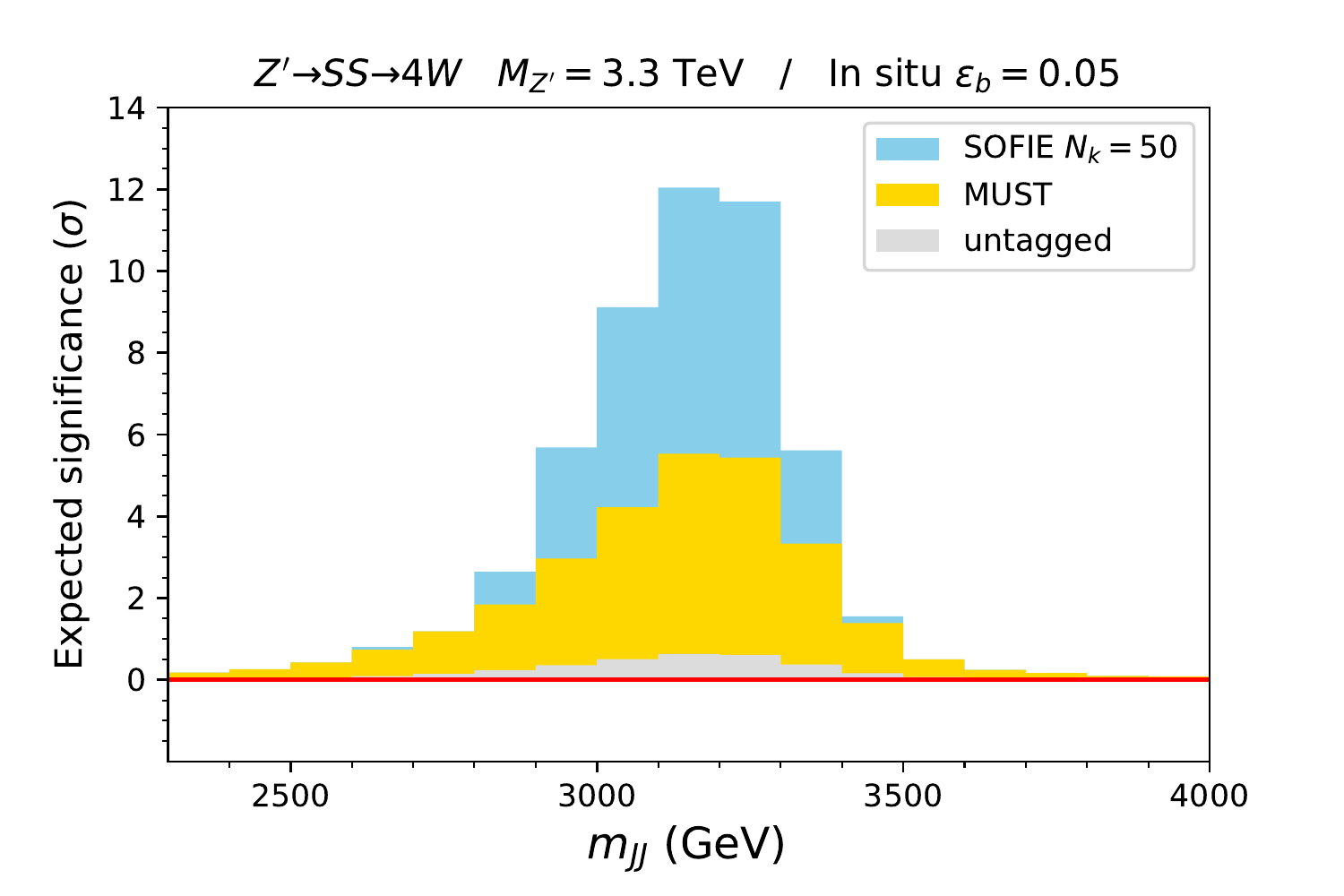}
\end{tabular}
\caption{Expected significance of the deviations per 100 GeV bin at three stages: (a) in the untagged event sample; (b) using only the MUST jet tagging; (c) using the full SOFIE anomaly detection method. Exclusive (left column) and in-situ (right column) mass decorrelation is used with $\varepsilon_b = 0.05$.}
\label{fig:signif-2body}
\end{center}
\end{figure*} 
\item Since the deviation $p_2$ is mostly independent of the deviation $p_1$ in the total number of events, we can sum both in quadrature to obtain the the combined deviation $p = (p_1^2 + p_2^2)^{1/2}$.
\end{enumerate}
The translation of this procedure to an actual measurement is straightforward. The number $n$ would correspond to the observed number of events with both jets tagged. The expected number of background events $n_\text{bkg}$ would be obtained by rescaling the measured number of events in the untagged sample, by the selected background efficiency, and $n_\text{sig} = n - n_\text{bkg}$, not necessarily positive due to statistical fluctuations. The mixed sample comparison would be done by comparing the $n$ observed events with $n$ events from the untagged sample, irrespectively of whether $n_\text{sig}$ is positive, obtaining a particular value of $T$. Needless to say, systematic uncertainties should be taken into account in the extraction of the observed deviations, and a more sophisticated combination of the deviations, also across different bins, could be performed.

For simplicity we choose here to keep the same value of $N_k$ across all $m_{JJ}$ bins, and present in the left column of Fig.~\ref{fig:signif-2body} our results using $N_k = 100$ for the 2.2 TeV signals and $N_k = 50$ for the 3.3 TeV signals. The individual values of $p_2$ for these and other values of $N_k$ can be found in appendix~\ref{sec:c}. For each $m_{JJ}$ bin, the gray bar shows the excess of events in the untagged sample, the yellow bar the pull $p_1 = n_\text{sig} / \sqrt{n_\text{bkg}}$ that results from the jet tagging alone, and the blue bar the combined deviation $p$.

The sensivity to $Z'$ signals is at least doubled by the use of the full SOFIE anomaly detection method, with respect to jet tagging alone. (Besides, in most cases $p_2 \gg p_1$, so the final sensitivity is mainly driven by $p_2$.) We also point out that the deviations $p$ in each bin are statistically independent, therefore their combination gives a much larger signal significance, around $19\sigma$ for $Z' \to 4W$ with $M_{Z'} = 3.3$ TeV. 
In the very demanding scenarios with $Z' \to 8b$ the overall significance improvement obtained is $s = 6.4$ and $s = 8.5$ for $M_{Z'} = 2.2$, $3.3$ TeV, respectively. In the two benchmarks with $Z' \to 4W$ we have $s = 17.0$, $s = 18.3$.

\subsection{In-situ mass decorrelation}
\label{sec:5.2}

The analysis follows the same steps explained in the previous subsection, first selecting $\varepsilon_b = 0.05$.\footnote{For the 2.2 TeV signals we resort to using event pools with $\varepsilon_b = 0.2$ due to insufficient statistics.} Results are presented in the right column of Fig.~\ref{fig:signif-2body},
only using the inclusive jet sample JS2 since JS1 is used for the training of the \gentx\ tagger and the mass decorrelation.  For the $Z' \to 8b$ signals with $M_{Z'} = 2.2, 3.3$ TeV the significance improvement is $s = 7.0$,  $s = 11.2$, respectively, and for the $Z' \to 4W$ signals we have $s = 19.3$, $s = 18.8$.

\begin{figure}[t]
\begin{center}
\begin{tabular}{c}
\includegraphics[width=9cm,clip=]{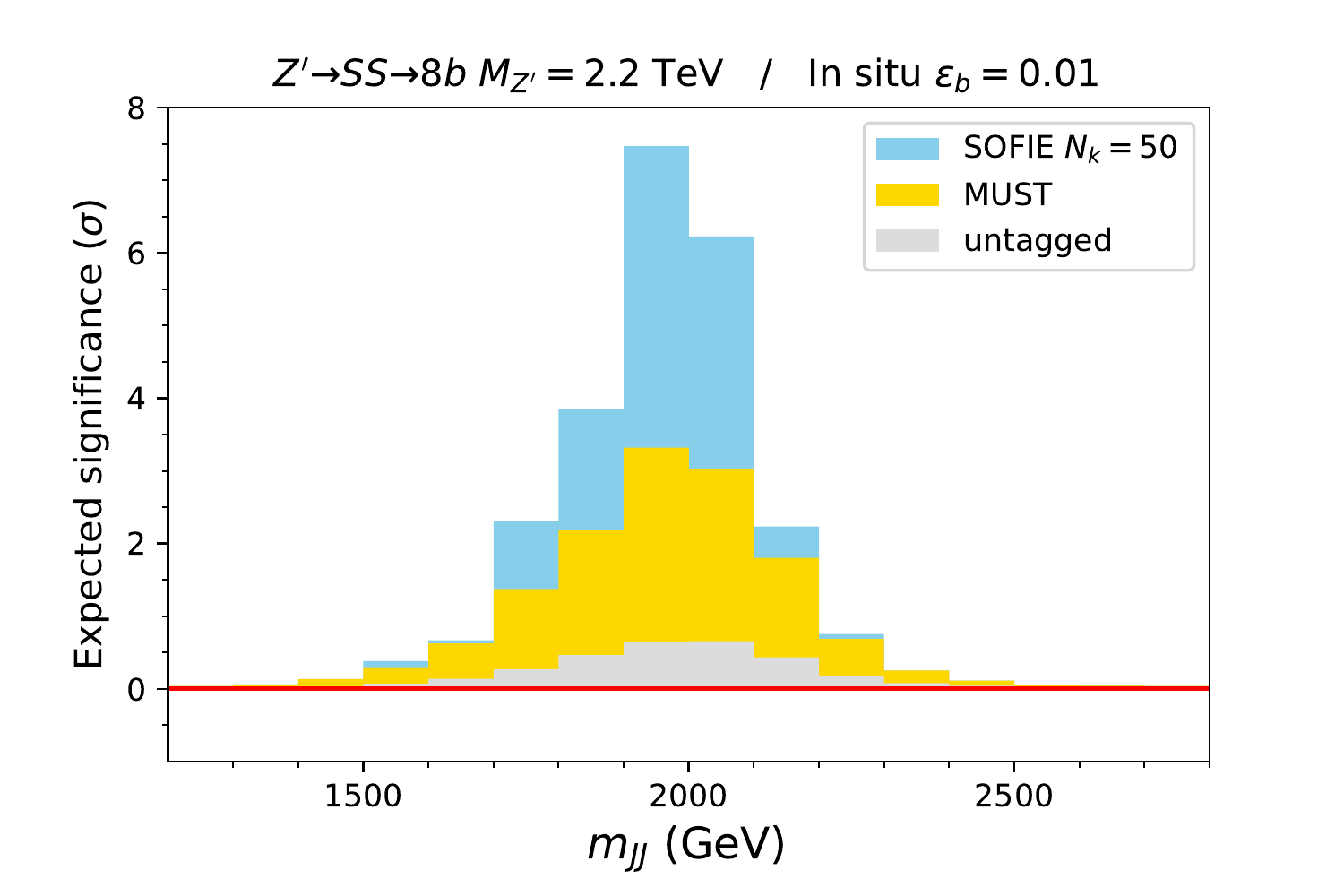} \\
\includegraphics[width=9cm,clip=]{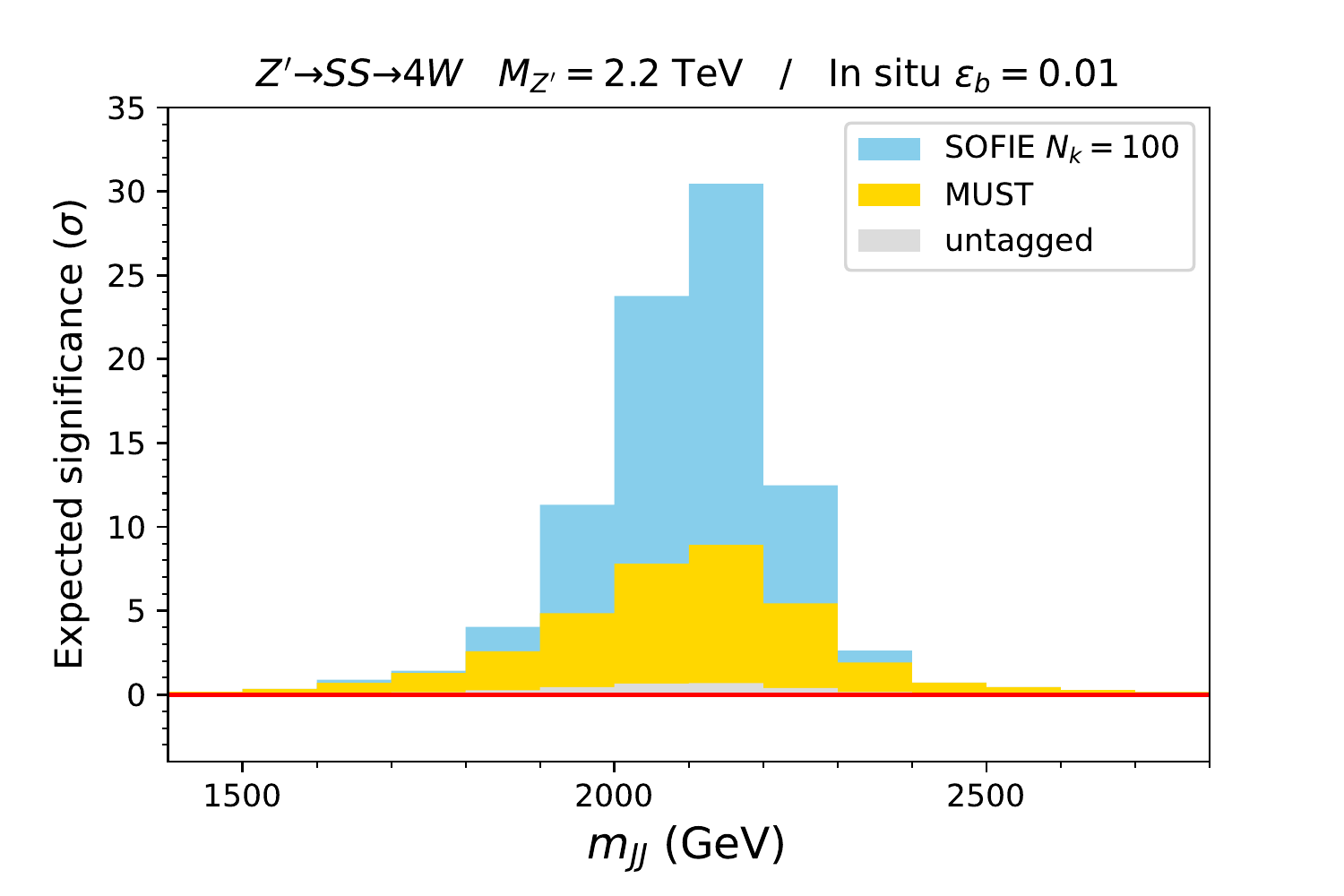}
\end{tabular}
\caption{Expected ignificance of the deviations per 100 GeV bin at three stages: (a) in the untagged event sample; (b) using only the MUST jet tagging; (c) using the full SOFIE anomaly detection method. In-situ mass decorrelation is used with $\varepsilon_b = 0.01$.}
\label{fig:signif-2body-Tgen-qu01}
\end{center}
\end{figure} 

With in-situ mass decorrelation we also investigate the performance of the SOFIE method with a more aggresive background reduction, setting $\varepsilon_b = 0.01$ (this was not possible with exclusive decorrelation because of the statistical limitation of the samples). As previously discussed and shown in Fig.~\ref{fig:ROC}, the tagging with $\varepsilon_b = 0.01$ gives a larger tagger significance improvement $s_T$, which is also the case for several other benchmark signals~\cite{Aguilar-Saavedra:2020uhm}. We keep the same luminosity $L = 20~\text{fb}^{-1}$ and restrict ourselves to the $M_{Z'} = 2.2$ TeV benchmarks. Results are presented in Fig.~\ref{fig:signif-2body-Tgen-qu01}. For $Z' \to 8b$ we find $s = 10.3$, compared to $s = 7.0$ obtained with $\varepsilon_b = 0.05$. For $Z' \to 4W$ the significance improvement reaches the impressive value $s = 39.4$, more than doubling the previously obtained value.

\section{Comparison with other methods}
\label{sec:6}

We test in this section the anomaly detection in a benchmark in which a heavy resonance decays into three-pronged jets, as considered in Ref.~\cite{Collins:2021nxn}. Specifically, we consider the benchmark with $W' \to F_1 F_2$, $F_1 \to uud$, $F_2 \to udd$ with $m_{F_1} = m_{F_2} = 300$ GeV. A sample of $2 \times 10^5$ events is generated using the card provided~\cite{card}. 

\begin{figure}[t]
\begin{center}
\begin{tabular}{c}
\includegraphics[width=9cm,clip=]{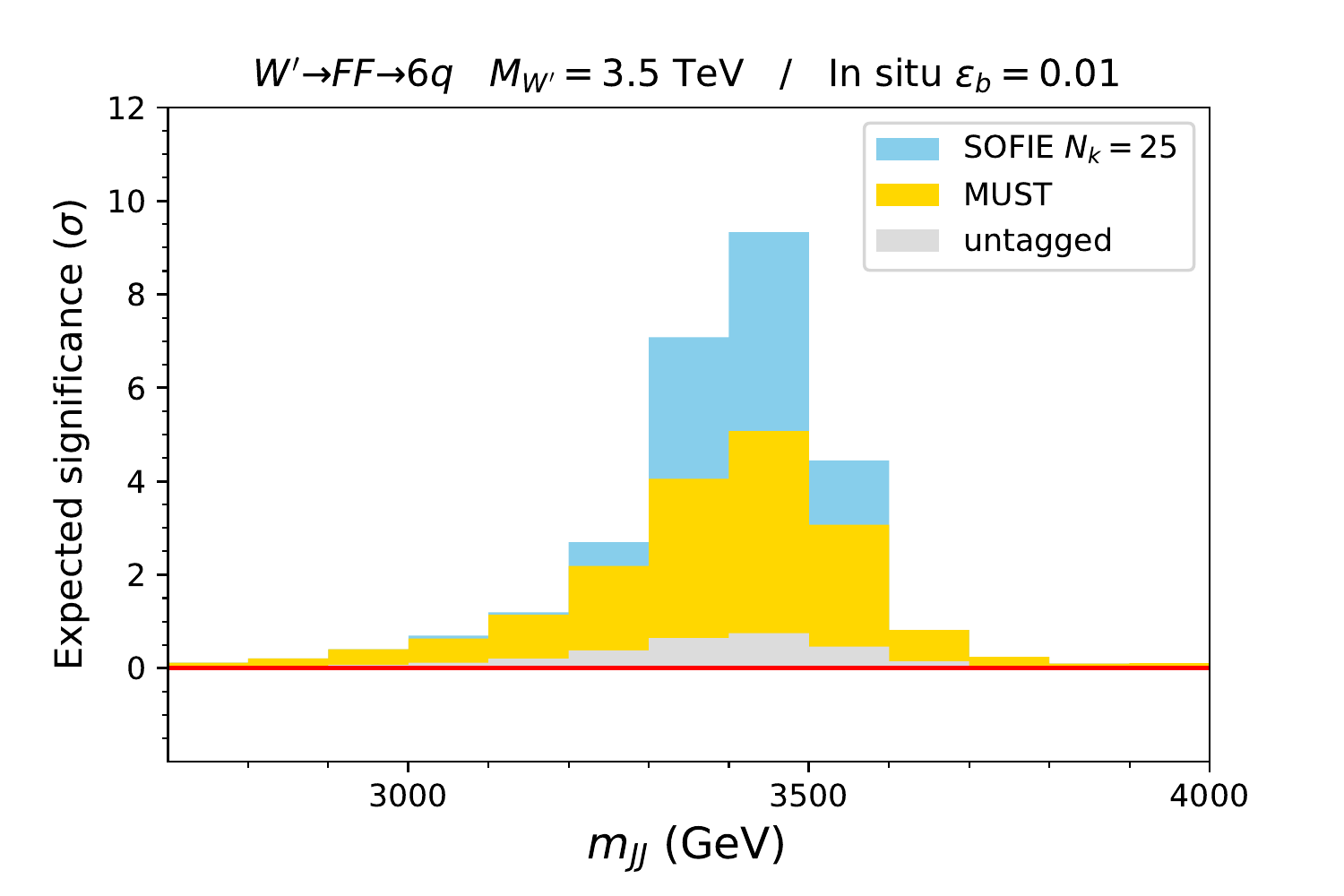}
\end{tabular}
\caption{Expected significance of the deviations per 100 GeV bin at three stages: (a) in the untagged event sample; (b) using only the MUST jet tagging; (c) using the full SOFIE anomaly detection method. In-situ mass decorrelation is used with $\varepsilon_b = 0.01$.}
\label{fig:signif-2body-3q}
\end{center}
\end{figure}

We first consider an example with a luminosity $L = 140~\text{fb}^{-1}$, normalising $n_\text{sig}/\sqrt n_\text{bkg} = 1$ before jet tagging as before. This normalisation implies $n_\text{sig}/ n_\text{bkg} = 1.1 \times 10^{-3}$. We use \gentx\ and in-situ decorrelation with $\varepsilon_b = 0.01$, and select a moderately small $N_k = 25$.\footnote{In this example the background event pools $\mathcal{P}_{B1}$ are made using events tagged with $\varepsilon_b = 0.05$, and for the scan with $\varepsilon_b = 0.1$, in order to increase the statistics.}  The results are presented in Fig.~\ref{fig:signif-2body-3q}. We find a significance improvement $s = 11.6$, around five times larger than the value obtained in Ref.~\cite{Collins:2021nxn} for CWoLa, and four times larger than with the autoencoder.

\begin{figure}[t]
\begin{center}
\begin{tabular}{c}
\includegraphics[width=9cm,clip=]{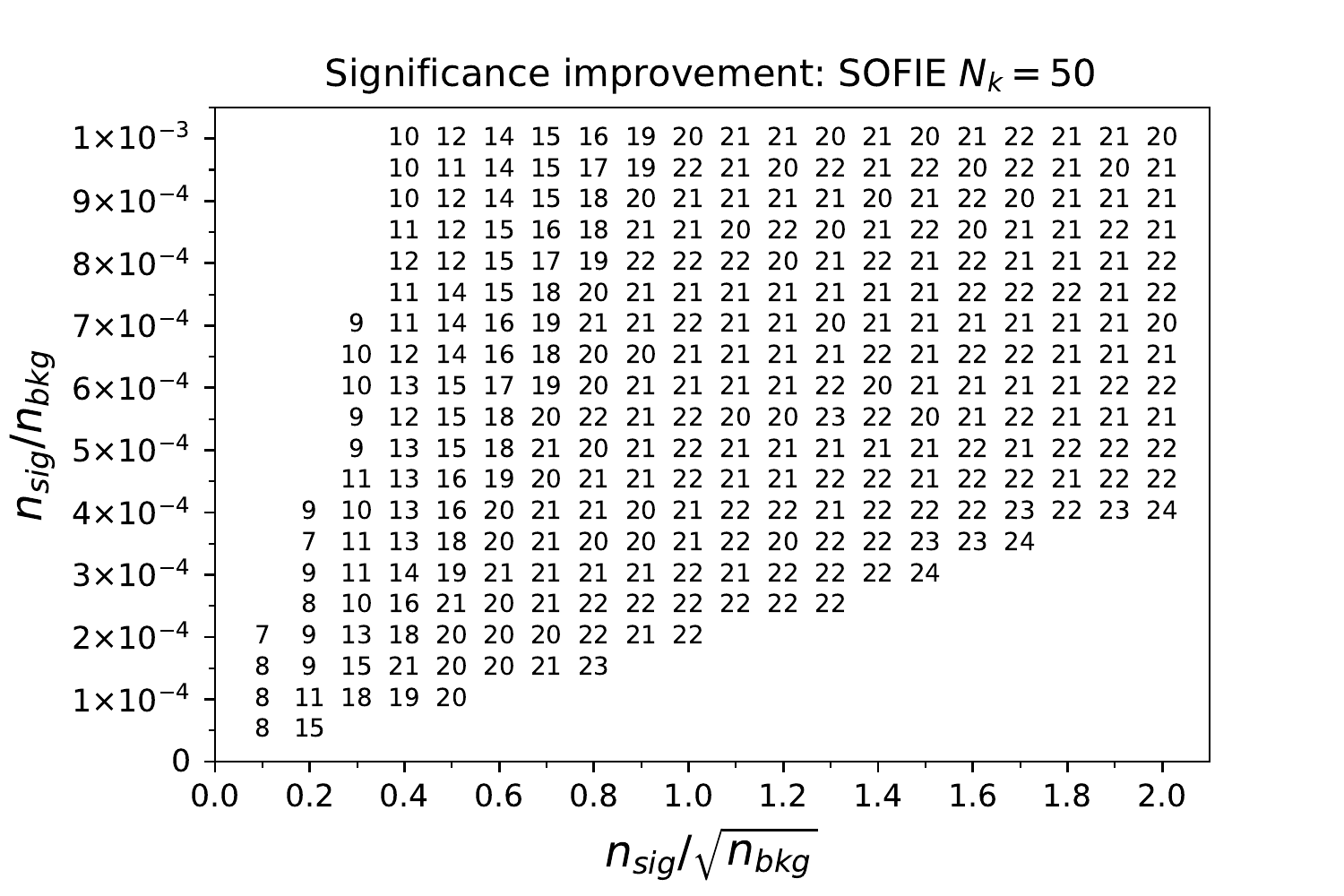} \\
\includegraphics[width=9cm,clip=]{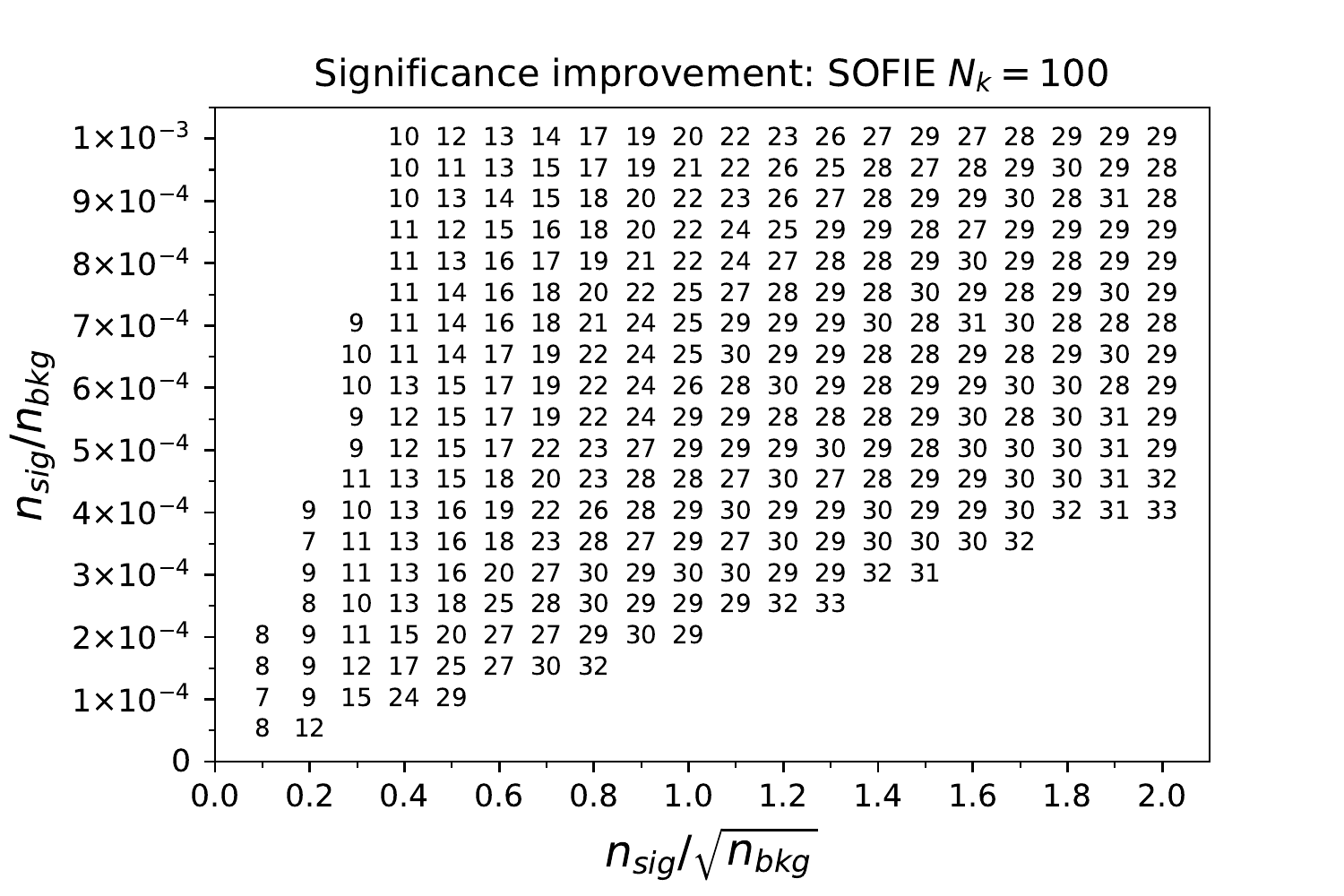} \\
\includegraphics[width=9cm,clip=]{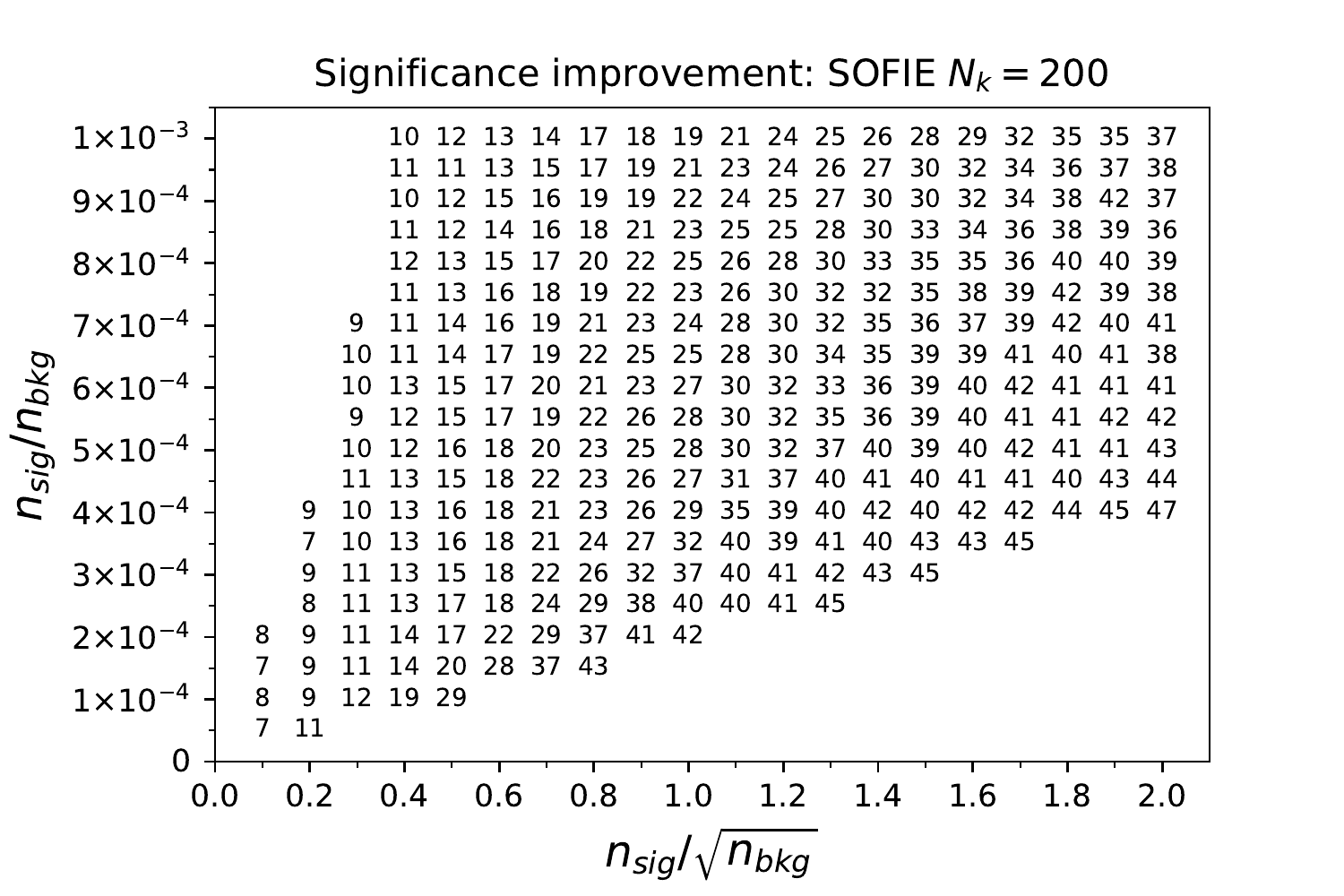}
\end{tabular}
\caption{Significance improvement as a function of $n_\text{sig}/\sqrt n_\text{bkg}$ and $n_\text{sig}/ n_\text{bkg}$, for in-situ mass decorrelation with $\varepsilon_b = 0.01$. }
\label{fig:scan}
\end{center}
\end{figure}

For better comparison with other results, we scan over different values of $n_\text{sig}/\sqrt n_\text{bkg}$ and $n_\text{sig}/ n_\text{bkg}$, but only concentrating on the bin with $m_{JJ} \in [3400,3500]$ GeV. In the scan we require $n \geq 20$ events and also $n \leq 2645$, which is the size of the background event pool $\mathcal{P}_{B1}$ used.
Results are presented in Fig.~\ref{fig:scan} for $N_k = 50$ (top), $N_k = 100$ (middle) and $N_k = 200$ (bottom). (For small samples we use a maximum value of $N_k = n/3$.) The upper left and down right corners of the plots are removed by the requirements $n \geq 20$ and $n \leq 2645$, respectively.

From this scan we observe that the significance improvement can be quite large, even for very small signal fractions in the sample, for which sideband-based methods do not perform well. For example, for signal ratios of $10^{-4}$ the significance enhancement can reach 29, a factor of $10-15$ larger than with CWoLa and the autoencoder. For larger signal ratios the significance improvement is larger than with those methods too, up to a factor of 50. This can be explained by the better jet discrimination using the supervised MUST tagger over simple variables (used by CWoLa) or an unsupervised tagger, together with the excellent performance of the mixed sample estimator to pinpoint mass bumps. The performance of the density estimation method ANODE~\cite{Nachman:2020lpy} is expected to be similar to that of CWoLa~\cite{Kasieczka:2021xcg}.

We finally note that for $N_k = 50$ we observe a performance saturation for larger samples (near the down right edge), showing that this value does not allow to exploit the full potential of the method. Indeed, the results with $N_k = 100$ and especially $N_k = 200$ are significantly better. On the other hand, in a handful of points around the down left corner, corresponding to small dataset, the results with $N_k = 50$ are better. This behaviour is not a problem, as nothing prevents experiments from using a couple of $N_k$ values, or a varying $N_k$.

\section{Extension to electroweak backgrounds}
\label{sec:7}

The anomaly search method presented can directly be applied to final states where the SM background involves jets that originate from either quarks or gluons. However, there might be cases where background contributions involving boosted hadronically-decaying $W$ or $Z$ bosons may not be neglected. In such case, the method can still be applied provided one includes this electroweak contribution (simulated by Monte Carlo) in the reference event sample, against which the data in the signal region is compared.

\begin{figure}[t]
\begin{center}
\begin{tabular}{c}
\includegraphics[width=9cm,clip=]{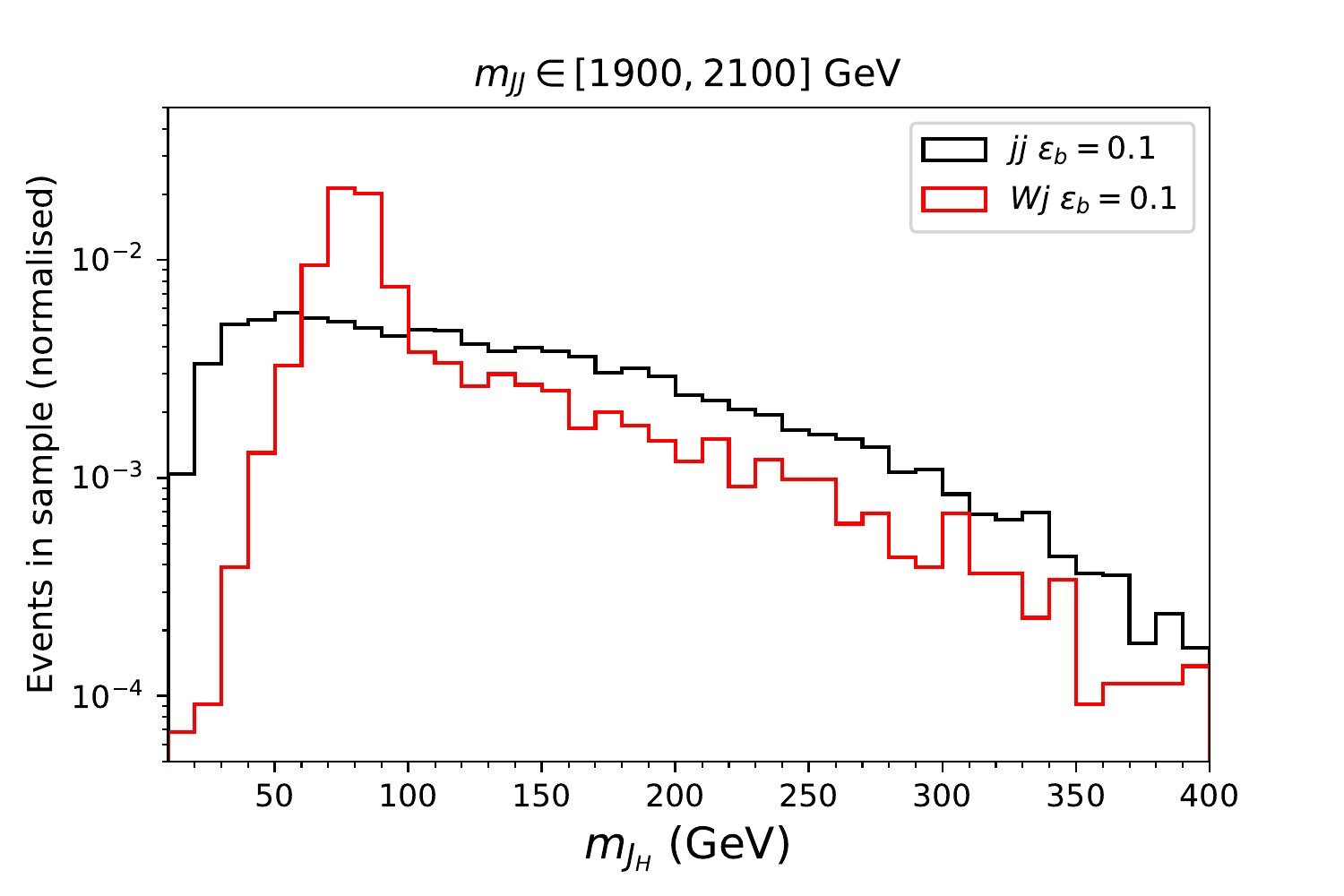} \\
\includegraphics[width=9cm,clip=]{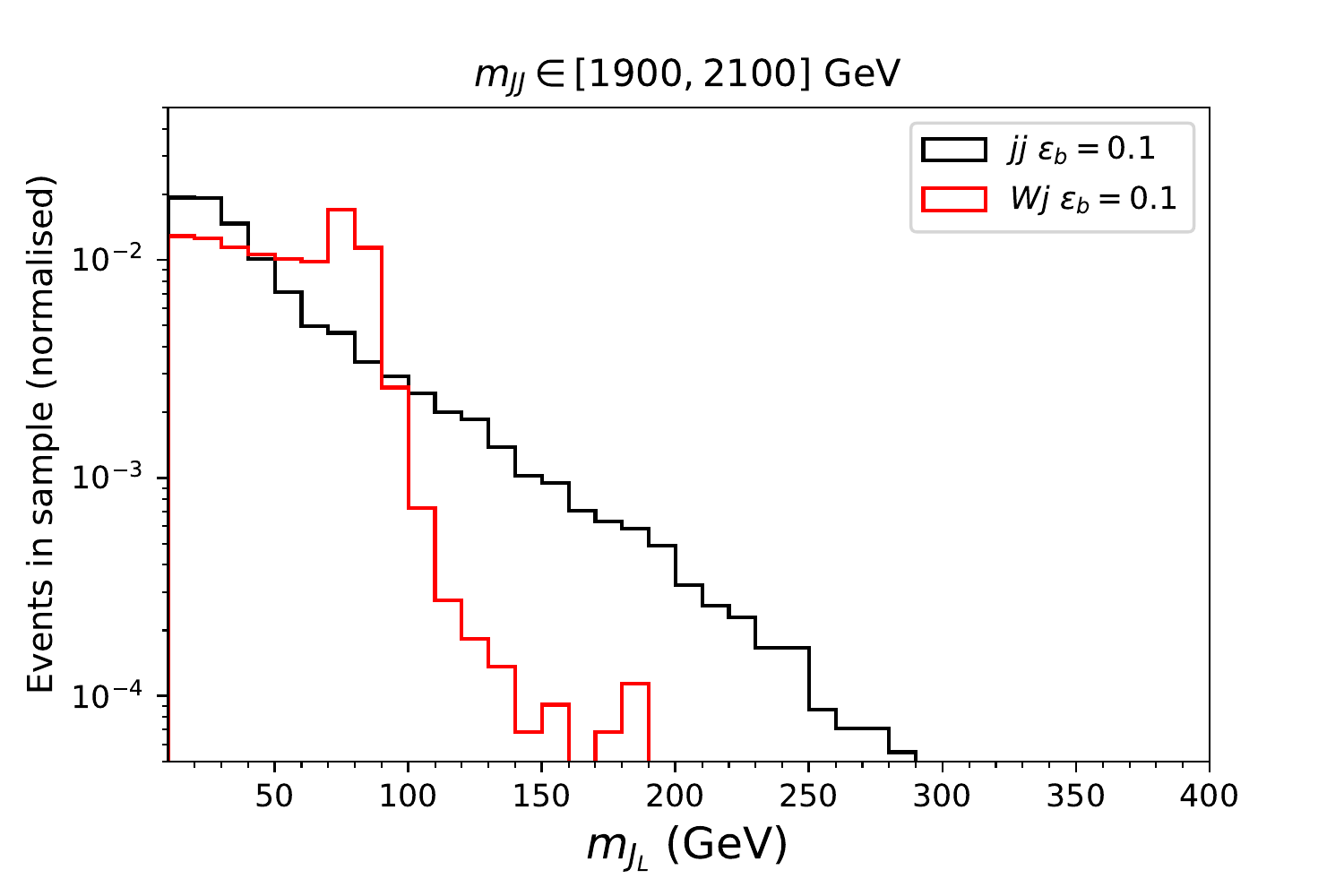}
\end{tabular}
\caption{Mass distribution of the two jets, ordered by mass, for $jj$ and $Wj$ events with invariant mass  in a narrow slice $m_{JJ} \in [1900,2100]$ GeV, after the application of the \gentx\ tagger and in-situ mass decorrelation with $\varepsilon_b = 0.1$.}
\label{fig:mJ-Wj}
\end{center}
\end{figure}

Let us take as example the dijet final state of section~\ref{sec:5} but injecting a $Wj$ contribution, with $W \to q \bar q$, two orders of magnitude larger than the actual one. In order to have sufficiently large event samples, we use the $\varepsilon_b = 0.1$ reference point for the \Gent\ tagger, with exclusive mass decorrelation, but of course the results are independent of this choice. We plot in Fig.~\ref{fig:mJ-Wj} the normalised distributions for the mass of the two jets, $j_H$ and $j_L$, ordered by mass, for events in the samples with $m_{JJ} \in [1900,2100]$ GeV. The distributions for $jj$ and  $Wj$ are radically different and for the latter the bump at the $W$ boson mass is clearly seen over the continuum distribution corresponding to the light jet. 

For our test we select events in this $m_{JJ}$ slice. For dijet events we use two event pools, with untagged and tagged jets. In actual experiments, the former would correspond to data from the reference sample (with a small contamination from other backgrounds, likely negligible) and the latter to  $jj$ events in the signal region. For $Wj$ events we use two event pools with both jets tagged. The first one would correspond to the Monte Carlo prediction, to be included in the reference sample, and the latter to $Wj$ events in the signal region.

\begin{figure}[t]
\begin{center}
\begin{tabular}{c}
\includegraphics[width=9cm,clip=]{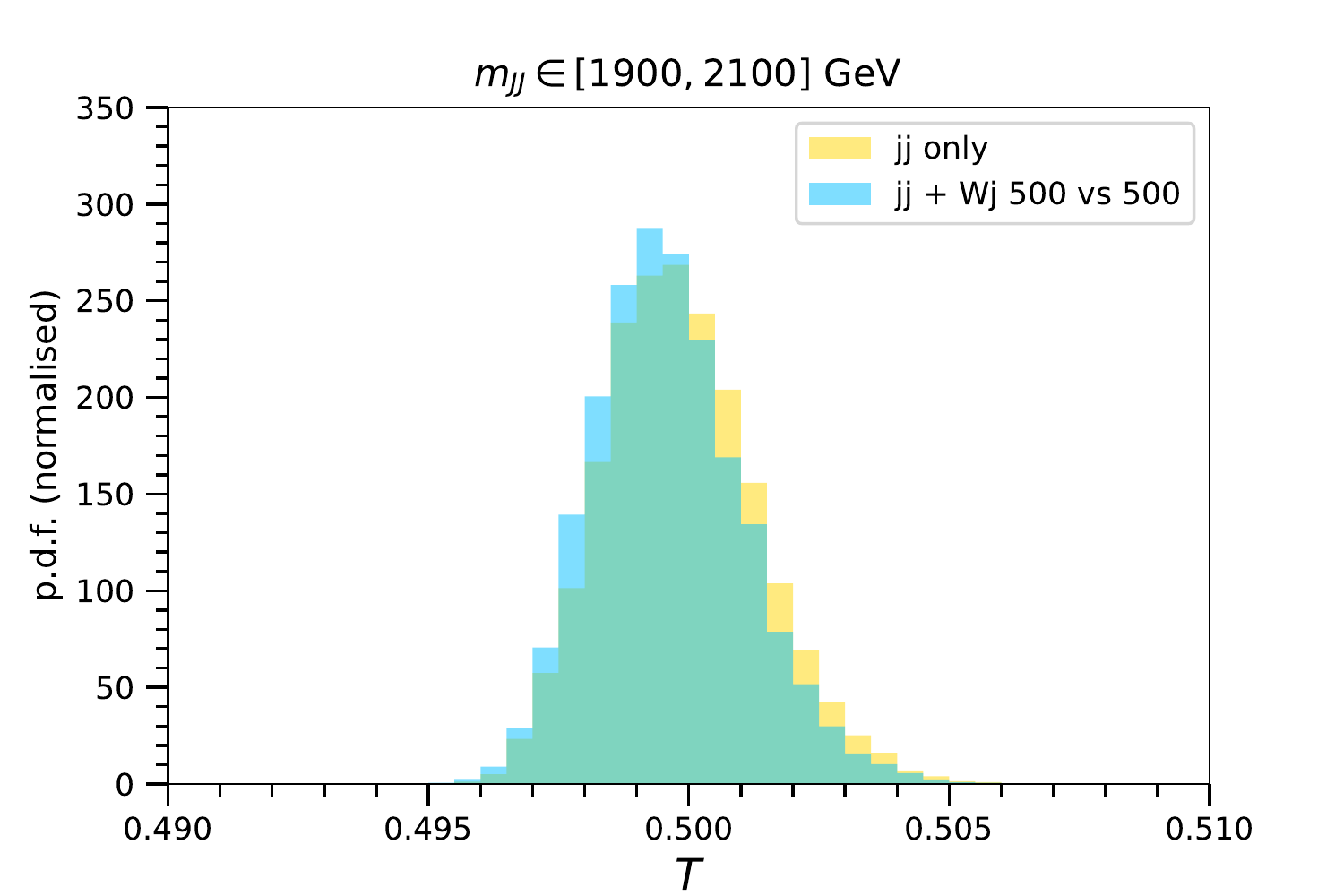} \\
\includegraphics[width=9cm,clip=]{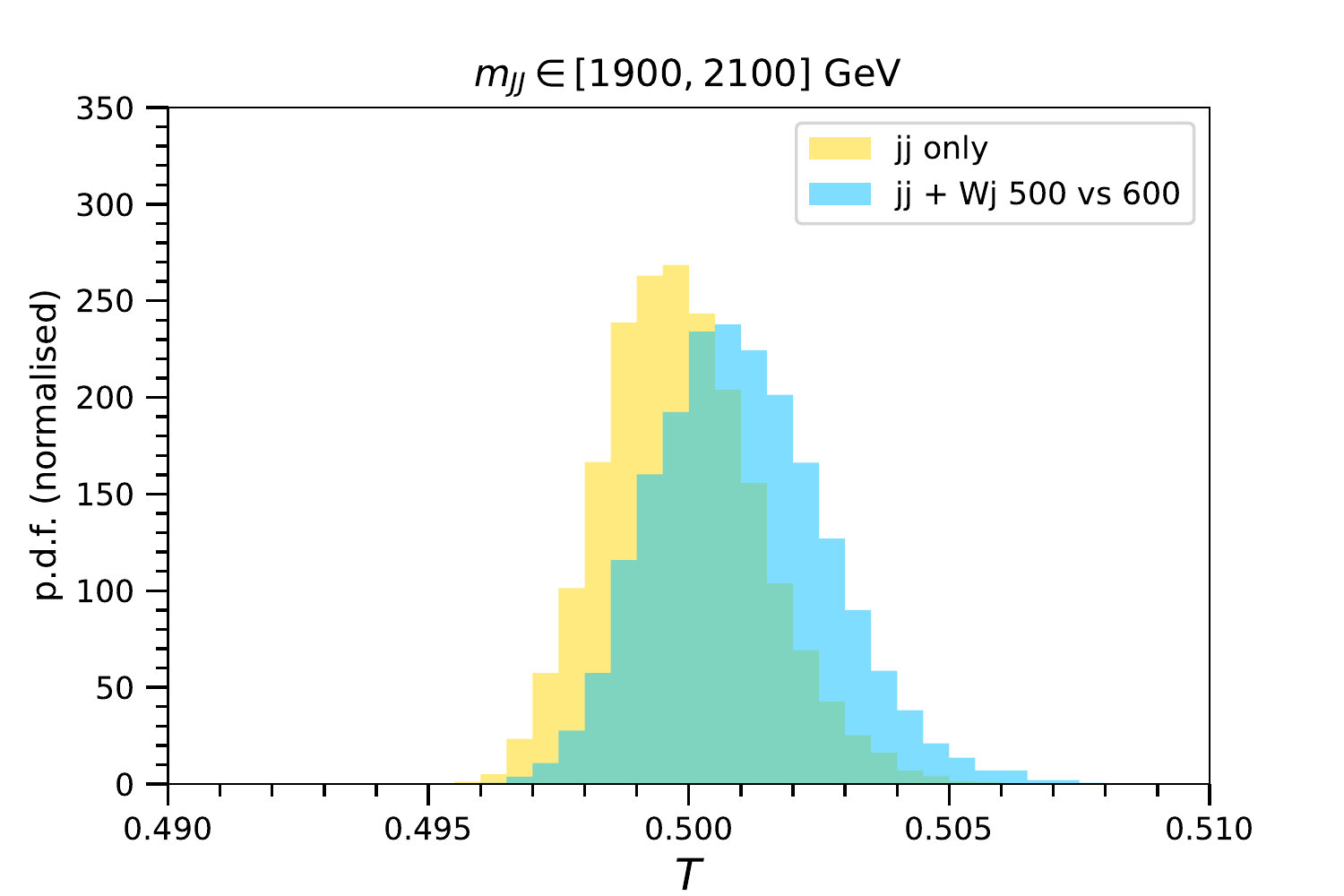}
\end{tabular}
\caption{Probability density functions for the estimator $T$, obtained when the two samples are drawn from the QCD jet pools (yellow) or also including $Wj$ events (blue). See the text for details.}
\label{fig:T-Wj}
\end{center}
\end{figure}

We use as variables the masses of the two jets, ordered by mass, as in section~\ref{sec:5}, and perform $N=10^4$ pseudo-experiments (a) drawing $n_A = n_B = 1000$ random $jj$ events from the untagged and tagged pools; (b) the same, but taking 500 $jj$ events from each pool, plus 500 $Wj$ events. The result is shown in Fig.~\ref{fig:T-Wj} (top). Both distributions agree between themselves, and with the expectation for the comparison between datasets with the same density distributions. (The small shift of the second distribution is merely statistical; a true difference in the density distributions would result in larger $\mu$ for the p.d.f. of $T$.)

A possible caveat to this procedure arises from the theoretical uncertainties in cross sections, and the statistical fluctuations that may change the background composition in the signal region with respect to the expectation from Monte Carlo. (Note, however, that the normalisation and even the shape of the samples containing $W/Z$ bosons can be tuned with measurements in the final states where the weak bosons decay leptonically.) We test the effect by comparing:
\begin{itemize}
\item[(a)] A reference sample with 500 untagged dijet events and 500 tagged $Wj$ events. 
\item[(b)] A sample with 400 tagged dijet events and 600 tagged $Wj$ events. This would correspond to data in the signal region, in which the fraction of $Wj$ events is $20\%$ larger than the expectation.
\end{itemize}
The resulting distribution is shown in Fig.~\ref{fig:T-Wj} (bottom), together with the distribution for $jj$ only. The mean of the distribution, $\mu = 0.501$, is shifted by a moderate $0.8\sigma$ from $\mu_0$. The effect is maximal when the expected fractions of $jj$ and $Wj$ are equal and, for example, if the fraction of $Wj$ events is 10\%, the shift of the mean is negligible.

An additional test has been performed by choosing, in each pseudo-experiment, a varying number of $Wj$ events obtained from a Gaussian distribution centered at the expected value of 500 events, with standard deviation $\sqrt{500}$, and keeping the total number of events fixed to $n = 1000$. The shift in $\mu$ turns out to be negligible as well. Therefore, from this exercise we can learn that the anomaly search method including additional electroweak backgrounds is quite robust against theoretical and statistical uncertainties in the sample composition.
The precise implementation the evaluation of uncertainties must be performed case by case.

\section{Discussion}
\label{sec:8}

The SOFIE anomaly detection method can be a useful addition to already existing proposals, and features two main advantages: (i) the high sensitivity when the signal to background ratio is small; (ii) its easy generalisation to include several kinematical quantities. 

We illustrate the potential of the method presented here in Table~\ref{tab:SIsumm}, summarising the significance improvement found for several benchmarks, using in-situ mass decorrelation. Unsupervised methods that use sidebands to spot differences between signal and background may have difficulties in scenarios where the signal is very small. We have compared in section~\ref{sec:6} our method with a couple of previous proposals, using a common benchmark $W' \to 6q$. A much large significance improvement is found with SOFIE for small signal fractions $n_\text{sig}/n_\text{bkg}$.  

\begin{table}[htb]
\begin{center}
\begin{tabular}{ccccc}
Signal   & Resonance mass                                       & $\varepsilon_b$ & $n_\text{sig}/ n_\text{bkg}$ & $s$ \\
$Z' \to 4W$ & 3.3 TeV  &  0.05                    & $2.4 \times 10^{-3}$ & 18.8 \\
$Z' \to 8b$  & 3.3 TeV  &  0.05                     & $1.9 \times 10^{-3}$ & 11.2 \\
$Z' \to 4W$ & 2.2 TeV  &  0.01                     & $5.9 \times 10^{-4}$ & 39.4 \\
$Z' \to 8b$  & 2.2 TeV   &  0.01                     & $5.2 \times 10^{-4}$ & 10.3 \\
$W' \to 6q$ & 3.5 TeV & 0.01                      & $1.1 \times 10^{-3}$ & 11.6 \\
\end{tabular}
\caption{Significance improvement $s$ in some of the examples analysed in sections~\ref{sec:5} and \ref{sec:6}.}
\label{tab:SIsumm}
\end{center}
\end{table}

Because the mass-decorrelated tagger preserves the kinematics, one can apply the SOFIE method for example to a final state with three particles, e.g. three massive jets, or a leptonically-decaying weak boson plus two jets, to search for triboson resonances~\cite{Aguilar-Saavedra:2015rna,Aguilar-Saavedra:2015iew,Agashe:2016kfr,Agashe:2017wss}. In this more complex final state, in addition to the jet masses, one could use as discriminant the invariant masses of pairs of two objects. 
One can easily realise that, with methods that use sidebands, the use of two-body invariant masses as variable is problematic, as the kinematical distributions may change from the potential signal region to the sidebands.

In this paper we do not address systematic uncertainties associated to the modeling. It is known that the description of the hadronisation and parton showering influences subjettiness observables, and this has some effect in the performance of a generic tagger, as shown in Ref.~\cite{Aguilar-Saavedra:2017rzt}. However, the MUST method significantly reduces this dependence: the degradation of the sensitivity to new physics signals is minimal~\cite{Aguilar-Saavedra:2022ejy}. In any case these uncertainties are not expected to create fake bumps in the SM background, as long as the mass decorrelation is calibrated on real data, or real data is used to tune the mass decorrelation performed with Monte Carlo simulation. The modeling of parton shower and hadronisation is also expected to improve, especially when independent measurements on quarks and gluons have become available~\cite{CMS:2021iwu}. We also note in passing that, since the extraction of the significance involves pseudo-experiments comparing random samples, the incorporation of systematic uncertainties in our method is easy, by changing parameters between pseudo-experiments. 

The application of the SOFIE method in actual searches would pinpoint kinematical regions where a significant deviation with respect to the SM prediction is found. A subsequent analysis of the kinematical variables used would show the origin of that deviation. In our example, examining the jet masses in the region of interest would show if there are bumps, and at which mass.

We finally remind the reader that the SOFIE method can only be applied to final states where at least one of the objects can be {\em tagged}. The method relies on selecting a subsample of `interesting' events from a larger `reference' sample, based on some quantity that discriminates usual objects from the more interesting ones. There are many final states and signatures that fall into this category. We have illustrated the method by applying it to massive boosted jets, in the simplest configuration (two boosted jets resulting from a heavy resonance decay). Further applications may include not only triboson resonances, but also searches for vector-like quarks, or in general any signal that involves massive boosted particles. And there are other types of objects, for example with displaced vertices, which can be subject to an analogous treatment in a search for long-lived particles.

\section*{Acknowledgements}

I thank  F.R. Joaquim and J. Seabra for previous colaboration in the MUST development, and J.H. Collins and D. Shih for advice in the generation of Monte Carlo events of Ref.~\cite{Collins:2021nxn}. This work has been supported by MICINN project PID2019-110058GB-C21 and by FCT project CERN/FIS-PAR/0004/2019.

\appendix
\section{Monte Carlo samples}
\label{sec:a}

In order to have sufficient Monte Carlo statistics across a wide range of energies, the backgrounds are generated in narrow slices of $p_T$ for some of the final state particles. (If there are only two particles, their $p_T$ is the same at LO and at the generation level.) When considering some kinematical distribution such as the dijet invariant mass, the different samples generated at the partonic level are combined with a weight proportional to the cross section. That is not required when performing sensitivity tests of the $T$ estimator in narrow $p_T$ intervals, however.

Dijet samples are generated with {\scshape MadGraph}, in 100 GeV bins of $p_T$, from $[200,300]$ GeV to $[2.2,2.3]$ TeV. (The last bin is $p_T \geq 2.2$ TeV for inclusive dijet production.) Several samples are generated for $gg$, $qq$ and inclusive dijet production, which are used for training of the taggers, mass decorrelation, and tests. The statistics of the different samples is the following:
\begin{itemize}
\item Gluon set 1 (GS1): $4 \times 10^5$ $gg$ events per 100 GeV slice of $p_T$
\item Gluon set 2 (GS2): the same size
\item Quark set 1 (QS1): $4 \times 10^5$ $qq$ events per 100 GeV slice of $p_T$
\item Quark set 2 (QS2): the same size
\item Jet set 1 (JS1): $8 \times 10^5$ $jj$ events per 100 GeV slice of $p_T$
\item Jet set 2 (JS2): $6 \times 10^5$ $jj$ events per 100 GeV slice of $p_T$
\end{itemize}
Both jets in the events are used if they satisfy the event selection constraints required. Additional test samples are used for tests of the tagger performance, which are the same ones used in Refs.~\cite{Aguilar-Saavedra:2017rzt,Aguilar-Saavedra:2020uhm} for better comparison. They are generated in the processes $pp \to Zg$, $pp \to Zq$, with $Z \to \nu \bar \nu$, with a lower cut on jet $p_T$ at the parton level. The samples used in this paper are:
\begin{itemize}
\item G1.0\_80: $p_T \geq 1$ TeV at parton level; 356000 gluon jets with $\ptj \geq 1$ TeV, $m_J \in [60,100]$ GeV.
\item Q1.0\_80: $p_T \geq 1$ TeV at parton level; 304000 quark jets with $\ptj \geq 1$ TeV, $m_J \in [60,100]$ GeV.
\item G1.5\_175: $p_T \geq 1.5$ TeV at parton level; 136000 gluon jets with $\ptj \geq 1.5$ TeV, $m_J \in [150,200]$ GeV.
\item Q1.5\_175: $p_T \geq 1.5$ TeV at parton level; 95000 quark jets with $\ptj \geq 1.5$ TeV, $m_J \in [150,200]$ GeV.
\item G1.5\_200: $p_T \geq 1.5$ TeV at parton level; 196000 gluon jets with $\ptj \geq 1.5$ TeV, $m_J \in [160,240]$ GeV.
\item Q1.5\_200: $p_T \geq 1.5$ TeV at parton level; 134000 quark jets with $\ptj \geq 1.5$ TeV, $m_J \in [160,240]$ GeV.
\end{itemize}
The numbers of jets quoted are those satisfying the constraint on jet mass and $p_T$ after simulation.

$Wj$ production is simulated for various tests, in 100 GeV bins of $p_T$, starting at $[200,300]$ GeV and with the last bin $p_T \geq 2.2$ TeV. The statistics is $2 \times 10^5$ events per bin. 

The $Z'$ signals in sections~\ref{sec:2} and \ref{sec:5} are generated with {\scshape MadGraph}. The model of Ref.~\cite{Aguilar-Saavedra:2019adu} is implemented in {\scshape Feynrules} ~\cite{Alloul:2013bka} and interfaced to {\scshape MadGraph} using the universal Feynrules output~\cite{Degrande:2011ua}. The signal samples contain $2 \times 10^{5}$ events. In the signals with $S_{1,2} \to WW$, only hadronic decays of the $W$ boson are considered.
 The $W'$ signal in section~\ref{sec:6} is generated with {\scshape Pythia}, according to the procedure followed in Ref.~\cite{card}.

\section{Effect of a fixed threshold}
\label{sec:b}

\begin{figure}[t]
\begin{center}
\begin{tabular}{c}
\includegraphics[width=9cm,clip=]{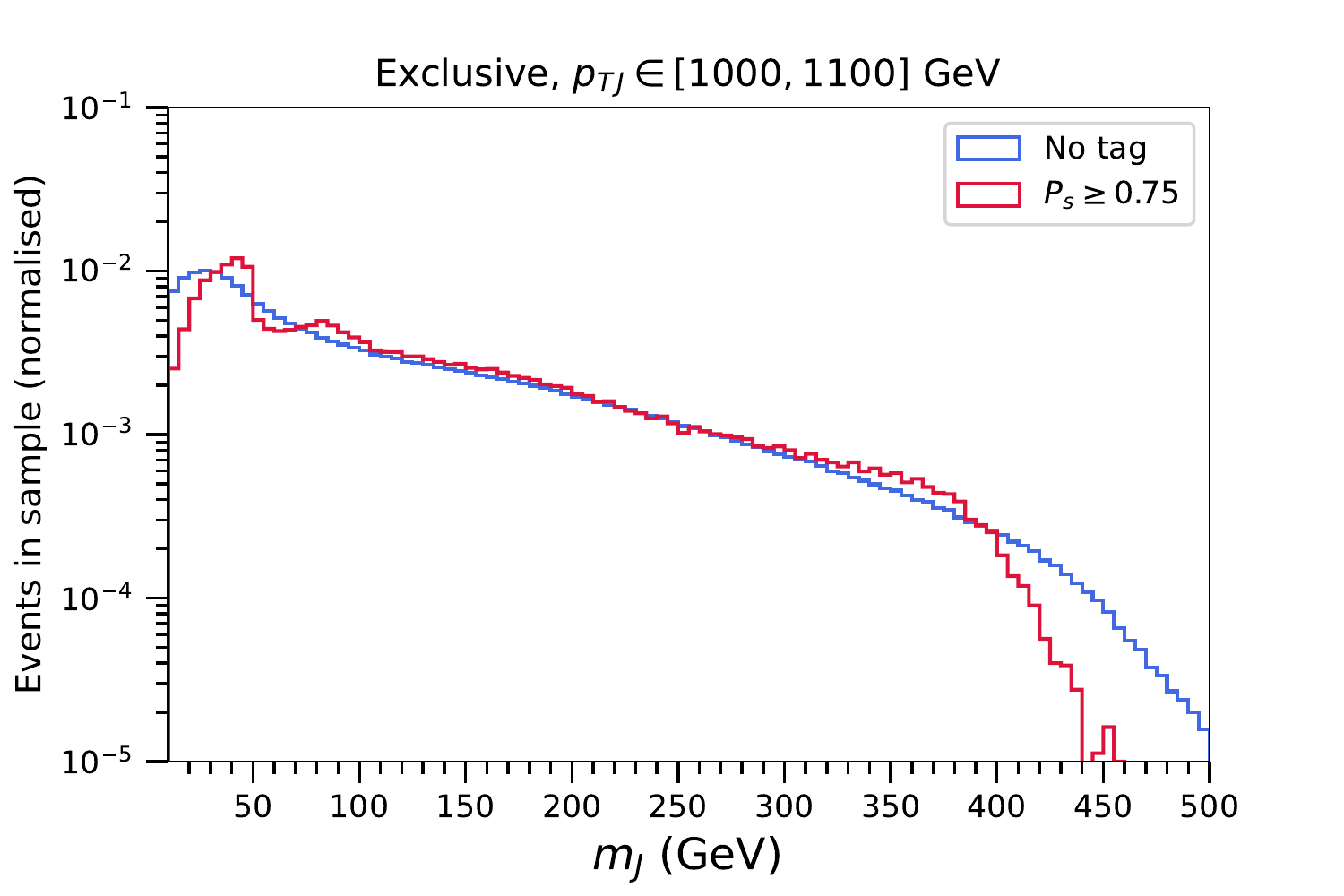} \\
\includegraphics[width=9cm,clip=]{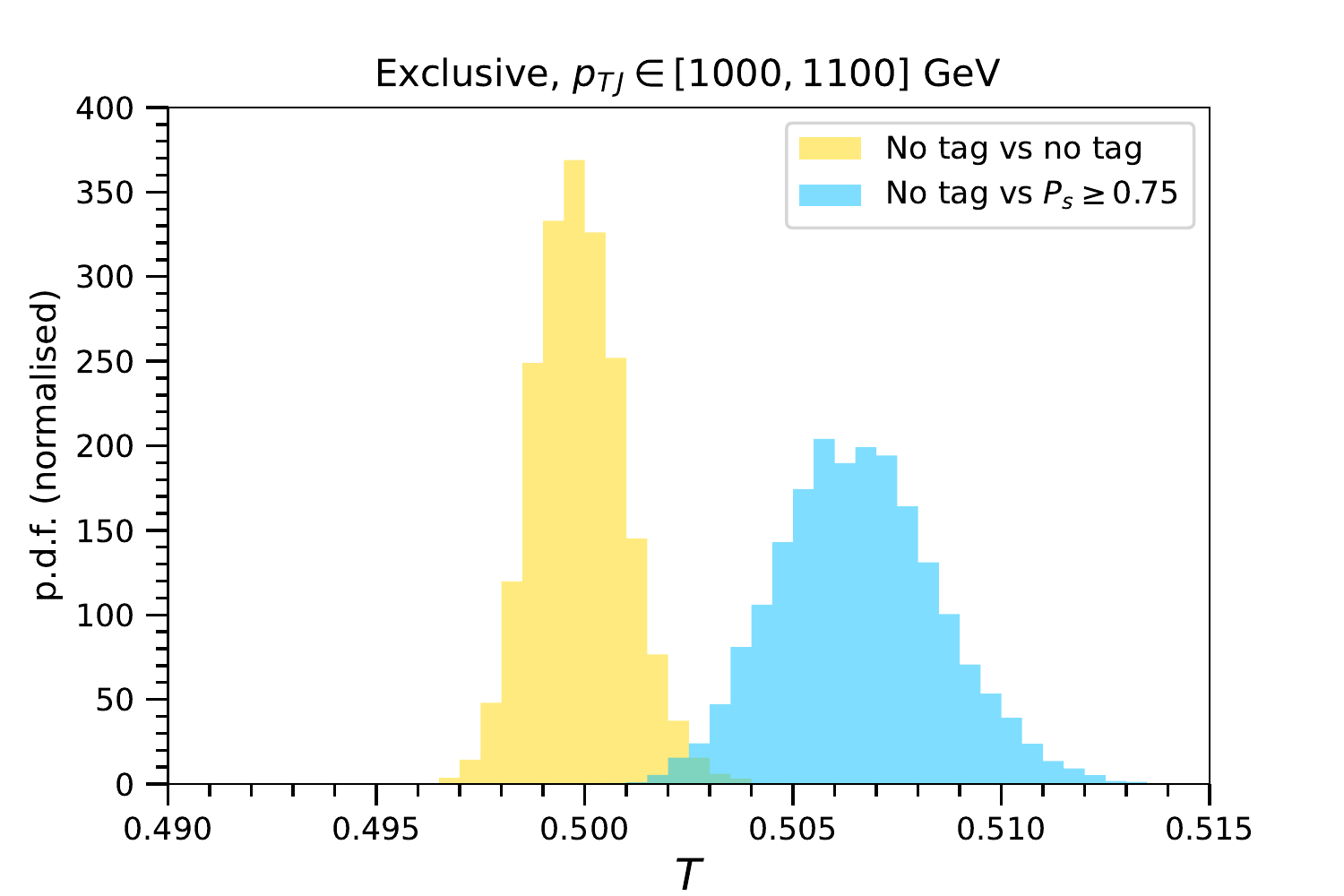}
\end{tabular}
\caption{Top: Jet mass distribution for QCD jets with $p_T \in [1000,1100]$ GeV, without cuts and after a fixed cut $P_s \geq 0.75$ with the \Gent\ tagger. Bottom: Probability density functions for the estimator $T$, obtained when the two samples are drawn from the untagged jet pool (yellow) or from the untagged pool of jets tagged with fixed threshold $P_s \geq 0.75$ (blue).}
\label{fig:mJF}
\end{center}
\end{figure}

A fixed cut on the NN output distorts the jet mass distribution even with a mass unspecific tagging, in which the jet mass and $p_T$ are variables that vary across wide ranges in the training sets. Concentrating on the tagger \Gent\, we display this effect in the top panel of Fig.~\ref{fig:mJF} for jets with $\ptj \in [1000,1100]$ GeV and a fixed cut $P_s \geq 0.75$. This cut approximately has an efficiency of 0.05 for the full sample. The distribution can be compared with the corresponding plot in Fig.~\ref{fig:dec3} for a mass-decorrelated tagging with about the same efficiency. The prominent peaks and dips in the low side of the distribution generate a large shift in the mean of the $T$ estimator, which can be seen in the bottom panel of Fig.~\ref{fig:mJF}, which amounts to $6.2\sigma$. Large deviations of $15.5\sigma$ and $4.7\sigma$ are obtained in the $[500,600]$ and $[2000,2100]$ bins applying the cut $P_s \geq 0.75$, which has an efficiency around 0.05 in these $p_T$ bins too, and similar figures are obtained for the \gentx\ tagger.

This exercise also puts in appropriate perspective the small $\lesssim 0.3\sigma$ shifts in the $T$ estimator found with the mass-decorrelated tagging in section~\ref{sec:4}. The estimator with $n = 2000$ and $N_k = 100$ is really sensitive to details of the distributions, and shifts of $0.3\sigma$ and smaller indicate that the distributions before and after tagging agree very well.

\section{Dependence of the sensitivity on $N_k$}
\label{sec:c}

\begin{figure*}[t!]
\begin{center}
\begin{tabular}{cc}
\includegraphics[width=9cm,clip=]{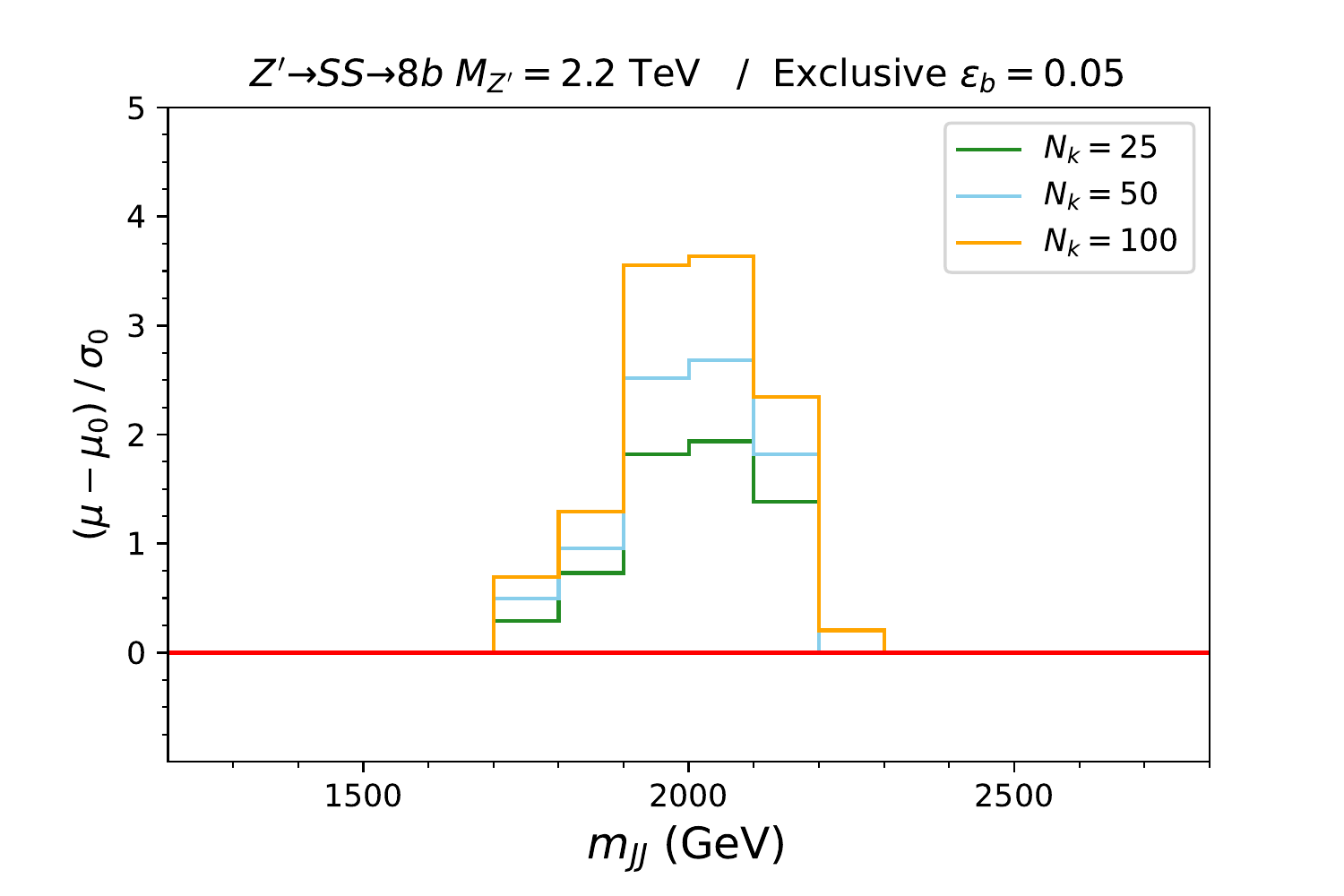} &
\includegraphics[width=9cm,clip=]{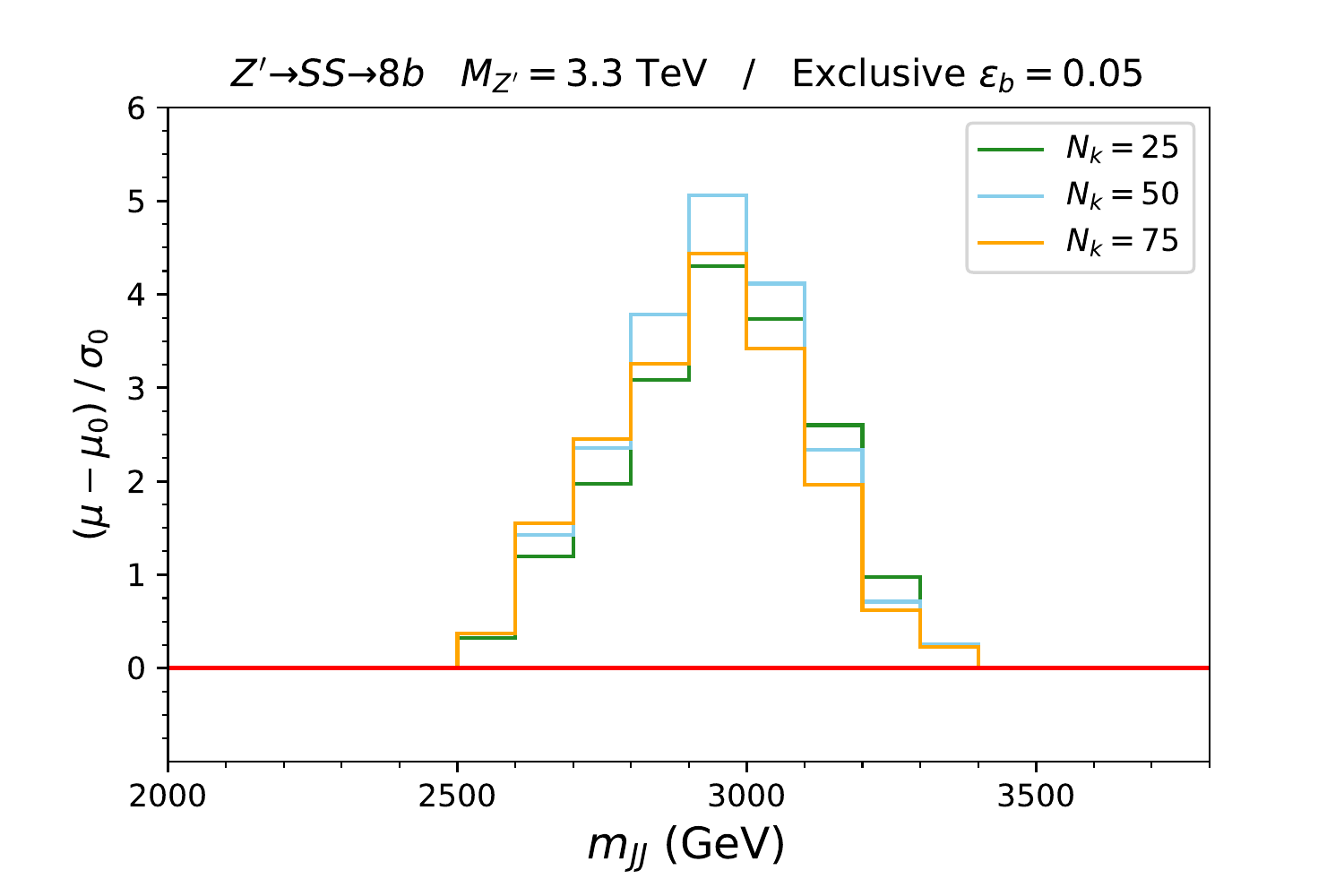} \\
\includegraphics[width=9cm,clip=]{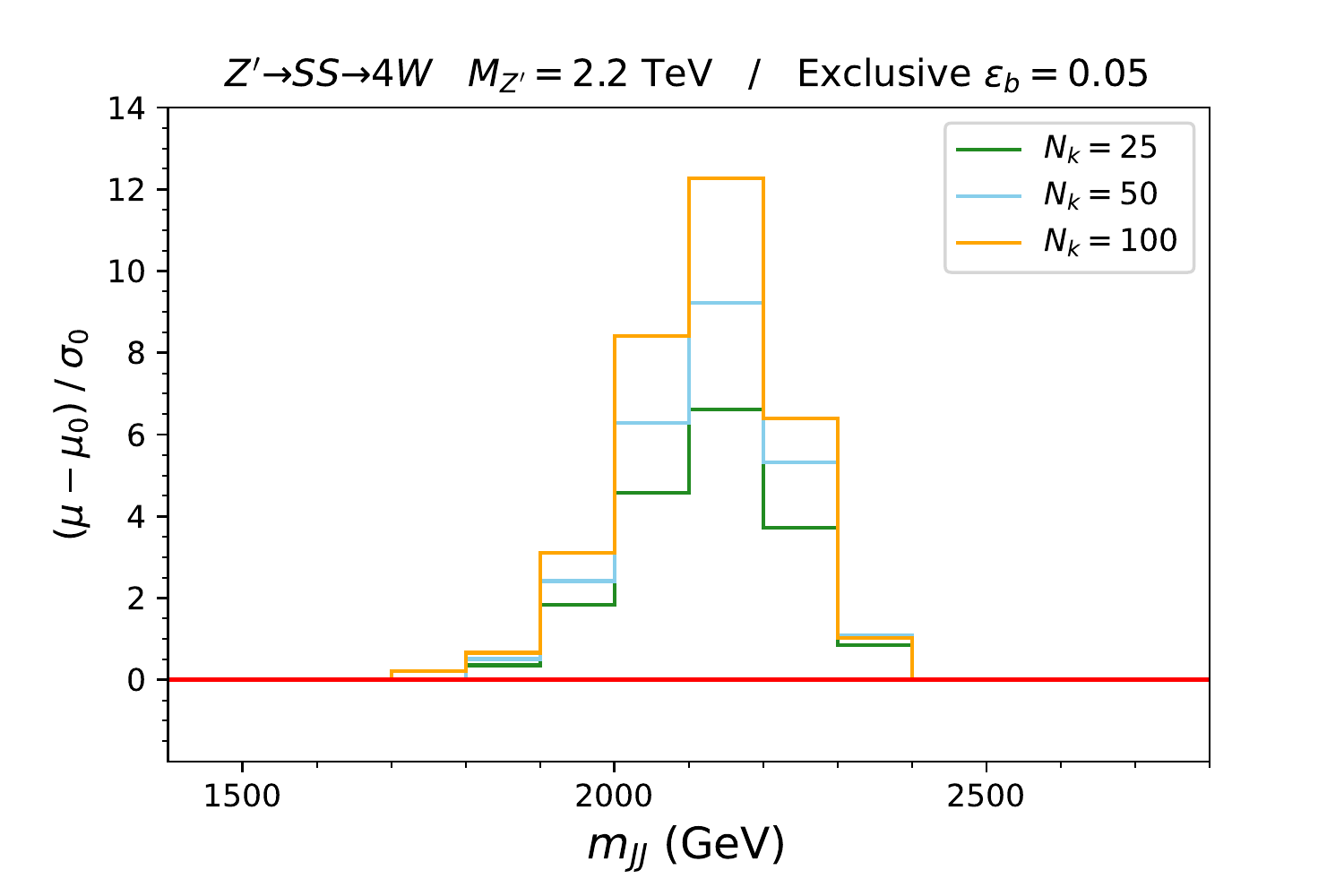} &
\includegraphics[width=9cm,clip=]{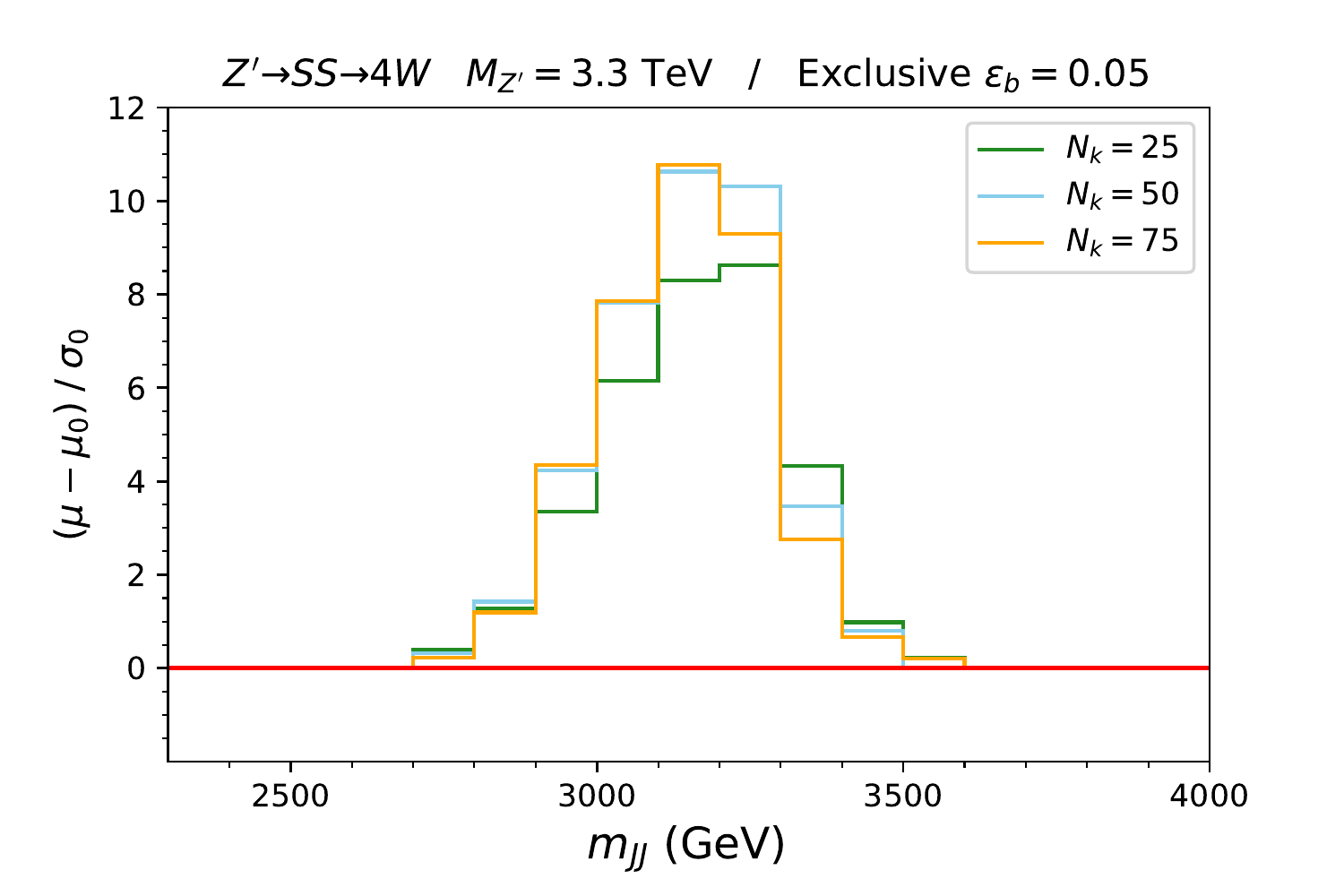} 
\end{tabular}
\caption{Expected sensitivity (per 100 GeV bin) to $Z'$ signals using the $T$ estimator alone in the four benchmarks described in section~\ref{sec:5}, for several values of $N_k$.}
\label{fig:pullT-2body}
\end{center}
\end{figure*}

The optimal value of $N_k$ to maximise the sensitivity to sample `contaminations', i.e. the presence of a signal with a density distribution different from the reference background, depends not only on the size of the sample, but also on the signal present. In our calculations of section~\ref{sec:5} we have used several fixed values of $N_k$ to obtain the expected sensitivity $p_2 = (\mu - \mu_0)/\sigma_0$, which we present in Fig.~\ref{fig:pullT-2body} for the four benchmark scenarios studied. There are a few points that deserve some discussion.
\begin{itemize}
\item For the 2.2 TeV signals he number of background events in the interesting region remains large after tagging, for example around 2400 events with $m_{JJ} \in [2,2.1]$ TeV. The signal, with the chosen normalisation, is around 100 events. Then, an improvement is expected with larger $N_k$.
\item Notably, even in bins where $n_\text{sig} < N_k$ there is some improvement in the sensitivity with larger $N_k$, namely in the scenario with $Z' \to 8b$, in the bins $[1.7,1.8]$ TeV and $[2.1,2.2]$ TeV. 
\item For the 3.3 TeV signals the number of background events in the interesting region is quite smaller, for example 170 events with $m_{JJ} \in [3,3.1]$ TeV. For the $Z' \to 4W$ signal the sensitivity almost saturates with $N_k = 50$, and for the $Z' \to 8b$ signal the sensitivity is worse with $N_k = 75$.
\end{itemize}
In conclusion, we find again that a larger $N_k$ often provides more sensitivity, up to some saturation point where the sensitivity decreases.
In actual searches, $N_k$ could be a selectable parameter (like the tagger background rejection) that could even be adjusted bin-by-bin, performing the search for a few $N_k$ values.

\end{document}